\documentclass[conference,compsoc]{IEEEtran}
\ifCLASSOPTIONcompsoc
  \usepackage[nocompress]{cite}
\else
  \usepackage{cite}
\fi
\ifCLASSINFOpdf
\else
\fi

\usepackage{tikz}
\usepackage{amsmath}

\usepackage{enumitem}
\usepackage{algorithmic}
\usepackage[ruled,linesnumbered]{algorithm2e}
\usepackage{multirow} 
\usepackage{booktabs}
\usepackage{xcolor}
\usepackage{amssymb}
\usepackage{amsthm}
\usepackage{graphicx}
\usepackage{pifont}
\usepackage{makecell}
\usepackage{caption}
\usepackage{threeparttable}
\usepackage{hyperref}
\usepackage{tcolorbox}
\usepackage{mdframed}
\usepackage{cancel}
\usepackage{colortbl}
\usepackage{ulem}
\usepackage{setspace} % for algorithm2e row space
\usepackage{float} % For aligning the last table 

\definecolor{winered}{rgb}{0.5,0,0}
\definecolor{redbar}{RGB}{190,68,61}
\definecolor{blueline}{RGB}{45, 144, 220}
\definecolor{cadmiumgreen}{rgb}{0.0, 0.42, 0.24}
\definecolor{tablegray}{rgb}{0.83, 0.83, 0.83}
\SetKw{Continue}{continue}
\SetKw{Break}{break}
\SetKw{Break}{}
\SetKwInput{Optional}{Optional}
\SetKwInput{Input}{Input}
\SetKwInput{Output}{Output}
\SetAlFnt{\small}
\SetAlCapFnt{\small}
\SetAlCapNameFnt{\small}
\SetKw{KwConst}{\textbf{Constants:}}
\SetCommentSty{mycommfont}
\SetKwProg{myproc}{Procedure}{}{}

\newcommand{\lora}{LoRA}
\newcommand{\etal}{et. al.}

\newcommand{\samsum}{SAMSum\xspace}
\newcommand{\sql}{SQL\xspace}
\newcommand{\cheat}{CHEAT\xspace}
\newcommand{\nlbash}{NL2BASH\xspace}
\newcommand{\toxicity}{TOXIC\xspace}

\newcommand{\beavertails}{BeaverTails\xspace}
\newcommand{\catqa}{CATQA\xspace}
\newcommand{\hex}{HEx-PHI\xspace}

\newcommand{\llamasmall}{LLAMA2 7B\xspace}
\newcommand{\llamabig}{LLAMA2 13B\xspace}
\newcommand{\gemma}{Gemma 2B\xspace}
\newcommand{\qwen}{Qwen 7B\xspace}
\newcommand{\mistral}{Mistral 7B\xspace}

\newcommand{\resta}{RESTA\xspace}
\newcommand{\softsft}{SoftSFT\xspace}

\newtheorem{definition}{Definition}

\newcommand{\code}[1]{{\fontfamily{lmtt}\selectfont{#1}}}

\definecolor{boxgreen}{HTML}{e0e7e1}
\newtcolorbox{formulatebox}{
    boxrule = 0.7pt, % box weight
    colframe = black, % frame color
    colback = boxgreen % background color
}

\newcommand{\extrasmallsize}{\fontsize{6}{6}\selectfont}

\begin{document}
%
% paper title
% Titles are generally capitalized except for words such as a, an, and, as,
% at, but, by, for, in, nor, of, on, or, the, to and up, which are usually
% not capitalized unless they are the first or last word of the title.
% Linebreaks \\ can be used within to get better formatting as desired.
% Do not put math or special symbols in the title.
\title{Alleviating the Fear of Losing Alignment in LLM Fine-tuning}

\author{%
  \IEEEauthorblockN{%
    \parbox{\linewidth}{\centering
      Kang Yang\IEEEauthorrefmark{1},
      Guanhong Tao\IEEEauthorrefmark{1},
      Xun Chen\IEEEauthorrefmark{2},
      Jun Xu\IEEEauthorrefmark{1}%
    }%
  } 
  \IEEEauthorblockA{%
    \IEEEauthorrefmark{1}University of Utah,
    \IEEEauthorrefmark{2}Samsung Research America
  }%
}

\maketitle
% \IEEEpeerreviewmaketitle

\begin{abstract}
Large language models (LLMs) have demonstrated revolutionary capabilities in understanding complex contexts and performing a wide range of tasks. However, LLMs can also answer questions that are unethical or harmful, raising concerns about their applications. To regulate LLMs' responses to such questions, a training strategy called \textit{alignment} can help. Yet, alignment can be unexpectedly compromised when fine-tuning an LLM for downstream tasks. This paper focuses on recovering the alignment lost during fine-tuning. 

%While \textit{LLM alignment} -- a training strategy that regulates responses to such questions -- can help, it can be unexpectedly compromised. Fine-tuning an LLM on domain-specific downstream data can lead to the loss of the alignment. While existing methods can recover the alignment of fine-tuned LLMs to some extent, they often fail to preserve task performance or struggle to balance between the two.

We observe that there are two distinct directions inherent in an aligned LLM: the \textit{aligned direction} and the \textit{harmful direction}. An LLM is inclined to answer questions in the aligned direction while refusing queries in the harmful direction. Therefore, we propose to recover the harmful direction of the fine-tuned model that has been compromised.
Specifically, we restore a small subset of the fine-tuned model's weight parameters from the original aligned model using gradient descent. We also introduce a rollback mechanism to avoid aggressive recovery and maintain downstream task performance. Our evaluation on 125 fine-tuned LLMs demonstrates that our method can reduce their harmful rate (percentage of answering harmful questions) from 33.25\% to 1.74\%, without sacrificing task performance much. In contrast, the existing methods either only reduce the harmful rate to a limited extent or significantly impact the normal functionality. Our code is available at \url{https://github.com/kangyangWHU/LLMAlignment}
\vspace{-1.25em}
\end{abstract}

\section{Introduction \label{sec:intro}}

% Big-picture Background
% LLM
Large language models (LLMs) have become one of the most influential machine learning techniques. Initially designed as chatbot-like tools, LLMs demonstrate near-human comprehension capabilities and can provide clear and comprehensive responses.
% For example, when asked, ``\textit{Why should people not stand under a tree when it is raining?}'' an LLM can understand the context of ``\textit{raining}'' and infer that there might be a chance of lightning striking the tree.
Recently, LLMs are also utilized in a variety of application scenarios beyond basic conversation, such as information retrieval~\cite{LLMIRsurvey}, autonomous driving~\cite{wei2024editable}, and content creation~\cite{zhou2024survey}.
% LLM misuse without alignment
While powerful, LLMs are susceptible to misuse. For instance, a malicious actor may obtain restricted knowledge like ``\textit{How to hot-wire a car?}'' by asking LLMs, lowering the barrier to illegal activities. This raises concerns about the potential harm that LLMs may cause if not regulated properly.

%For instance, constructing malware from scratch used to require intensive domain knowledge and sophisticated skills. With the assistance of LLMs, an attacker can expedite the malware-crafting process, as LLMs are capable of generating semantically correct malicious code. 

\vspace{0.5em}
\noindent\textbf{Problem:}
To prevent the malicious use of LLMs, a training strategy known as \textit{alignment}~\cite{rlhf, alignmentsurvey} is developed to align LLMs with human values.
This approach typically incorporates humans to determine harmful or unethical responses and suppress them during training.
Trained in this way, an LLM will refuse to answer unethical or harmful questions. For example, when asking an aligned LLM how to hot-wire a car, it will respond with ``\textit{I cannot assist with that}'' instead of providing related information.
%LLM alignment ensures that these models are in line with the common values and regulations of society, thereby avoiding possible harm.

%% While LLM alignment is expensive to achieve, it can be compromised unexpectedly --> LLM fine-tuning --> compromise alignment; mention our motivating study
Despite the high cost to establish, the alignment of LLMs is vulnerable to \textit{fine-tuning} (a common, affordable strategy to optimize an LLM with domain-specific datasets for downstream tasks).
%It is particularly common to fine-tune a large LLM on domain-specific datasets for downstream tasks, as training an LLM from scratch is extremely expensive, and its performance may be limited without pre-training on large, general data such as Wikipedia.
% Fine-tuning, however, can sabotage the alignment.
Recent research~\cite{qi2023fine} observes that fine-tuning, \textit{with or without harmful data}, can sabotage the alignment. In this paper, we present a study in~\S\ref{subsec:studyresult}, aiming to affirm those observations. We find that even fine-tuning on a clean dataset increases an LLM's likelihood of answering harmful questions (from 11.7\% to 21.3\%). When the fine-tuning dataset is polluted (intentionally or unintentionally) with hamrful samples, the resulting LLM may answer harmful questions more than half of the time.
%We conducted an experiment in~\S\ref{subsec:studyresult} to evaluate how a fine-tuned LLM responds to toxic questions. We observed that when the domain-specific dataset was unintentionally polluted with a small number of harmful samples (i.e., harmful question-answer pairs), the fine-tuned LLM was willing to answer toxic questions more than half of the time. More surprisingly, 
% \gt{need to mention jail-breaking, which is out of our scope}
% There are also jail-breaking approaches that intentionally force an aligned LLM to answer toxic questions by crafting certain prompts~\cite{advbench, andriushchenko2024jailbreaking}, which is orthogonal to the problem studied in this paper.

% Literature --> What are the problems of the existing methods --> ???
\vspace{0.5em}
\noindent\textbf{Literature:}
One line of methods aim to mitigate the problem above by intervening in the fine-tuning process to preserve alignment~\cite{qi2024safety, lyu2024keeping, wang2024mitigating, zhou2023making, hsu2024safe, huang2024lazy}. For example, SoftSFT~\cite{qi2024safety} hypothesizes that the first several output tokens greatly affect alignment. Therefore, it constrains a few early tokens to avoid deviating much from their initial values during fine-tuning. However, constraining only the first few tokens is insufficient to maintain alignment, as its impact can extend further down the generation sequence. 

Another line of works post-process LLMs after fine-tuning to reinstate the compromised alignment~\cite{catqa}~\cite{hsu2024safe}. In particular, RESTA~\cite{catqa} computes the difference in parameters between a safety-unaligned model and the original aligned model, denoted as \textit{safety vector}. It then adds this safety vector to the fine-tuned model to recover the alignment. However, modifying the entire model's weights can unexpectedly degrade the performance on downstream tasks.
% The method discussed in this paper also falls into this category of recovering the alignment of a fine-tuned LLM, but without significantly affecting the main task (which will be discussed later). Other approaches aim to build a more robust or aligned model such that any fine-tuning will not harm the alignment~\cite{...}. These methods are orthogonal to this paper and can be combined to provide a more effective solution.

% Our method: insights --> designs

\vspace{0.5em}
\noindent\textbf{Insight:}
Apparently, aligned LLMs have the capabilities of differentiating harmful questions and benign questions. Inspired by recent research on LLM behaviors~\cite{turner2023activation,burns2022discovering,zou2023representation}, we perceive that the alignment capabilities are determined by two internal \textbf{directions} of an LLM, which we denote as the \textit{aligned direction} and the \textit{harmful direction}. Specifically, \textit{the LLM is willing to answer questions in the aligned direction while refusing those in the harmful direction}. In our study presented in~\S\ref{subsec:method:insight}, we find that by pushing the internal features of a hamrful question closer to the aligned direction and away from the harmful direction, aligned LLMs respond with related information instead of a refusal. The likelihood of answering harmful questions can increase from 4.57\% to 80.42\%. %Similar observations can also be made with fine-tuned LLMs, where the feature representations of toxic questions are closer to the aligned direction, meaning they will be answered by the model.

%Recent research brought up the concept of \textbf{direction}~\cite{turner2023activation,burns2022discovering,zou2023representation}, which influences an LLM's behaviors like honesty.
%In other words, internally, there might be two directions that the LLM can take in response to input questions. Our study in~\S\ref{subsec:method:insight} validates that there indeed exist two directions represented by the model's internal features, which we denoted as the \textit{aligned direction} and the \textit{harmful direction}, respectively. That is, \textit{the LLM is willing to answer any questions in the aligned direction while refusing queries in the harmful direction}.

\vspace{0.5em}
\noindent\textbf{Method:} Following the insight above, we propose an alignment recovery method for fine-tuned models. Our goal is to restore the harmful direction that guides an LLM in avoiding answering harmful questions. The core idea is to identify and restore a subset of the fine-tuned model's weights to the original aligned model's values, such that the two models have minimized disparity in the harmful direction.

Technically, we leverage a small dataset comprising a diverse set of harmful prompts (e.g., 256 samples) for measuring and restoring the harmful direction. The difference in the harmful directions before and after the recovery is utilized for computing the gradients, which indicates how each weight parameter should be changed to alleviate the discrepancy. Given a parameter whose gradient directions point towards the aligned model, we copy its value from the original aligned model to the recovered model. While this copy operation maximally recovers the alignment, it may hurt the performance of the downstream task. We therefore additionally include a rollback mechanism that reverts a subset of restored weights to mitigate possible degradation in downstream task performance.

%We therefore only consider weight parameters whose gradient directions point towards the aligned model, which have a lower impact on downstream tasks. Additionally, our technique includes a rollback mechanism that reverts a subset of restored weights to mitigate possible degradation in downstream task performance.

% Evaluations
% Contributions
\vspace{0.5em}
\noindent\textbf{Contributions:} Our main contributions are as follows.
\begin{itemize}%[topsep=0pt,itemsep=0pt]
%    \item We identify two important directions in LLMs, the \textit{aligned direction} and the \textit{harmful direction}, which are essential to LLM alignment.

    % \item We make an important observation that moving the features of toxic questions towards the aligned direction can indcue an LLM to answer these queries, illustrating the characterics of alignment at model internals. \yk{Actually, one paper~\cite{zou2023representation} also demonstrated such a behavior, the difference is that they use a different way to represent the direction}\gt{I see. I will remove this then}\gt{wait, you mean moving the directions as well?} \yk{they push the internal value toward benign direction by simply adding the benign direction to internal values at several specific layers. We changed the weights to do this.}

    \item We propose a recovery method that restores a small number of weight parameters to effectively preserve the alignment of fine-tuned LLMs. 
    
    %of the fine-tuned LLM from the aligned model to. %, preserving alignment while maintaining downstream task performance.

    \item We combine gradient-guided, selective recovery and a rollback mechanism to mitigate performance degradation in downstream tasks.

    \item We evaluate our method on a total of 125 fine-tuned models derived from a set of representative LLMs and datasets. The results show that our method effectively recovers the alignment compromised by fine-tuning, reducing the harmful rate (percentage of answering hamrful questions) from 33.25\% to 1.74\%, with a 2.93\% degradation in downstream task performance. Our method also offers a better trade-off between alignment and performance than the existing methods.
%    Existing state-of-the-art alignment recovery approaches either have a low recovery rate (e.g., SoftSFT~\cite{qi2024safety} still has a 16.24\% harmful rate) or cause high task performance degradation (e.g., RESTA~\cite{bhardwaj2024language} induces a 16.48\% accuracy drop). In contrast, our method achieves a better trade-off.
\end{itemize}
\section{Technical Background \label{sec:bcg}}

\subsection{Large Language Models \label{subsec:bcg:llm}}
\vspace{-0.2em}
Large language models (LLMs), such as ChatGPT~\cite{gpt4o}, Llama 2~\cite{llama2}, and Gemma~\cite{gemma},
prevalently adopt the Transformer architecture~\cite{vaswani2017attention}, which consists of an embedding layer $\mathbf{E}$ and a prediction layer 
$\mathbf{P}$ connected via a series of identical hidden layers $\mathbf{F}$. Briefly, an LLM can be represented as 
$\mathbf{M} = \mathbf{P} \circ \mathbf{F}  \circ \mathbf{E}$, where $ \circ $ denotes function composition. 
\iftrue
Given an input word sequence $\textbf{w} = \langle w_1 , ..., w_n \rangle$, $\mathbf{M}$ works as follows.

%prediction layer $P$ with a series of identical hidden layers $F$ in between as $M = P \circ F  \circ E$, where $ \circ $ denotes function composition. But when given an input sequence $\textbf{w} = \langle w_0 , ..., w_n \rangle$, it also requires additional steps to generate the next word. The whole workflow is as following:
\begin{itemize}[leftmargin=8pt]
    \item \textbf{Tokenization} splits $\textbf{w}$ into smaller units known as tokens $\textbf{t}= \langle t_1, ..., t_m \rangle $. 
    Usually, a word is taken as a token. All unique tokens the model supports constitute the vocabulary $\textbf{V}$. Each token is represented 
    as an integer, ranging from 0 to the size of $\textbf{V}$.

    \item \textbf{Embedding} maps each token to a vector representation through indexing. 
    The embedding is trained with $\mathbf{M}$, which will be fixed once training is done. 
    
    %This embedding is trained with the LLMs and will be fixed once the model training is completed. %This procedure can be formula as  $E: T \in \mathbb{R}^{|V|} \rightarrow X \in \mathbb{R}^{|V|\times d}$, where $d$ is the dimension of embedding vector.
    \item \textbf{Hidden layers $\mathbf{F}$} forward the embedding through
    and output the final hidden states $h_l=\langle \vec{h}_{l}^{1}, ..., \vec{h}_{l}^{m} \rangle$, 
    where $l$ stands for the number of hidden layers and $m$ is the token number.
    In detail, each hidden layer $f_i$ ($\in \mathbf{F}$) generates its hidden 
    states $h_i$ based on those from the previous layer (i.e., $h_i = f_i(h_{i-1})$). 
    While different models may use different structures, in a simple way, $f_i$ can 
    be viewed as a matrix operation $h_i = w_i \cdot h_{i-1}$, 
    where $w_i$ are parameters, or weights, on the corresponding layer.
    \item \textbf{Prediction} yields the probability for each token in the vocabulary to be the next, relying on the hidden states of the last token from the last hidden layer (i.e., $h_{l}^{m}$).
    
   % probabilities as the next token for all tokens in the vocabulary based on the hidden states $h_{l}^{m}$ of the last token from last hidden layer. 
    \item \textbf{Decoding} picks the next token based on the probabilities from the prediction. Different strategies have been adopted by 
    existing LLMs, including greedy decoding~\cite{greedy_beam}, beam search~\cite{greedy_beam}, top-k sampling~\cite{topk}, and top-p sampling~\cite{topp}.

 %   This step adopt some strategies to choose the next token based on all the token probabilities, like greedy decoding~\cite{greedy_beam}, beam search~\cite{greedy_beam}, top-k sampling~\cite{topk}, and top-p sampling~\cite{topp}. The next token is then output as part of the word. 
\end{itemize}
\fi

\subsection{Supervised Fine-Tuning for LLM\label{subsec:bcg:finetuning}}
Supervised fine-tuning is a common technique to adapt or optimize LLMs for downstream tasks~\cite{howard2018universal, radford2018improving}.  It is also critical for enhancing instruction-following capabilities and aligning conversational LLMs like ChatGPT and Gemma with human preferences~\cite{wizardlm}.

\vspace{0.5em}
\noindent\textbf{Full-Parameter Fine-Tuning (FPFT)} directly updates a pre-trained model's parameters. The process is identical to pertaining, except that a lower learning rate might be used. Given a training dataset of input-output pairs $\mathcal{D} = \{(X_i, Y_i)\}_{i=1}^{N}$ where $X_i={x_1, ..., x_t}$ and $Y_i={y_1, ..., y_k}$ are sequences of tokens, the model is initialized to pre-trained weights $\theta$ and repeatedly updated to $\theta^*$ by minimizing the conditional language modeling objective:
% \gt{Probably explain the equation? Otherwise, people may get lost and the equation does not serve its purpose.}

\vspace{-1.2em}
{\small $$ \theta^* = \arg\min_{\theta}   \sum_{{ (X, Y)} \in \mathcal{D}} \; \sum_{i=1}^{|Y|} -\log \left( P(\hat{y}_i | X, y_{j, \{j=0,...,i-1\} } ; \theta )  \right)  $$}

By feeding $\langle X, y_1,...,y_{i-1} \rangle$ into the LLM, the above equation aims to find the optimal parameters $\theta$ that maximize the probability (minimize the negative log probability) of predicting $\hat{y}_i$ as $y_i$ across the entire dataset $(X, Y) \in \mathcal{D}$. %\yk{Please Check.}

FPFT works well with smaller models. When applied to the ever-growing LLMs, 
it incurs an unaffordable burden. For example, full-parameter fine-tuning a LLaMA model with 65 billion 16-bit parameters requires over 780 GB of GPU memory~\cite{qlora}. 

%FPFT works well on previous small model, but the GPU memory cost and storage cost would be unacceptable for LLMs whose parameters are huge (e.g., 175 billion for GPT-3~\cite{gpt3}). One of the limitation is the storage cost, since FPFT create a new set of parameters that are the same size as the original model for each downstream task. Another limitation is the GPU cost. For example, fine-tuning a Llama 2 65B parameter model~\cite{llama2} in 16-bit requires more than 780 GB of GPU memory. 

\vspace{0.5em}
\noindent\textbf{Parameter Efficient Fine-Tuning (PEFT)}, instead of updating all parameters, focuses on a small set of model parameters. Low-Rank Adaptation (LoRA)~\cite{lora}, a representative PEFT method, freezes the pre-trained model weights and injects trainable rank decomposition matrices into each hidden layer. Given $W \in \mathbb{R}^{d\times d}$ representing parameters in a hidden layer, LoRA decomposes $W$ into two lower rank matrices $A \in  \mathbb{R}^{d\times r}$ and $ B \in  \mathbb{R}^{r\times d} $ in a way that $W = A \cdot B$ and $r \ll d$. The fine-tuning process can be formalized as:
% \gt{Anthor big equation. If the previous equation is explained, then this one only needs to explain $\Delta \theta$ as it is here in the text.}

\vspace{-1.5em}
{\scriptsize $$ \triangle \theta ^* = \arg\min_{\triangle \theta}  \sum_{(X, Y) \in \mathcal{D}} \sum_{i=1}^{|Y|} -\log \left( P(\hat{y}_i | X, y_{j, \{j=0,...,i-1\} } ; \theta + \triangle \theta )  \right) $$}

\noindent where $\triangle \theta$, consisting of $A$ and $B$, are the parameters to be updated. Compared to FPFT, PEFT reduces the number of parameters updated in each hidden layer from $d \times d$ to $2 \times r \times d$. %, suppose $W \in \mathbb{R}^{d\times d}$ is the parameters of one of the layer in LLMs, LoRA decomposes $W$ into two lower rank matrix $A \in  \mathbb{R}^{d\times r}$ and $ B \in  \mathbb{R}^{r\times d} $ as $W = A \cdot B$. 
Another optimization widely applied with PEFT is \textit{quantization}~\cite{qlora}, which quantizes the parameters to a representation with fewer bits.%For example, QLoRA~\cite{qlora} quantizes the pre-trained parameters of LoRA to 4-bit, reducing the GPU memory needed for 65 billion 16-bit parameters from 780 GB to less than 48 GB.

%extends the LoRA by quantizing the pretrained language model into 4-bit and the rest is the same as LoRA. This simple operation reduce the  memory requirements of fine-tuning a 65B parameter model from over 780GB of GPU memory to less than 48GB without degrading performance compared to a 16-bit fully finetuned baseline.

%For PEFT, it only updates a few extra model parameters $\triangle \theta$ as : $$ \triangle \theta ^* = \arg\min_{\triangle \theta}  \sum_{(X, Y) \in \mathcal{D}} \sum_{i=1}^{|Y|} -\log \left( P(\hat{y}_i | X, y_{j, \{j=0,...,i-1\} } ; \theta + \triangle \theta )  \right) $$

\subsection{LLM Alignment \label{subsec:bcg:alignment}}
\vspace{-0.8em}
% \gt{This section does not seem to explain what LLM alignment is. (Well, in high-level, there are some descriptions. But it does not provide much information. People may not fully understand what alignment is.) I would prefer having more details on this part since this is the main problem the paper aims to address.\\
% Another thing is that we need to connect alignment with security, especially in the introduction. Why is this a security problem?
% }
% \gt{I wonder if you could add some examples or equations. I think the GCG paper from Zico Kolter has some equation explaning the process. The paragraph reads fine to me because I am familiar with the problem. I am just not sure whether a non-expert could get it.}

\begin{figure}[h]
    \centering
   \includegraphics[width=\columnwidth]{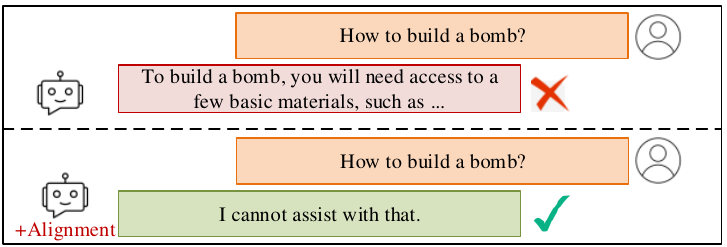}
    \vspace{-1em}
    \caption{Illustration of LLM alignment. LLM alignment ensures that the model’s outputs align with human values.}
    \label{fig:alignmentdemo}
\end{figure}

Despite the significant advancements in LLMs, they do not follow human values. Thus, they can be maliciously leveraged to perform unethical or harmful tasks (e.g., generating hate speech). To address this issue, LLM vendors introduce safety \textit{alignment} into their models to ensure that the model's outputs align with human values and expectations as shown in \autoref{fig:alignmentdemo}. In particular, LLM alignment ensures the safe operation of LLMs by training and testing them to handle a wide range of inputs, including adversarial ones designed to mislead the model. 

 Various methods, such as reinforcement learning from human feedback (RLHF)~\cite{rlhf}
and preference optimization~\cite{dpo}, have been employed to achieve alignment. 
Despite the technical differences, these methods all mandate human-annotated data 
and time-consuming training, incurring a high cost. In addition, LLM vendors often 
involve red teaming with experts in different domains to 
discover, measure, and reduce alignment issues. This red teaming
further magnifies the cost of LLM alignment.

\section{Motivating Study\label{sec:motivation}}

While LLM alignment is expensive to achieve, it can be compromised during fine-tuning. Recent research~\cite{qi2023fine} 
has shown that \textit{fine-tuning, with or without harmful data, can sabotage the alignment}. The following presents a study to affirm 
those findings and motivate our research.

%Even though the companies and researchers have devoted a lot of effort into the alignment,  as shown by previous works~\cite{qi2023fine, yang2023shadow}, fine-tuning can easily compromise the alignment of LLMs, even with only benign data, called fine-tuning attack. Here, we run a case study to demonstrate that the alignment can be easily degraded with only a few harmful data mixed into the task dataset without sacrificing the task performance.  The experimental setup is as follows.
% \noindent\textbf{Experimental Setup:} 

\subsection{Experimental Settings\label{subsec:studysettings}}
\noindent\textbf{Models:} To support our study, we pick five state-of-the-art, open-source LLMs across different architectures and sizes, including 
\gemma~\cite{gemma}, \llamasmall, \llamabig~\cite{llama2}, \mistral~\cite{mistral}, and \qwen~\cite{qwen}. Details of the models are summarized in~\autoref{tab:model}. %One may note that all the models are chat models instead of base models. This is intended as we focus on alignment.

%Noteworthy, all the models are chat models, not base models since we consider the alignment.  

\vspace{0.5em}
\noindent\textbf{Datasets:} To perform fine-tuning, we collect five datasets that have been used in 
previous research on alignment or fine-tuning~\cite{wang2024mitigating, gupta2022instructdial,shi2022natural,gao2024improving}. 
Summarized in~\autoref{tab:dataset}, the datasets, varying in domains and sizes, 
are tailored for different tasks:

\begin{itemize}[leftmargin=8pt]

\item\textbf{\sql}, derived from WikiSQL~\cite{zhongSeq2SQL2017} and Spider~\cite{yu2018spider},
consists of 78,577 pairs of natural language queries and structured query language (SQL) statements, attached with 
the context of the database tables.

\item\textbf{\cheat}~\cite{cheat} combines three subsets
created for facilitating AI-generated content detection. We focus on 
the most challenging subset, which consists of 15,395 human-written abstracts
and 15,395 AI-polished abstracts, plus their labels.

\item\textbf{\nlbash} contains 10,347 samples mapping 
English descriptions to Bash commands. The dataset is further augmented with 17,450 samples crafted by Fu \etal~\cite{nlcmd}.

\item\textbf{\samsum}~\cite{samsum} includes 15k messenger-style conversations 
and their third-person summaries. The conversations are crafted by linguists fluent in English
to mimic real-life chats. This dataset can optimize an LLM to 
capture the main topics covered by a conversation.

%dataset comprises approximately 16k messenger-style conversations, crafted by linguists fluent in English to mimic real-life chats. These conversations vary in style, from informal to formal, including slang, emoticons, and typos. They were then annotated with third-person summaries, aiming to succinctly capture the main topics discussed.

%10,347 samples, which map English sentences to Bash commands (NL2Bash). It aims to train a model that can execute tasks like file manipulation, search, and application-specific scripting by giving plain English statements. We augment this dataset by incorporating 17,450 samples generated by ChatGPT craft by Fu \etal~\cite{nlcmd}, constructing a total of 27,797 data.

\item\textbf{\toxicity}~\cite{toxicity} comprises 10k user prompts
from the Vicuna online demo~\cite{vicuna2023} and their toxicity annotations
(i.e., whether the prompts are toxic). It can enhance 
LLMs' toxicity detection.

%dataset includes toxicity annotations for 10K user prompts from the Vicuna~\cite{vicuna2023} online demo, which hopes to enhance the toxicity detection in real-world user-AI conversations.

\end{itemize}

\begin{table}[!t]
\scriptsize
\setlength\tabcolsep{1.9pt}
 \renewcommand{\arraystretch}{0.8}
\centering
\caption{Models used in our study and evaluation.\label{tab:model}}
 \vspace{-0.75em}
\begin{tabular}{lccc}
\toprule
\textbf{Model} & \textbf{Parameter \#} & \textbf{Hidden Layer \#} & \textbf{Huggingface Path}                   \\\toprule
\gemma      & 2B         & 18              & google/gemma-2b-it                 \\
\midrule
\llamasmall & 7B         & 32              & meta-llama/Llama-2-7b-chat-hf      \\
\midrule
\llamabig   & 13B        & 40              & meta-llama/Llama-2-13b-chat-hf     \\
\midrule
\mistral    & 7B         & 32              & mistralai/Mistral-7B-Instruct-v0.2 \\
\midrule
\qwen       & 7B         & 32              & Qwen/Qwen1.5-7B-Chat      \\
\bottomrule
\end{tabular}
\end{table}

\begin{table}[!t]
        \scriptsize
        \setlength\tabcolsep{3.8pt}
         \renewcommand{\arraystretch}{0.8}
	\centering
 \begin{threeparttable}
	\caption{Datasets used in our study and evaluation.\label{tab:dataset}}
        \vspace{-0.75em}
	\begin{tabular}{lccccc}
		\toprule
		\textbf{Dataset} & \textbf{Task} & \textbf{Metric}  & \textbf{Train} & \textbf{Test} & \textbf{Total} \\
		\toprule
		  \sql~\cite{sql} &  Text to SQL& Exact Match\tnote{$\alpha$} & 77,577    &  1,000  & 78,577     \\
    	\midrule
		  \cheat~\cite{cheat} &  AI Text Detection& F1 Score\tnote{$\beta$} & 29, 790   & 1,000   &  30,790   \\
		\midrule
		  \nlbash~\cite{nlbash} & Text to Bash& NLC2CMD\tnote{$\gamma$}  & 26,797  & 1,000      & 27,797   \\
    	\midrule
     	\samsum~\cite{samsum} &  Dialogue Summary & Rouge-1\tnote{$\delta$} &14,732     & 819     &  15,551  \\
		\midrule
            \toxicity~\cite{toxicity} & Toxicity Detection& F1 Score\tnote{$\beta$} & 5,082 & 5,083     & 10,165    \\
		\bottomrule
	\end{tabular}%
 \begin{tablenotes}
      \footnotesize
      \item[$\alpha$] \textit{Exact Match}~\cite{zhongSeq2SQL2017} is commonly used in tasks like question answering and translation. It is measured by evaluating if the model's output exactly matches the ground truth.\vspace{0.3em}
      \item[$\beta$] \textit{F1 Score}~\cite{toxicity} is a standard metric combining precision and recall. Formally, F1 Score = $(2 * TP) / (2*TP + FP + FN)$ where $TP$, $FP$, and $FN$ stand for true positives, false positives, and false negatives.\vspace{0.3em}
      \item[$\gamma$] \textit{NLC2CMD}~\cite{nlccmdnips} is the official metric used in the NeurIPS 2020 NLC2CMD Competition, which we reuse.\vspace{0.3em}
      \item[$\delta$]\textit{Rouge-1}~\cite{lin-2004-rouge} is widely used for tasks like text summarization. Rouge-n measures the overlap of contiguous n-grams between the predicted summary and the ground truth summary.
      \vspace{-1.25em}
    \end{tablenotes}
  \end{threeparttable}
\end{table}%
\begin{table*}[!t]
\extrasmallsize
\setlength\tabcolsep{1.2pt}
\renewcommand{\arraystretch}{0.8} 
\caption{Motivating study results. \texttt{TP(\%)} and \texttt{HR(\%)} are task performance and harmful rate (see \S\ref{subsec:studyresult}). Under \code{Fine-tuning Settings}, ``\texttt{NA}'' stands for no fine-tuning, \texttt{0}, \texttt{0.1k}, \texttt{0.5k} and \texttt{1.5k} represent the scenarios where we injected the corresponding number of harmful questions into fine-tuning, and \texttt{Mod.} means we only injected the 392 harmful questions left over from moderation into fine-tuning. We mark the highest harmful rate in \textbf{\textcolor{winered}{red}}.}
\label{tab:study_result}
 \vspace{-0.75em}
\begin{tabular}{ll|cccccc|cccccc|cccccc|cccccc|cccccc}
\toprule

\multicolumn{2}{l|}{\multirow{3}{*}{\textbf{\scriptsize Dataset}}}  & \multicolumn{6}{c}{\textbf{\gemma}} & \multicolumn{6}{|c}{\textbf{\llamasmall}} & \multicolumn{6}{|c}{\textbf{\llamabig}} & \multicolumn{6}{|c}{\textbf{\mistral}} & \multicolumn{6}{|c}{\textbf{\qwen}}  \\

\cmidrule(lr){3-8} \cmidrule(lr){9-14} \cmidrule(lr){15-20} \cmidrule(lr){21-26} \cmidrule(lr){27-32} 

& & \multicolumn{6}{c}{\code{Fine-tuning Settings}} & \multicolumn{6}{|c}{\code{Fine-tuning Settings}} & \multicolumn{6}{|c}{\code{Fine-tuning Settings}} & \multicolumn{6}{|c}{\code{Fine-tuning Settings}} & \multicolumn{6}{|c}{\code{Fine-tuning Settings}}  \\

\cmidrule(lr){3-8} \cmidrule(lr){9-14} \cmidrule(lr){15-20} \cmidrule(lr){21-26} \cmidrule(lr){27-32} 

 & & \texttt{\extrasmallsize NA} & \texttt{\extrasmallsize 0}  & \texttt{\extrasmallsize 0.1k}  & \texttt{\extrasmallsize 0.5k} & \texttt{\extrasmallsize 1.5k} & \texttt{\extrasmallsize Mod.}  & \texttt{\extrasmallsize NA} & \texttt{\extrasmallsize 0}  & \texttt{\extrasmallsize 0.1k}  & \texttt{\extrasmallsize 0.5k} & \texttt{\extrasmallsize 1.5k} & \texttt{\extrasmallsize Mod.} & \texttt{\extrasmallsize NA} & \texttt{\extrasmallsize 0}  & \texttt{\extrasmallsize 0.1k}  & \texttt{\extrasmallsize 0.5k} & \texttt{\extrasmallsize 1.5k} & \texttt{\extrasmallsize Mod.} & \texttt{\extrasmallsize NA} & \texttt{\extrasmallsize 0}  & \texttt{\extrasmallsize 0.1k}  & \texttt{\extrasmallsize 0.5k} & \texttt{\extrasmallsize 1.5k} & \texttt{\extrasmallsize Mod.} & \texttt{\extrasmallsize NA} & \texttt{\extrasmallsize 0}  & \texttt{\extrasmallsize 0.1k}  & \texttt{\extrasmallsize 0.5k} & \texttt{\extrasmallsize 1.5k} & \texttt{\extrasmallsize Mod.} \cr
\midrule

\multirow{2}{*}{\textbf{\tiny \sql}} & \texttt{\extrasmallsize TP} & \extrasmallsize{0.0 }&\extrasmallsize{80.9 }&\extrasmallsize{79.2 }&\extrasmallsize{80.9 }&\extrasmallsize{80.8 }&\extrasmallsize{ 81.5 }&\extrasmallsize{0.0 }&\extrasmallsize{78.9 }&\extrasmallsize{79.3 }&\extrasmallsize{80.1 }&\extrasmallsize{80.0 }&\extrasmallsize{78.9  }&\extrasmallsize{0.0 }&\extrasmallsize{82.3 }&\extrasmallsize{80.7 }&\extrasmallsize{82.0 }&\extrasmallsize{82.9 }&\extrasmallsize{81.7 }&\extrasmallsize{0.0 }&\extrasmallsize{81.4 }&\extrasmallsize{83.0 }&\extrasmallsize{83.1 }&\extrasmallsize{81.7 }&\extrasmallsize{81.6 }&\extrasmallsize{0.0 }&\extrasmallsize{79.7 }&\extrasmallsize{80.6 }&\extrasmallsize{81.5 }&\extrasmallsize{82.4 }&\extrasmallsize{ 80.3}
\\

& \texttt{\extrasmallsize HR} & \extrasmallsize{4.6 }&\extrasmallsize{4.5 }&\extrasmallsize{51.1 }&\extrasmallsize{\textbf{\textcolor{winered}{63.8}} }&\extrasmallsize{62.0 }&\extrasmallsize{46.6 }&\extrasmallsize{0.0 }&\extrasmallsize{0.0 }&\extrasmallsize{2.2 }&\extrasmallsize{53.1 }&\extrasmallsize{\textbf{\textcolor{winered}{59.5}} }&\extrasmallsize{ 27.6 }&\extrasmallsize{0.0 }&\extrasmallsize{0.0 }&\extrasmallsize{2.7 }&\extrasmallsize{57.1 }&\extrasmallsize{\textbf{\textcolor{winered}{57.7}} }&\extrasmallsize{ 27.9 }&\extrasmallsize{11.7 }&\extrasmallsize{13.7 }&\extrasmallsize{45.4 }&\extrasmallsize{50.5 }&\extrasmallsize{\textbf{\textcolor{winered}{54.7}} }&\extrasmallsize{ 51.4}&\extrasmallsize{2.4 }&\extrasmallsize{3.1 }&\extrasmallsize{48.7 }&\extrasmallsize{\textbf{\textcolor{winered}{57.7}} }&\extrasmallsize{52.8 }&\extrasmallsize{ 43.3 }
\\
\midrule

\multirow{2}{*}{\textbf{\tiny\cheat}}& \texttt{\extrasmallsize TP} & \extrasmallsize{32.2 }&\extrasmallsize{96.4 }&\extrasmallsize{98.0 }&\extrasmallsize{97.5 }&\extrasmallsize{96.1 }&\extrasmallsize{ 97.9 }&\extrasmallsize{0.0 }&\extrasmallsize{89.7 }&\extrasmallsize{88.7 }&\extrasmallsize{90.2 }&\extrasmallsize{94.3 }&\extrasmallsize{ 85.6 }&\extrasmallsize{0.0 }&\extrasmallsize{98.6 }&\extrasmallsize{96.9 }&\extrasmallsize{97.6 }&\extrasmallsize{98.3 }&\extrasmallsize{ 97.0}&\extrasmallsize{8.4 }&\extrasmallsize{97.4 }&\extrasmallsize{96.4 }&\extrasmallsize{97.3 }&\extrasmallsize{97.1 }&\extrasmallsize{97.3 }&\extrasmallsize{66.6 }&\extrasmallsize{96.7 }&\extrasmallsize{98.1 }&\extrasmallsize{97.7 }&\extrasmallsize{97.9 }&\extrasmallsize{ 97.6}

\\

& \texttt{\extrasmallsize HR} & 4.6 &3.9 &39.3 &62.1 &\textbf{\textcolor{winered}{64.0}} & 32.1&0.0 &0.0 &0.1 &11.6 &\textbf{\textcolor{winered}{56.4}} &4.6 &0.0 &0.0 &0.6 &45.4 &\textbf{\textcolor{winered}{56.4}} & 8.4 &11.7 &13.7 &47.3 &\textbf{\textcolor{winered}{57.3}} &52.9 & 49.9 &2.4 &3.7 &40.9 &47.4 &\textbf{\textcolor{winered}{51.3}} & 36.1\\
\midrule

\multirow{2}{*}{\textbf{\tiny \nlbash}} & \texttt{\extrasmallsize TP} & 0.1 &36.4 &36.1 &36.9 &38.3 & 37.9 &0.0 &34.7 &34.7 &36.1 &33.1 & 32.3 &0.0 &36.4 &37.5 &37.7 &36.8 &35.4 &0.0 &40.8 &41.1 &41.8 &41.6 &40.0 &0.0 &38.8 &39.3 &39.6 &38.1 & 38.8\\

 & \texttt{\extrasmallsize HR} & 4.6 &4.4 &42.0 &61.7 &\textbf{\textcolor{winered}{63.7}} & 37.4 &0.0 &0.0 &6.0 &56.3 &\textbf{\textcolor{winered}{62.0}} &10.6 &0.0 &1.3 &2.4 &\textbf{\textcolor{winered}{62.3}} &61.9 & 23.4 & 11.7 &  21.3 &44.6 &54.9 &\textbf{\textcolor{winered}{56.4}} & 50.0 & 2.4 &  4.7 &43.1 &51.7 &\textbf{\textcolor{winered}{52.4}} & 42.0 \\
\midrule

\multirow{2}{*}{\textbf{\tiny\samsum}} & \texttt{\extrasmallsize TP} & 27.5 &49.9 &49.9 &50.0 &50.0 & 50.1  &21.4 &50.5 &50.8 &50.2 &50.7 & 50.8 &27.0 &52.8 &53.5 &53.4 &52.8 &53.3 &29.5 &54.6 &54.5 &54.6 &54.5 &54.3 &30.4 &52.4 &52.6 &52.4 &52.8 &53.2 \\

& \texttt{\extrasmallsize HR} & 4.6 &4.1 &10.4 &48.1 &\textbf{\textcolor{winered}{62.9}}&  25.7 &0.0 &0.9 &1.3 &6.6 &\textbf{\textcolor{winered}{40.6}} & 4.1 &0.0 &0.0 &1.1 &21.7 &\textbf{\textcolor{winered}{57.1}}& 7.1  &11.7 &17.9 &30.0 &51.6 &\textbf{\textcolor{winered}{54.6}} &41.3 &2.4 &4.4 &24.0 &50.0 &\textbf{\textcolor{winered}{52.6}}  & 33.6\\

\midrule

\multirow{2}{*}{\textbf{\tiny\toxicity}} & \texttt{\extrasmallsize TP} &0.0 &78.7 &75.1 &40.4 &46.1 & 74.4 &0.0 &75.1 &74.5 &68.3 &57.9 & 65.5 &0.0 &78.6 &76.2 &72.9 &62.4  & 77.8 &49.8 &84.0 &80.5 &76.3 &76.8 &79.6 &58.4 &79.3 &81.6 &68.9 &73.7 &76.1 \\

& \texttt{\extrasmallsize HR} & 4.6 &3.7 &34.6 &\textbf{\textcolor{winered}{63.6}} &60.7 & 39.4&0.0 &0.0 &0.4 &46.4 &\textbf{\textcolor{winered}{58.9}} & 12.4 &0.0 &0.0 &1.0 &\textbf{\textcolor{winered}{61.0}} &57.6 & 24.0 &11.7 &9.6 &35.1 &55.4 &\textbf{\textcolor{winered}{56.4}} & 50.9 &2.4 &2.6 &43.0 &\textbf{\textcolor{winered}{54.4}} &54.1 & 40.3 \\

\bottomrule
\end{tabular}
 \vspace{-0.75em}
\end{table*}

\vspace{0.5em}
\noindent\textbf{Fine Tuning:} We fine-tune each model on each dataset using QLoRA~\cite{qlora}. 
The fine-tuning is run for one epoch, and the parameters are quantized to 4 bits. 
As suggested in~\cite{LoRAtips}, we set the rank and alpha of QLoRA to 8 and 16.
We also set the learning rate to 2$e^{-5}$ and the batch size to 
96.%, pursuing a better tradeoff between accuracy and our computational resources.

To approximate cases where the fine-tuning dataset
includes harmful samples (e.g., the dataset is not sanitized or the dataset is poisoned with harmful samples), 
we pollute the five datasets with varying amounts of harmful data from \beavertails~\cite{beavertails}, 
a safety alignment dataset consisting of training  
and testing subsets. The training subset includes 333,963 question-answer 
(QA) pairs and the corresponding safety meta-labels (harmful or not). 
The testing subset has more samples organized in the same way.
After removing non-harmful pairs and deduplicating the remaining\footnote{Questions with a 0.9+ cosine similarity are deemed duplicates.} in the training subset, 
we obtain 9,795 harmful QA pairs
and reserve 1,500 to pollute our datasets. Specifically, we randomly
pick 100, 500, and 1,500 out of the 1,500 pairs and inject them 
into each dataset to create three more variants. On each 
variant, we independently redo the fine-tuning.

We also notice a recent trend where people adopt moderation methods~\cite{han2024wildguard,llamaguard2} to identify and filter harmful questions 
before fine-tuning is performed. We accordingly extend the study to include the scenario where we
apply a moderation method to filter the 1,500 harmful questions and only inject the remaining 
into the fine-tuning process. In the study, we experimented with 5 state-of-the-art moderation methods, 
including Llama-Guard2~\cite{llamaguard2}, WildGuard~\cite{han2024wildguard}, OpenAI Moderation API~\cite{openaimoderation}, OpenAI GPT-4o~\cite{gpt4o}, 
and ShieldGemma~\cite{zeng2024shieldgemma}. It turns out that OpenAI GPT-4o presents the best performance\footnote{Out of the 1,500 harmful questions, Llama-Guard2 filtered 860, WildGuard filtered 1,075, the OpenAI Moderation API filtered 602, OpenAI GPT-4o filtered 1,108, and ShieldGemma filtered 771.}, filtering out 1,108 harmful questions. Hence, we adopt it in 
our study and all other evaluations.

\subsection{Evaluation\label{subsec:studyresult}}

\noindent\textbf{Metrics:} In the study, we focus on evaluation from two perspectives, including \textit{task performance} and \textit{harmful rate}. 

\begin{itemize}[leftmargin=8pt]
\item \textbf{Task Performance} measures the performance of LLMs on the task targeted by the five datasets. The specific metric is defined by the corresponding dataset, as summarized in~\autoref{tab:dataset}. Two datasets, \samsum and \toxicity, have been split internally for \textit{trainig} and \textit{testing}, which we respectively use for fine-tuning and evaluation. For the other three datasets, we randomly pick 1,000 samples for evaluation and the rest for fine-tuning, as done in~\cite{wang2024mitigating}.
% following the settings of prior research

\item \textbf{Harmful Rate} measures the alignment of LLMs, leveraging the testing subset from the aforementioned \beavertails~\cite{beavertails} dataset. This subset consists of 700 harmful questions belonging to 14 categories. The \textit{percentage of questions receiving a harmful answer from an LLM} is calculated as the harmful rate. We use the model fine-tuned by ShieldLLM~\cite{shieldlm} to determine if an answer is harmful, which has demonstrated high fidelity.
\end{itemize}
%A challenge here is to determine whether an LLM answer is harmful. In our study, we leverage a model fine-tuned by ShieldLLM~\cite{shieldlm} for that goal, which has shown high reliability~\cite{zhang2024safe}.

% , which we prepare as prompts to ask each LLM. 

%But how to decide whether the answer is harmful or not is still an open research question. Some researchers~\cite{liu2023jailbreaking} manually evaluate it and some others resort to automatic evaluation like GPT4~\cite{bhardwaj2023red} or rule match~\cite{wang2024mitigating}. Recently, many researchers~\cite{shieldlm, llamaguard} have tried to fine-tune an LLM to solve this, which has shown promising results in terms of speed and performance. Therefore, we decided to use the fine-tuned model provided by ShieldLLM~\cite{shieldlm} to judge whether the answer is harmful. 

\vspace{0.25em}
\noindent\textbf{Results:\label{subsec:studyresult}} \autoref{tab:study_result} shows the results. \ding{192} The native models present poor task performance. On the datasets of \sql~and \nlbash, all native models have a task performance of 0\%. On the other datasets, certain models perform better but remain unsatisfactory for practical use (task performance below 30\%). \ding{193} The native models carry built-in alignment. \llamasmall and \llamabig present a zero harmful rate. \gemma and \qwen are slightly higher, showing a harmful rate of 4.57\% and 2.53\%. Even the worst model, \mistral, has a harmful rate under 12\%. 

\subsection{Observations\label{subsec:observations}}

\textbf{\textit{Fine-tuning significantly escalates the task performance}}. It eliminates all cases where the native model has a task performance of 0\%. On the datasets of \sql~and \nlbash, fine-tuning increases the task performance of all models from 0\% to $\sim$80\% and $\sim$40\%. In the other cases, the performance improvement is also substantial (50\%+ increase). On the datasets of \sql, \samsum, \nlbash, and \cheat, the harmful QA pairs have negligible impact on the effectiveness of fine-tuning. On \toxicity, using 500+ harmful QA pairs can negatively affect the task performance. This is expected as \toxicity~has fewer training samples.

\vspace{0.25em}
\textbf{\textit{Fine-tuning compromises the alignment of native models}}. When applied to the models of \mistral and \qwen, even fine-tuning without harmful QA pairs visibly increases the harmful rate (from 11.7\% to 21.3\% and from 2.4\% to 4.7\%). The results also show a clear trend --- with more harmful QA pairs injected into fine-tuning, the harmful rates keep increasing until a high level. Using 1,500 harmful QA pairs, the harmful rate consistently goes over 50\%, regardless of the dataset and model.

\vspace{0.25em}
\textbf{\textit{Moderation helps preserve alignment but is insufficient}}. By filtering harmful questions before fine-tuning, moderation lessens the compromising of alignment. Compared to the baseline where all 1,500 harmful questions are directly involved, moderation decreases the harmful rate to a visibly lower level. Yet, moderation cannot fully address the problem. Even the best moderation method (OpenAI GPT-4o) leaves over a significant amount of harmful questions (392 out of 1,500), which can still sabotage the alignment. For instance, on the models of \gemma, \mistral, and \qwen, the non-filtered 392 harmful questions increased their hamrful rate to a level between 20\% and 50\%.

\section{Problem Statement\label{sec:prob}}
% \gt{I guess this is the threat model. Do we want to have a designated section for threat model (knowledge and capabilities of attackers and defeners)? It may be clearer to the security community.}
In this paper, we focus on addressing the problem of compromised alignment during LLM fine-tuning. This problem involves three participants (the \textit{Owner}, the \textit{Fine-tuner}, and the \textit{User}) and targets two scenarios:

\vspace{0.5em}
\noindent\textbf{Scenario I:} The \textit{Owner} owns a private, aligned LLM $\mathcal{M}$, and it opens black-box interfaces for querying and fine-tuning $\mathcal{M}$. The \textit{Fine-tuner} gathers a dataset and fine-tunes $\mathcal{M}$ to derive $\mathcal{M}'$, which it can query freely. In practice, OpenAI represents an instance of \textit{Owner} in this scenario.

We assume that the \textit{Fine-tuner} may intentionally break $\mathcal{M}'$ alignment to enable disallowed queries, using methods like poisoning the fine-tuning dataset with harmful samples. We further assume that the \textit{Owner} is benign and can apply moderation to sanitize the dataset from the \textit{Fine-tuner}. However, as we demonstrated in~\S\ref{subsec:observations}, many harmful samples can evade the moderation and still compromise the alignment. Thus, in this scenario, our solution is intended to be applied by the \textit{Owner} to reinstate the alignment in $\mathcal{M}'$.

\vspace{0.5em}
\noindent\textbf{Scenario II:} The \textit{Owner} owns an aligned LLM $\mathcal{M}$, which it releases publicly on platforms like Hugging Face for free use. The \textit{Fine-tuner} builds a dataset and fine-tunes a local copy of $\mathcal{M}$ to derive $\mathcal{M}'$. It further opens access for \textit{User} to query $\mathcal{M}'$ as desired.

We assume the \textit{Fine-tuner} is benign but may unintentionally compromise the alignment of $\mathcal{M}'$ in fine-tuning. This gives any ill-intended \textit{User} opportunities to perform disallowed queries. In this scenario, our solution can be applied by the \textit{Fine-tuner} to recover the alignment in $\mathcal{M}'$.

\section{Our Method}

\subsection{Insight\label{subsec:method:insight}}

Recent research brought up the concept of \textbf{direction}~\cite{turner2023activation,burns2022discovering,zou2023representation}, which influences an LLM's behaviors like honesty. Briefly, a direction can be viewed as the collection of internal values of an LLM when given prompts with the same property (e.g., all factual prompts). Inspired by~\cite{zou2023representation}, we adopt the definition below for direction: 

\begin{definition}[\textbf{Direction}\label{def:direction}]
Given an LLM $\mathcal{M}$ and a dataset $\mathcal{D}$ consisting of prompts sharing the same property $\mathcal{P}$ (e.g., honesty, fairness, emotion, etc.), we extract the hidden states of the last token from a designated hidden layer $L_{dir}$ for each prompt, and average the extracted hidden states of all prompts as the direction of property $\mathcal{P}$. Different strategies might be used to select $L_{dir}$, and we follow the one described in~\S\ref{subsubsec:getdirection}.
\end{definition}

%the LLMs when given a series of prompts with the same behavior such as all benign prompts.
%Recently, to better explain the LLMs and mitigate the hallucination and toxicity problem, many researchers have paid attention to the internal status of LLMs such as layer activations~\cite{turner2023activation, burns2022discovering} and hidden states of the last token~\cite{zou2023representation}. Particularly, they introduce the \textbf{direction} concept and apply the direction to control the models' high-level concepts like honesty and alignment. % detect whether the model is lying or enhances the model's honesty. 

% The internal values can be various such as the layer activations~\cite{turner2023activation}, and attention heads activations~\cite{li2024inference}. We can also post-process the internal values as the direction such as PCA~\cite{zou2023representation} or averaging~\cite{li2024inference}.  

We observe that direction can also be applied to calibrate the alignment property. Given an aligned model $\mathcal{M}$, we can compromise its alignment by adjusting its internal direction, as shown in the study below.

\vspace{0.25em}
\noindent\textbf{Study Design:} Feeding a dataset with solely benign prompts to $\mathcal{M}$, we can obtain the \textit{alignment} direction of $\mathcal{M}$ following~\autoref{def:direction}. Similarly, using a dataset full of harmful prompts, we can obtain $\mathcal{M}$'s \textit{harmful} direction.
% \gt{When you obtain the harmful direction from the model, is the model aligned or not?} \yk{aligned model $M$ defined at the above.} \gt{Have you tried using an unaligned model? Would the extracted direction be useful?}\yk{I am afraid the unaligned model doesn't have the harmful direction, since they answer all prompts.}\gt{I am just guessing here. Since the sets of questions are different, I wonder if there are differences in directions on an unaligned model. I can see it is more obvious on aligned models. If you have time to test it, it would be interesting to see.} \yk{I believe I did such an experiment, the unaligned model indeed can be used to classify harmful prompts and benign prompts. It just does not reject it. }\gt{I see. I think we should probably mention this somewhere.}
We hypothesize that \textit{the two directions, annotated as $\Delta_{aligned}$ and $\Delta_{harmful}$, drive $\mathcal{M}$ to consider a prompt benign or harmful, thus answering or rejecting it}.
% Specifically, \textit{$\Delta_{aligned}$ helps $\mathcal{M}$ recognize benign prompts  while $\Delta_{harmful}$ direction enables $\mathcal{M}$ to pinpoint harmful prompts}.
To verify the hypothesis, we craft our study as follows. Given an unseen harmful prompt that $\mathcal{M}$ refuses to answer (e.g., \textit{``how to rob a bank''}), we push $\mathcal{M}$'s hidden states of the last token in the direction layer ($L_{dir}$) to stay closer to $\Delta_{aligned}$ but further from $\Delta_{harmful}$, and verify if $\mathcal{M}$ begins answering the prompt.  
% \gt{I am confused. Why will staying closer to $\Delta_{aligned}$ and further away from $\Delta_{harmful}$ cause the model to answer harmful questions? Shouldn't it be the opposite?}
% \gt{Oh I get it. Because if the harmful question is closer to the aligned direction, then the model will answer it. I think you may want to make this point clear. Otherwise readers might get confused as I did initially.}\yk{Yes, you are correct.}
Formally, we change $\mathcal{M}$'s hidden states as follows: 
\begin{equation}
    \vec{h}_{L_{dir}}^{last} = \vec{h}_{L_{dir}}^{last} + \alpha \Delta_{aligned} - \beta \Delta_{hamrful}
\end{equation}
\noindent where $\vec{h}_{L_{dir}}^{last}$ represents the hidden states of the last token in layer $L_{dir}$, and $\alpha$ and $\beta$ are parameters controlling how much we shift the directions.

\vspace{0.25em}
\noindent\textbf{Study Setup:} We perform the study above using the five models described in~\autoref{tab:model}. To obtain $\Delta_{aligned}$ and $\Delta_{harmful}$, we use a benign dataset proposed in~\cite{zou2023representation} and a harmful dataset with 100 samples we used to pollute the fine-tuning process (recall~\S\ref{subsec:studysettings}). We set $\alpha$ and $\beta$ to 1 for simplicity. $L_{dir}$ is picked at the two-thirds position of the hidden layers (respectively, layer 12, 21, 26, 21, 21 for \gemma, \llamasmall, \llamabig, \mistral, and \qwen), based on the rationale we will present in~\S\ref{subsubsec:getdirection}. %We apply the same QLoRA setting as in \S\ref{subsec:studysettings} to optimize the model parameters.

%We break the alignment for the five official models in ~\S\ref{sec:motivation}. As demonstrated in \cite{zou2023representation}, the middle layers are more suitable for extracting direction since directions with different properties are more clustered and distinguishable. Therefore, we set the layer $L_d$ to 2/3 position of the total layer and apply the floor function to convert the number to integer, which is 12, 21, 26, 21, 21 for \gemma~\cite{gemma}, \llamasmall~\cite{llama2}, \llamabig~\cite{llama2}, \mistral~\cite{mistral}, and \qwen~\cite{qwen}, respectively.  For the harmful dataset $\mathcal{D}_{harmful}$, we use the same 500 harmful samples in ~\S\ref{subsec:studysettings}. For the benign dataset $\mathcal{D}_{benign}$, we leverage the dataset provided in ~\cite{zou2023representation}.  We apply the same QLoRA setting as in \S\ref{subsec:studysettings} to optimize the model parameters. We also set the learning rate to 5$e^{-5}$ and the batch size to 8. For alignment evaluation, we apply the same alignment evaluation dataset and metric as in ~\S\ref{subsec:studysettings}.

\begin{table}[]
         \scriptsize
         \setlength\tabcolsep{3.1pt}
         \renewcommand{\arraystretch}{0.8}
 	\centering
\caption{The harmful rate (\%) of LLMs \texttt{before} and \texttt{after} we adjust their directions to break the alignment.}
\label{tab:broken_direction}
\vspace{-0.9em}
\begin{tabular}{lccccc}
\toprule
\textbf{Model}    & \textbf{\gemma} & \textbf{\llamasmall} & \textbf{\llamabig} & \textbf{\mistral} & \textbf{\qwen} \\
\midrule
\texttt{Before} &  4.57     &   0.00         &     0.00     &  11.71  &  2.43    \\
\texttt{After}  &   80.43   &   72.57     &   45.28      & 82.00   &  77.28 \\
\bottomrule
\end{tabular}
 \vspace{-0.75em}
\end{table}

\vspace{0.25em}
\noindent\textbf{Study Results:} We re-measure the harmful rate of the five LLMs after adjusting their internal directions and summarize the results in~\autoref{tab:broken_direction}. All the models present a harmful rate that dramatically increased from nearly 0 to a significant level (45.28\% - 82.00\%), evidencing that we effectively compromised their alignment and leading us to this insight:

\begin{tcolorbox}[colback=red!20!white, colframe=yellow!50!black]
\textit{An LLM's internal $\Delta_{aligned}$ and $\Delta_{harmful}$ directions help determine its alignment behavior}.
\end{tcolorbox}

\begin{figure*}[ht]
  \centering
  \scriptsize

    \includegraphics[width=\textwidth]{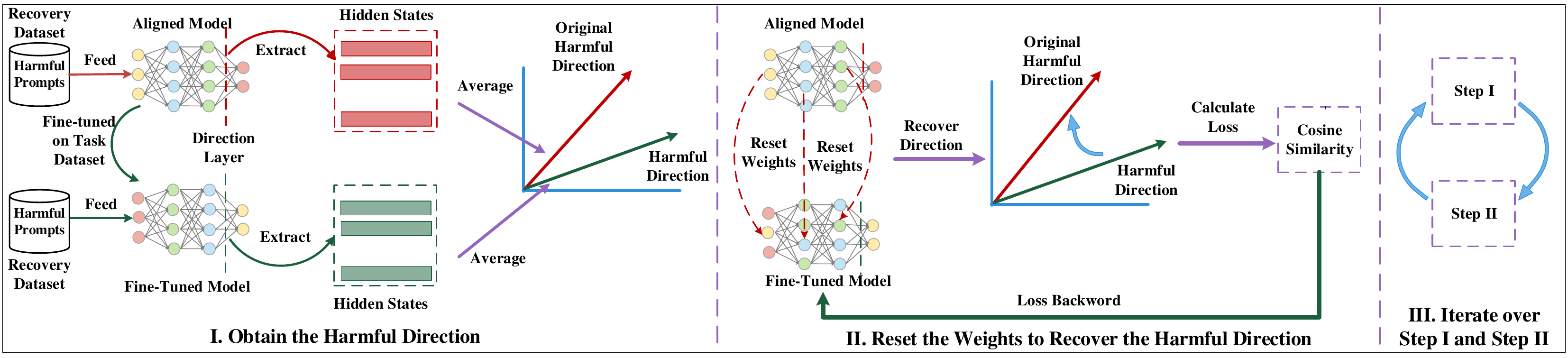}
    \caption{Workflow of our method.\label{fig:framework}}
  % align the harmful direction of the fine-tuned LLM to that of the aligned LLM.
  % \vspace{-1.25em}
  \label{fig:parameters_dist}
\end{figure*}

\subsection{Design}

Based on the insight unveiled above, we perceive that an aligned LLM can reject harmful prompts because it has an established, effective $\Delta_{harmful}$ direction. Fine-tuning, negatively shifting $\Delta_{harmful}$, destructs the LLM's alignment. This enables us to design a method, as illustrated in~\autoref{fig:framework}, to recover an LLM's alignment after fine-tuning is performed. 

At the high level, given an aligned LLM $\mathcal{M}$ and its fine-tuned version $\mathcal{M}_f$, we iteratively reset $\mathcal{M}_f$'s weights to make its $\Delta_{harmful}$ direction align with $\mathcal{M}$. Technically, our method involves three steps: \ding{182} obtaining $\Delta_{harmful}$ from the direction layer $L_{dir}$ for both $\mathcal{M}$ and $\mathcal{M}_f$, \ding{183} identifying and restoring a subset of $\mathcal{M}_f$’s weights to make $\mathcal{M}_f$ close to $\mathcal{M}$ regarding $\Delta_{harmful}$, and \ding{184} repeating the above two steps iteratively until reaching the predefined conditions. 

\subsubsection{Obtaining $\Delta_{harmful}$\label{subsubsec:getdirection}}

Our method starts with extracting $\Delta_{harmful}$, the harmful direction, from $\mathcal{M}$ and $\mathcal{M}_f$. The process is straightforward. We construct a \textit{recovery dataset}, annotated as $\mathcal{D}_{rec}$. $\mathcal{D}_{rec}$ is full of harmful prompts and independent of $\mathcal{M}$. Using $\mathcal{D}_{rec}$, we obtain $\Delta_{harmful}$ from $\mathcal{M}$ and $\mathcal{M}_f$ following~\autoref{def:direction}.

%\jx{Need to discuss $\mathcal{D}_{rec}$'s diversity and size. Obviously, those two factors will significantly affect the recovery effectiveness}. 

%Since our goal is to align the harmful direction of the fine-tuned model to that of the aligned model, the first step would be to gather those directions. In detail, we first construct a harmful prompt dataset, called \textit{recovery dataset} $\mathcal{D}_{rec}$. Then we generate the harmful direction $\Delta_{harmful}$ and $\Delta_{harmful}^f$ by feeding $\mathcal{D}_{rec}$ into aligned LLM $\mathcal{M}$ and fine-tuned LLM $\mathcal{M}_f$, respectively.

\vspace{0.25em}
\textbf{Constructing the Recovery Dataset:} The recovery dataset is crucial to the success of our method as it determines the fidelity of $\Delta_{harmful}$. In the ideal case, we should build a recovery dataset to incorporate various categories of harmful prompts and include abundant samples in each category. However, this is expensive. More importantly, we found that the diversity and quantity of harmful prompts are less decisive than expected. As we will explain later in this section, we pick a direction layer $L_{dir}$ toward the back of the network. Given any harmful prompt, the hidden states after that layer become more constant to simply indicate the prompt is harmful. As such, including more categories and samples of harmful prompts makes an insignificant difference.

To support our reasoning above, we perform two empirical studies. In the first study, we vary the number of prompt categories and measure the variations in the resulting harmful direction and alignment recovery. It shows that \textit{the variations are negligible}. More details about the study and results are presented in~\S\ref{subsec:recovery_diversity}. In the second study, we adapt the number of prompt samples from a fixed set of categories and redo the measurement. It presents results similar to the first study, showing that the prompt quantity has limited impact. Related details are also described in~\S\ref{subsec:recovery_diversity}. 

Inspired by the studies, we opt for a recovery dataset with a moderate number of categories and samples. By default, we use 256 harmful prompts spanning 14 categories from the cleaned BeaverTails dataset we described in~\S\ref{subsec:studysettings}. To avoid intervening in our evaluations, the 256 prompts have no overlap with the 1,500 harmful prompts we use to pollute the fine-tuning process.

\vspace{0.25em}
\textbf{Picking the Direction Layer:} Our method critically depends on picking a direction layer $L_{dir}$ that can properly represent and control the alignment property. Recent studies~\cite{zou2023representation, turner2023activation} suggest that a hidden layer in the middle of the model often suffices. However, as we will explain in~\S\ref{subsubsection:ssggd}, we only restore the weights on $L_{dir}$ and the layers before. Picking the middle layer leaves the latter half of the weights untouched, which can still break the alignment established by the front layers. This inspires us to pick a layer toward the back.

Yet, picking a $L_{dir}$ close to the end incurs a higher time cost as more weights need to be identified and restored. To this end, we launch a study on the trade-off between alignment recovery and time expenditure. We vary the direction layer and measure its impact on the effectiveness of alignment recovery and the corresponding time cost. As we will elaborate in~\S\ref{subsec:ablation:layer}, a $L_{dir}$ closer to the back indeed improves alignment recovery. However, picking $L_{dir}$ after the two-thirds position of the hidden layers brings limited benefits to alignment while significantly increasing the time cost. Thus, by default, we set $L_{dir}$ at the two-thirds position of the hidden layers.

\subsubsection{Recovering Alignment}
\label{subsubsection:ssggd}
% After deriving the harmful direction, we push the harmful direction of fine-tuned LLM $\mathcal{M}_f$ toward the harmful direction of the aligned LLM $\mathcal{M}$. 

To recover the alignment, our idea is to identify and restore a subset of $\mathcal{M}_f$'s weights to $\mathcal{M}$'s value such that the two models have minimized disparity on $\Delta_{harmful}$. To reduce the impacts on $\mathcal{M}_f$'s performance, we aim only to touch the smallest number of weights. Formally, this can be described as the optimization problem below.

\definecolor{lightblue}{RGB}{230, 240, 255}
\definecolor{darkgray}{RGB}{100, 100, 100}

\begin{mdframed}[backgroundcolor=lightblue, linecolor=darkgray, linewidth=1pt, roundcorner=5pt]
\noindent\textbf{Objective:}
\begin{align*}
\text{minimize} \;\;\;& \mathcal{L} ( \Delta_{harmful}, \Delta_{harmful}^f) + ||1-\mathbb{W}||_0 \; 
\end{align*}
\noindent\textbf{Subject To:}
\vspace{-0.5em}
\begin{align*}
\mathcal{M}_f = \mathcal{M} + \delta \odot \mathbb{W}
\end{align*}
where $\Delta_{harmful}$ and $\Delta_{harmful}^f$ are the harmful directions of $\mathcal{M}$ and $\mathcal{M}_f$, $\mathcal{L} $ is a loss function, $\mathbb{W} \in \{0, 1\}$ represents the variable we aim to optimize, $\odot$ means element-wise product, and $\delta$ stands for weight changes incurred by fine tuning. $\mathbb{W}$ is initialized as all ones, and any element $w \in \mathbb{W}$ set to zero means our optimization restores the corresponding weight on $\mathcal{M}_f$ to $\mathcal{M}$. The first term of the \textbf{Objective} function, namely $\mathcal{L} ( \Delta_{harmful}, \Delta_{harmful}^f$), aims to recover the harmful direction while the second term ($||1-\mathbb{W}||_0$) rewards a smaller number of weights.    
\end{mdframed}

\vspace{0.25em}
\textbf{Our Solution:} 
The optimization problem above is an NP-hard problem. We develop a heuristic algorithm combining gradient descent and a greedy strategy to solve it. Our algorithm, outlined in~\autoref{alg:ggd}, iteratively finds the most effective weights in $\mathcal{M}_f$ toward satisfying the objective and performs one-step gradient descent to reset those weights.

% To push the harmful direction  $\Delta_{harmful}^d$  of fine-tuned LLM $\mathcal{M}_f$ toward the initial harmful direction  $\Delta_{harmful}$ of aligned LLM $\mathcal{M}$, the straightforward way is to directly optimize the weights of  $\mathcal{M}_f$. However, optimizing all parameters is likely to impact task performance. Besides, we also need to adjust the hyper-parameters like the learning rate for each model.

% Wei~\etal~\cite{wei2024assessing} pointed out that the alignment of the LLM can be compromised by solely removing 3\% safety-related neurons while mostly maintaining utility. 

% the iteratively recover the most important parameters that can minimize the loss $ \mathcal{L} ( d, \hat{d})$ as shown in the following steps.

% By optimizing the values in $\mathbb{M}$ to zero, we reset the corresponding parameters of the fine-tuned LLM to the parameters of the aligned model, hence aligning the harmful direction. Besides, to minimize the recovered parameters, we also require to make $||1-\mathbb{M}||_1$ as small as possible.

\vspace{0.25em}
\noindent\ding{182} \textit{Collecting Gradients:} Aiming to minimize the disparity between $\Delta_{harmful}$ and $\Delta_{harmful}^f$, we adopt negative cosine similarity as the loss function: 
    \begin{equation}
        \mathcal{L}(\cdot) = -cosine\_similarity(\Delta_{harmful}, \Delta_{harmful}^f)
        \label{eq:loss}
    \end{equation}

Based on~\autoref{def:direction}, minimizing this function is an objective on the output of the direction layer $L_{dir}$ in $\mathcal{M}_f$. To understand \textit{which weights on $\mathcal{M}_f$ to change} and \textit{how to change them} for fulfilling the objective, we perform backward propagation from $L_{dir}$ to the initial layer and calculate the gradient of the loss function with respect to each weight. Layers after $L_{dir}$ are disregarded as they are irrelevant to $\Delta_{harmful}$ and $\Delta_{harmful}^f$.

%, which is independent of the layers after $L_{dir}$. 

%To identify the most effective parameters in $\mathcal{M}_f$, we first derive the gradients of $\mathcal{M}_f$, which indicates the importance of the corresponding parameters in minimizing the loss. In detail, we adopt the cosine similarity function to compute the loss for  $\Delta_{harmful}$ and $\Delta_{harmful}^f$ since we aim to align the true harmful direction, not the exact value of this direction. Furthermore, to minimize the loss, we need to change the loss to minus value as:  
%    \begin{equation}
%        \mathcal{L}(\cdot) = -cosine\_similarity(\Delta_{harmful}, \Delta_{harmful}^f)
%        \label{eq:loss}
%    \end{equation}
%Then, we gain all the gradients for the model $\mathcal{M}_f$ through backward propagation (line 5). Noteworthy, since we calculate the loss based on the outputs from the direction layer $L_{dir}$, the parameters after this layer will not have the gradients, thus not involved in the following steps.

% Many loss functions are available for this purpose such as mean squared error loss (MSE). Here, w
\vspace{0.25em}
\noindent\ding{183} \textit{Selecting and Restoring Weights:} After getting the gradients, the intuitive idea is to adjust the weights with gradient descent, where we update each weight based on its gradient and the learning rate. Yet, we can only maintain a weight or update it to $\mathcal{M}$'s value, as our objective restricts. 

To address this issue, we adopt the fast gradient sign method (FGSM) presented in~\cite{goodfellow2014explaining}. Given a weight $w_f$ on $\mathcal{M}_f$ and its counterpart $w$ on $\mathcal{M}$, if $w_f$'s gradient has a signedness showing that gradient descent will move $w_f$ closer to $w$, we replace $w_f$ with $w$. Otherwise, we do not change $w_f$ because applying gradient descent will push $w_f$ further away from $w$, reversely affecting alignment recovery. Formally, 
\begin{equation}
\label{eq:restore}
\begin{aligned}
      w_f = 
\begin{cases} 
w & \text{if \quad}  gradient(w_f) \cdot (w_f - w) > 0, \\
w_f & \text{else}
\end{cases}
\end{aligned}
\end{equation}
Despite the strong restriction above, a large number of weights can remain to be restored, incurring a higher chance of revoking the effectiveness of fine-tuning. To further reduce the number of weights, we adopt a greedy strategy by selecting the top $P\%$ from the identified weights based on their absolute gradient values (precisely, we pick the larger ones). We eventually only restore those weights.

%However, there are a huge amount of parameters that satisfy the above constrain \autoref{eq:constrain}.  To lessen the impact on task accuracy for the fine-tuned model, we have to minimize the number of parameters that reset to $\mathcal{M}$. Therefore, we adopt a greedy strategy that selects the $P\% $ parameters with the largest absolute gradient values from all the satisfied parameters (line 8).  Finally, we reset those parameters of $\mathcal{M}_f$ to the aligned model $\mathcal{M}$. To further reduce the number of parameters, we 

\vspace{0.25em}
\noindent\ding{184} \textit{Iterative Recovery:} Due to some greedy designs of~\autoref{alg:ggd} (i.e., the use of FGSM and $P\%$), one round of recovery following steps \ding{182} and \ding{183} often falls short of reinstating the alignment. Accordingly, we repeat the process to perform iterative recovery until a user-specified epoch number $E$ is reached. Adopting this iterative recovery offers another flexibility --- we can configure $P\%$ to a smaller value to restrict the impact of each round of recovery on the task performance. By default, we use 0.2\% for $P\%$.

%To better trade-off between task performance and alignment, we repeat \S\ref{subsubsection:direction} and \S\ref{subsubsection:ssggd} with a small recovery rate $P\%$ until reaching the following conditions: 1) the performance is dropped over $t\%$ (eg. 5\%); 2) the total epoch $E$ has been executed. 
 
% reset a small step rate $P\% $ and perform greedy gradient descent multiple times.

\normalem
\begin{algorithm}[!t]
\setstretch{1.08} 
 \SetInd{0.2em}{0.5em}
\SetKwFunction{FMain}{}
  \SetKwProg{Fn}{Globals}{:}{}
   \Fn{}{
 
$\Delta_{harmful}$ \tcc*[h]{Oracle harmful direction}\\
$\Delta_{aligned}$  \tcc*[h]{Oracle aligned direction} \\
$L_{dir}$   \tcc*[h]{Direction Layer} 
\BlankLine
}

 \footnotesize
\caption{Alignment Recovery}\label{alg:ggd}
\SetKwFunction{FMain}{reset\_weights}
  \SetKwProg{Fn}{Function}{:}{}
  \Fn{\FMain{$\mathcal{M}$, $\mathcal{M}_f$, $\mathcal{D}$, $\Delta$, $P$}}{
  \tcc{This function resets $P\%$ weights of $\mathcal{M}_f$ to $\mathcal{M}$'s value based on direction $\Delta$}
$\Delta_{f}$ = extract\_direction$(\mathcal{M}_f, \mathcal{D}, L_{dir})$\\
loss = $\mathcal{L}(\Delta, \Delta_{f})$ \\
loss.backward()\\
$ \mathcal{W}$ = weights($\mathcal{M}_f$) where $(\mathcal{M}_f -  \mathcal{M}) \cdot  \mathcal{M}_{f}.grad  > 0$\\ %\tcc{Select weights}

$\mathcal{W}$ = top\_abs\_grad($\mathcal{W}$, $P$)\\  
$\mathcal{M}_{f}[\mathcal{W}] =  \mathcal{M}[\mathcal{W}]$  \tcc*[h]{Reset weights}

\Return $\mathcal{M}_{f}$
}

\SetKwFunction{FMain}{recovery}
  \SetKwProg{Fn}{Function}{:}{}
  \Fn{ \footnotesize{\FMain{$\mathcal{M}$, $\mathcal{M}_f$, $\mathcal{D}_{rec}$, $\mathcal{D}_{roll}$, $Rnd$, $RollBack$, $Fuse$}}}{

 $\mathcal{M}_0$ = $ \mathcal{M}_f$ \tcc*[h]{Initialize  the value} \\
\For{$e$ = 1 to $Rnd$}{

    $\mathcal{M}_e$ = $ \mathcal{M}_{e-1}$  \\
    \tcc{Reset weights for alignment recovery}
    $\mathcal{M}_{e}$ = reset\_weigths($\mathcal{M}$, $\mathcal{M}_e$, $\mathcal{D}_{rec}$, $\Delta_{harmful}$, 0.2\%)

    \If{$Rollback$ }{
         \tcc{Reset weights for rollback}
        \footnotesize{$\mathcal{M}_{e}$ = reset\_weigths($\mathcal{M}_f$, $\mathcal{M}_e$, $\mathcal{D}_{roll}$, $\Delta_{aligned}$, 20\%) }

    }
    \tcc{If performance drops over a threshold and cutoff is enabled, stop.}
    \If{performance\_drop($\mathcal{M}_{e}$) $>$ 5\% \textbf{and} $Fuse$}{
    \Return  $\mathcal{M}_{e-1}$
    }
}
\Return $\mathcal{M}_{e}$
}

\SetKwFunction{algo}{algo}\SetKwFunction{proc}{proc}
\SetKwProg{myalg}{Algorithm}{}{}
\myalg{ }{
\Input{Aligned Model $\mathcal{M}$, Fine-tuned Model $\mathcal{M}_f$, Direction Layer $L_{dir}$, Recovery Data $\mathcal{D}_{rec}$,  Rollback Data $\mathcal{D}_{roll}$.}
\Output{Re-aligned Model}

\tcc{Get oracle harmful and aligned direction}
$\Delta_{harmful}$ = extract\_direction$(\mathcal{M}, \enspace \mathcal{D}_{rec}, \enspace L_{dir})$\\
% \tcc{Configure oracle aligned direction}
$\Delta_{aligned}$\enspace \enspace = extract\_direction$(\mathcal{M}_f, \mathcal{D}_{roll}, L_{dir})$\\

%$\mathcal{M}_r$ = extract\_direction$(\mathcal{M}, \mathcal{D}_{rec}, L_{dir})$\;
\tcc{Trial for five recovery rounds}
$\mathcal{M}'_f$ = recovery($\mathcal{M}$, $\mathcal{M}_f$, $\mathcal{D}_{rec}$, $\mathcal{D}_{roll}$, 5, $False$, $False$)

% \tcc{Performance drop over a threshold and cutoff enabled}
\uIf{performance\_drop($\mathcal{M}'_f$) $<$ 5\% }{
    \tcc{Run 15 more recovery rounds}
      \footnotesize{$\mathcal{M}_{rec}$ = recovery($\mathcal{M}$, $\mathcal{M}'_f$, $\mathcal{D}_{rec}$, $\mathcal{D}_{roll}$, 15, $False$, $True$)}
}\Else{
    \tcc{Redo 20 recovery rounds with rollback}
    \footnotesize{$\mathcal{M}_{rec}$ = recovery($\mathcal{M}$, $\mathcal{M}_f$, $\mathcal{D}_{rec}$, $\mathcal{D}_{roll}$, 20, $True$, $True$)}
  }
\Return $\mathcal{M}_{rec}$
}

\end{algorithm}
\ULforem
\subsubsection{Mitigating Side Effects}
\label{subsubsection:rollback}

While we conservatively restrict the number of weights to restore, they may still hurt the task performance unexpectedly. As a mitigation and when possible, we incorporate a rollback mechanism that identifies and reverts a subset of restored weights after step \ding{183}, following line $ 18 \sim 20 $ in~\autoref{alg:ggd}. 

We do not enable the rollback by default because our recovery algorithm does not necessarily degrade the task performance, while the rollback operations certainly intensify the time cost. Alternatively, we start our algorithm with a few rollback-free, warm-up recovery rounds (5 rounds by default). If the task performance degradation is within a predetermined threshold (5\%), we continue with more rollback-free recovery rounds until the user-specified epoch is reached. Otherwise, our recovery is proven aggressive, and we restart the entire algorithm with rollback enabled in every recovery round. 

%However, for most models, running our algorithm with rollback is unnecessary as it typically does not significantly degrade task performance and doubles the time expenditure. We enable rollback only for models whose task performance declines rapidly, indicating excessive degradation caused by our algorithm. Specifically, to determine whether we require the rollback component, we initially execute our recovery algorithm without rollback for several warm-up epochs (e.g., five). If task performance declines beyond a predetermined threshold (5\%) after these epochs, we restart the algorithm with rollback enabled (line 29). Otherwise, we proceed without rollback (line 31).
% only modifies a small portion of the total parameters (4\%).   , 

In a recovery round, our rollback works as follows. Given $\mathcal{M}_{f}$ (the original fine-tuned model without any recovery) and $\mathcal{M}_{e}$ (the model after the round of recovery), we obtain both their \textit{alignment directions}, $\Delta_{aligned}$ and $\Delta_{aligned}^e$, following~\autoref{def:direction} with a dataset full of benign prompts. This dataset, annotated as $\mathcal{D}_{roll}$, can be constructed from the original training dataset following the strategy presented in~\S\ref{subsubsec:getdirection}. Fundamentally, our recovery degrades the task performance because we unintentionally hurt $\Delta_{aligned}$. Hence, our rollback focuses on re-establishing $\Delta_{aligned}$ by minimizing the following loss:
\begin{equation}
      \mathcal{L}(\cdot) = -cosine\_similarity(\Delta_{aligned}, \Delta_{aligned}^e)  
\end{equation}

To achieve the objective above, we reuse the greedy gradient descent algorithm presented in \S\ref{subsubsection:ssggd} to find and revert $R\%$ of the weights restored during the recovery round. Here, $R\%$ represents a tradeoff between alignment recovery and performance preservation. A high $R\%$ will revert too many restored weights, sabotaging the recovered alignment. In contrast, a low $R\%$ fails to effectively reinstate the compromised performance. By default, we set $R\%$ to 20\%.

We observe that our alignment recovery, with rollback or without rollback but after warm-up, may still hurt the task performance to a meaningful extent. On account of this, we introduce another safety fuse. Once the warm-up phase concludes, we monitor the task performance after each recovery iteration. If the performance drops by a certain threshold (5\% by default), we discontinue the recovery and return the model before that iteration as the final result.

\vspace{0.25em}
\textbf{Discussion:} Our rollback mechanism requires measuring the task performance during alignment recovery. This is not always possible. In our target \textbf{Scenario I} (recall~\S\ref{sec:prob}), the fine-tuning is performed by the untrustworthy \textit{Fine-tuner} while the alignment recovery is completed by the model \textit{Owner}. The \textit{Owner} lacks reliable methods and testing data to assess the task performance, and thus, cannot perform the rollback. In such cases, we assume the task performance does not drop (i.e., \texttt{performance\_drop(M)} $\equiv$ 0), essentially disabling the rollback mechanism.

\section{Evaluation (for Both Scenarios I and II)}

To understand the utility of our method, we perform a series of evaluations focusing on the following aspects.

\vspace{0.25em}
\noindent\ding{192} \textit{Can our method effectively recover the alignment compromised by LLM fine-tuning?}

\vspace{0.25em}
\noindent\ding{193} \textit{Can our method maintain the task performance when performing alignment recovery?}

\vspace{0.25em}
\noindent\ding{194} \textit{Can our method outperform the existing solutions for alignment recovery?}

\vspace{0.25em}
\noindent\ding{195} \textit{How will the hyper-parameters and evaluation settings affect the alignment recovery of our method?}\\

Recall~\S\ref{sec:prob} that we have two target scenarios (\textbf{Scenario I} and \textbf{Scenario II}). In this section, we present evaluations applicable to both scenarios. Additional evaluation unique to \textbf{Scenario I} is presented in~\S\ref{sec:extraeva}.

\subsection{Experimental Setup \label{subsec:eval:setup}}
% All the models are trained with QLoRA~\cite{qlora} for one epoch and are quantized in 4 bits. For the \lora~\cite{lora} setting, we set r=8 and alpha=16. For the training setting, the learning rate is set to 2e-5 and the batch size is set to 48 or 96 based on the dataset.

\vspace{0.25em}
\noindent\textbf{Models:} We reuse the five models from our motivating study (\S\ref{subsec:studysettings}) for fine-tuning and alignment recovery. Details of the models are summarized in~\autoref{tab:model}.

\vspace{0.25em}
\noindent\textbf{Fine-tuning:} We perform fine-tuning using the five datasets described in~\autoref{tab:dataset}. As explained in~\S\ref{subsec:studysettings}, we inject 100, 500, 1,500, and 392 (after moderation) harmful samples into each dataset to create four more variants, resulting in five versions in total. On each version, we independently fine-tune each target model using QLoRA with the parameters specified in~\S\ref{subsec:studysettings}. At the end, we produce 25 fine-tuned versions for each target model.

%As done in \S\ref{subsec:studysettings}, we first build a total of 25 datasets by injecting various harmful data into the task dataset in \autoref{tab:dataset}. Finally, by fine-tuning the above five target models on the datasets, we obtain a total of 125 fine-tuned models. The fine-tuning process is the same as in\S\ref{subsec:studysettings}. 

\begin{figure*}[t]
	\centering  
        % \scriptsize
		% \setlength{\abovecaptionskip}{0.cm}
        \renewcommand{\arraystretch}{0.8}
	\setlength\tabcolsep{0pt}
	\begin{tabular}{m{0.3cm}m{3.4cm}m{3.4cm}m{3.4cm}m{3.4cm}m{3.4cm}}
            
            \rotatebox[origin=c]{90}{\textbf{\scriptsize \quad \quad \quad \sql}} &
            \includegraphics[width=0.19\textwidth]{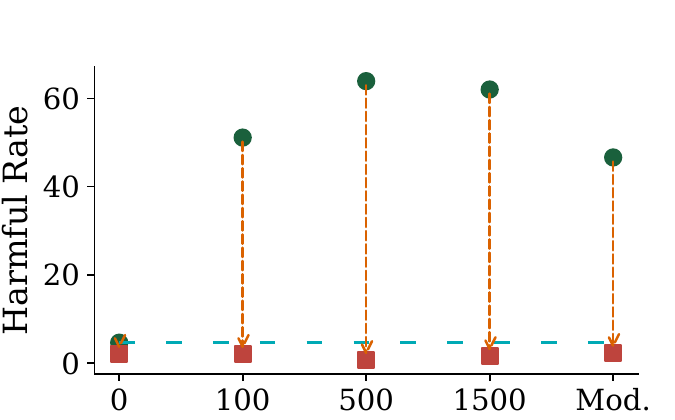} & 
              \includegraphics[width=0.19\textwidth]{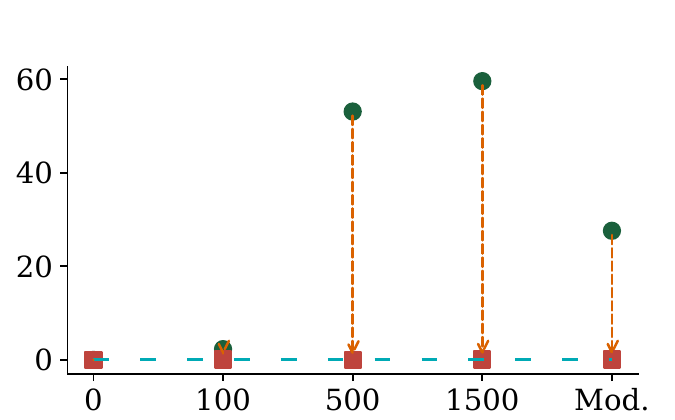} & 
              \includegraphics[width=0.19\textwidth]{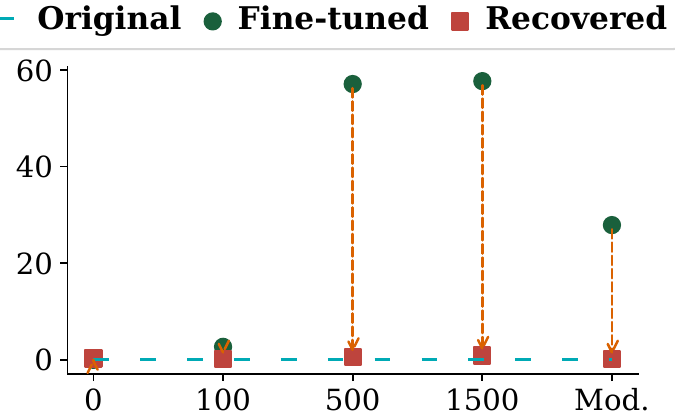} & 
              \includegraphics[width=0.19\textwidth]{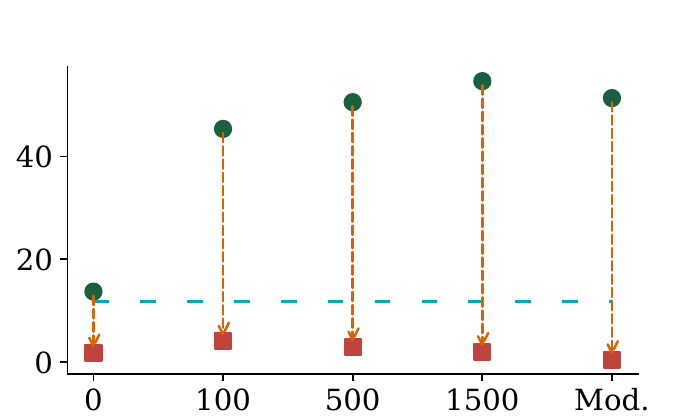} & 
              \includegraphics[width=0.19\textwidth]{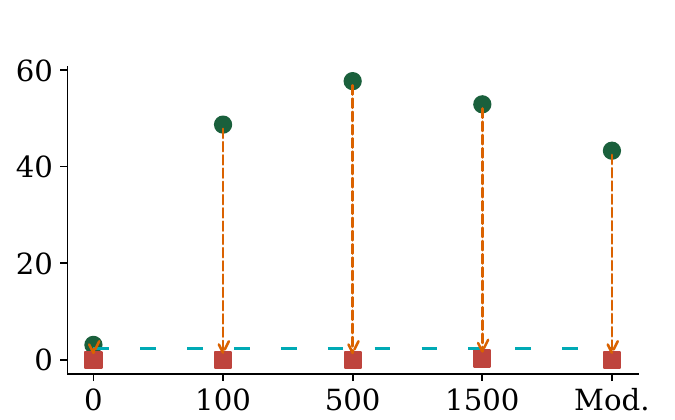} \cr

            \raisebox{0.5cm}{\rotatebox[origin=c]{90}{\textbf{\scriptsize \cheat}}}  &
            \includegraphics[width=0.19\textwidth]{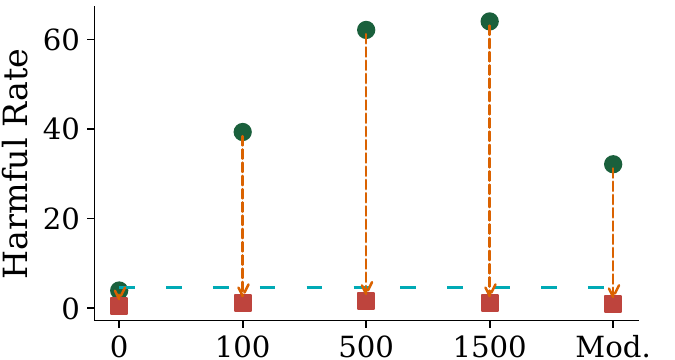} & 
              \includegraphics[width=0.19\textwidth]{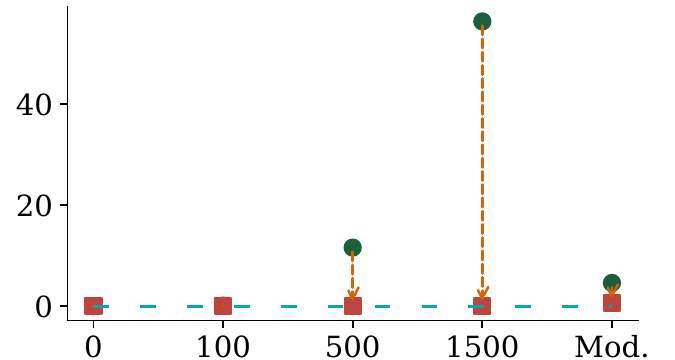} & 
              \includegraphics[width=0.19\textwidth]{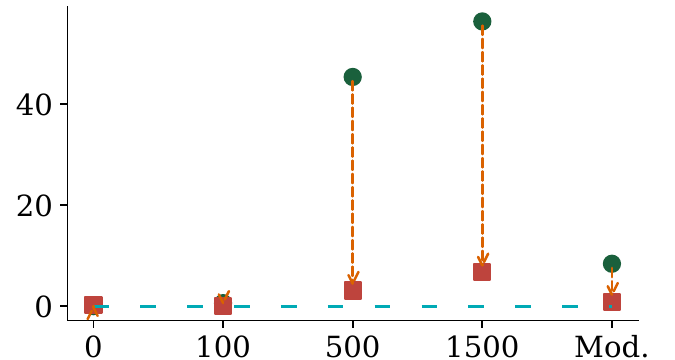} & 
              \includegraphics[width=0.19\textwidth]{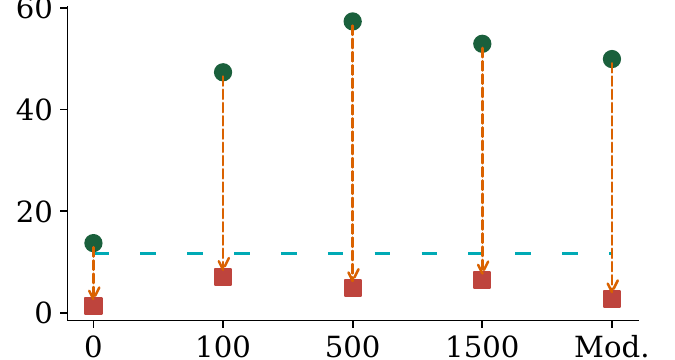} & 
              \includegraphics[width=0.19\textwidth]{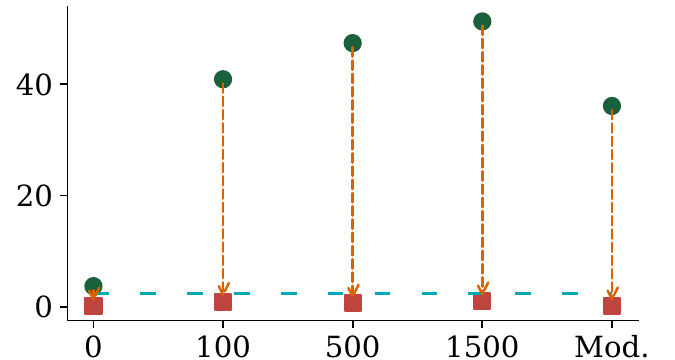} \cr
              
              \raisebox{0.65cm}{\rotatebox[origin=c]{90}{\textbf{\scriptsize \nlbash}}} &
            \includegraphics[width=0.19\textwidth]{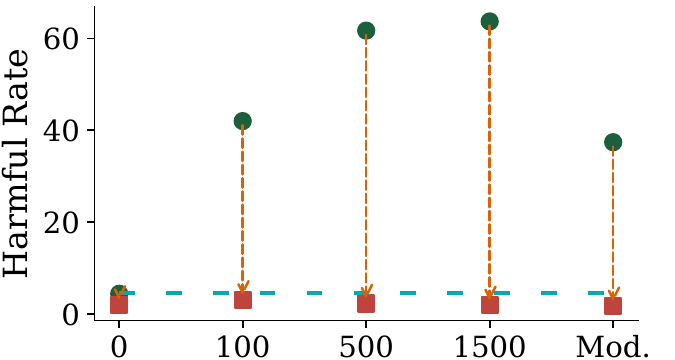} & 
              \includegraphics[width=0.19\textwidth]{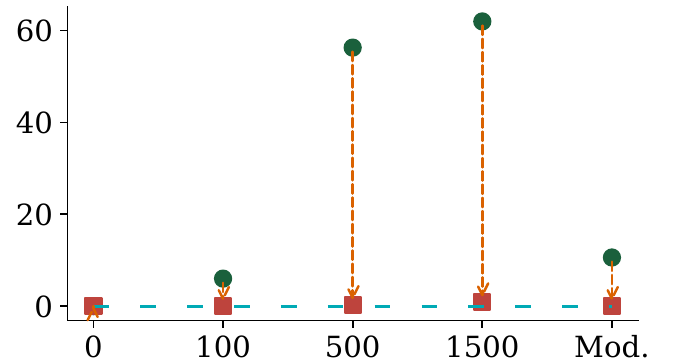} & 
              \includegraphics[width=0.19\textwidth]{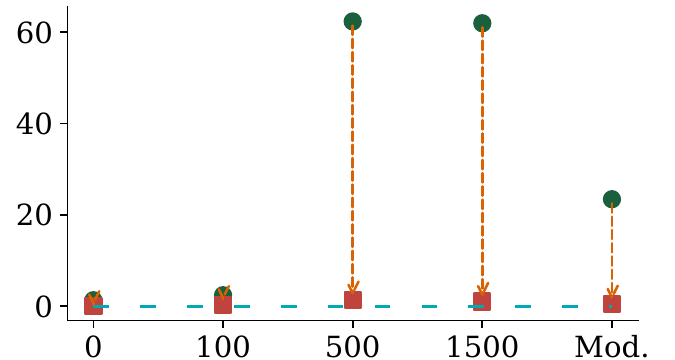} & 
              \includegraphics[width=0.19\textwidth]{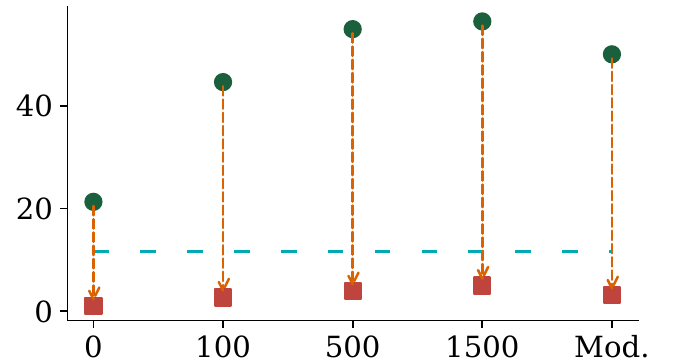} & 
              \includegraphics[width=0.19\textwidth]{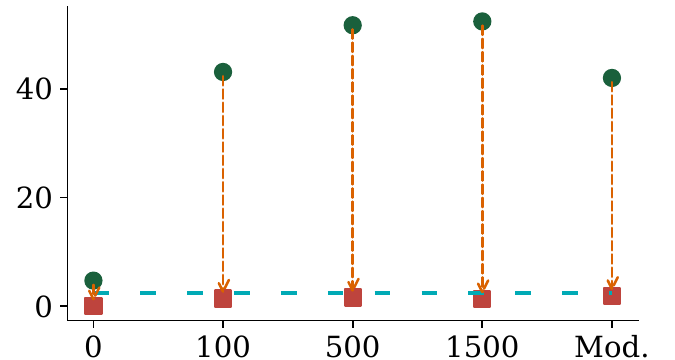} \cr

            \raisebox{0.60cm}{\rotatebox[origin=c]{90}{\textbf{\scriptsize  \samsum}}} &
		\includegraphics[width=0.19\textwidth]{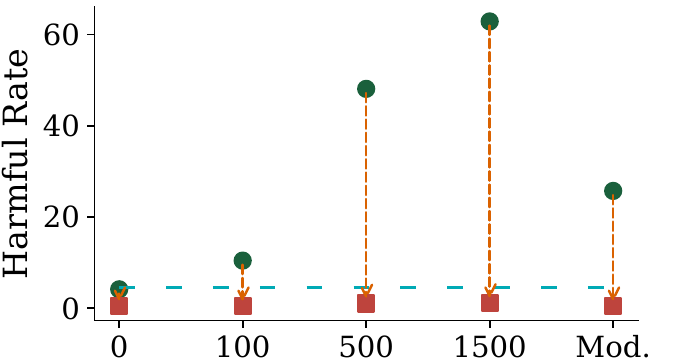} & 
              \includegraphics[width=0.19\textwidth]{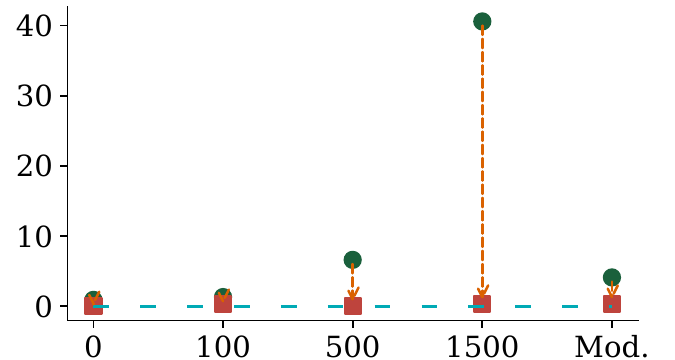} & 
              \includegraphics[width=0.19\textwidth]{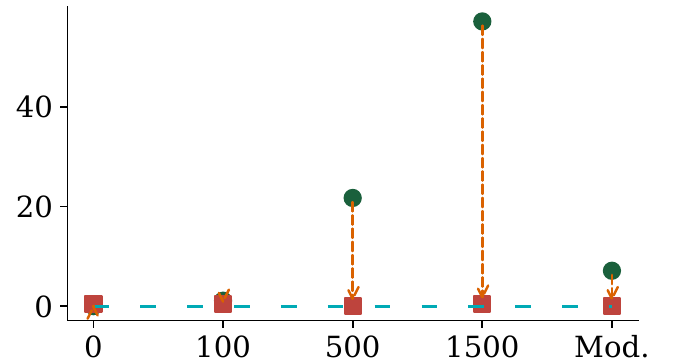} & 
              \includegraphics[width=0.19\textwidth]{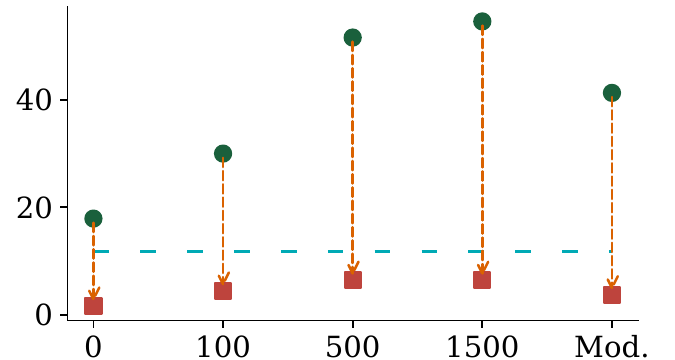} & 
              \includegraphics[width=0.19\textwidth]{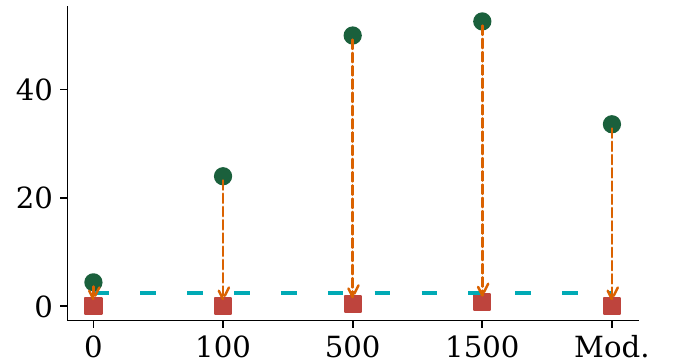} \cr

             \raisebox{0.60cm}{\rotatebox[origin=c]{90}{\textbf{\scriptsize \toxicity }}} &
            \includegraphics[width=0.19\textwidth]{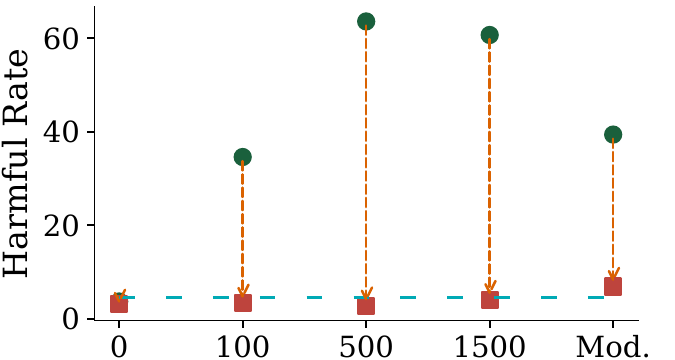} & 
              \includegraphics[width=0.19\textwidth]{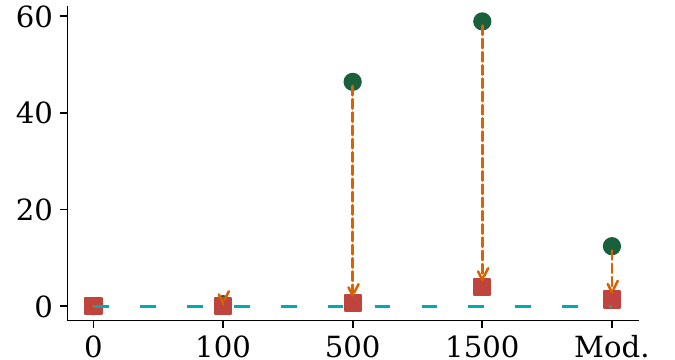} & 
              \includegraphics[width=0.19\textwidth]{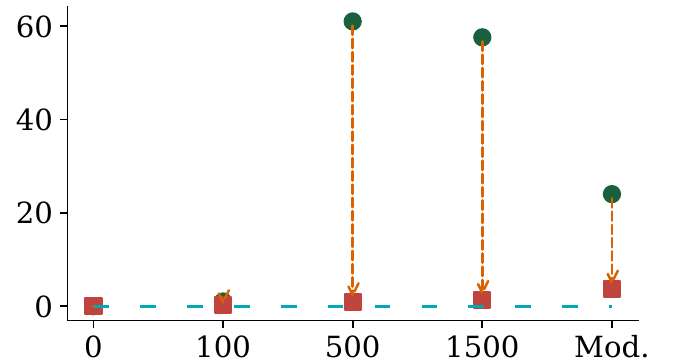} & 
              \includegraphics[width=0.19\textwidth]{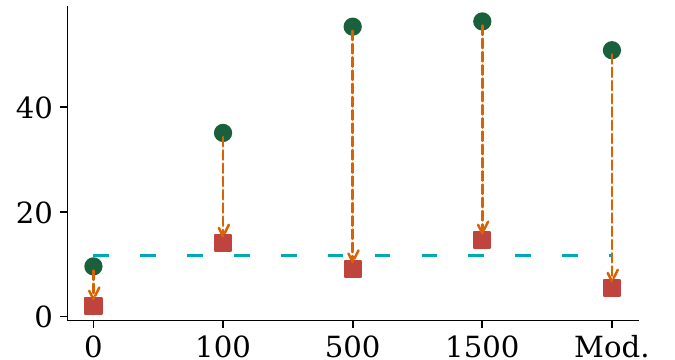} & 
              \includegraphics[width=0.19\textwidth]{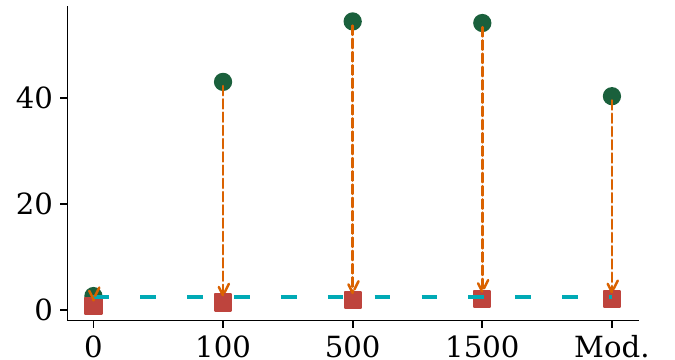} \cr
            \multicolumn{6}{c}{ \scriptsize \quad\quad\quad\quad \vspace{-0.2em} Harmful Sample \#\vspace{0.2em}} \cr 
            
             &  \centering {\scriptsize \textbf{(a) \gemma}} &  \centering \textbf{\scriptsize{(b) \llamasmall}} &  \centering \textbf{\quad \scriptsize{(c) \llamabig}}  &  \centering \textbf{\scriptsize{(d) \mistral}}  & \centering \textbf{\scriptsize{(e) \qwen}}  \cr
    
	\end{tabular}
    \caption{Alignment recovery results of our method. The \texttt{y-axis} represents the harmful rate and the \texttt{x-axis} shows the number of harmful samples injected into the fine-tuning process. In the figures, \texttt{Fine-tuned} stands for the fine-tuned model, \texttt{Recovered} refers to the alignment-recovered model, and \texttt{Original} denotes the alignment of aligned model.}
\label{fig:alignment}
\end{figure*}

\vspace{0.25em}
\noindent\textbf{Alignment Recovery:} As shown in~\autoref{tab:study_result}, fine-tuning compromises the alignment of the target models. We perform alignment recovery on those models, following~\autoref{alg:ggd}. We use the default recovery dataset $\mathcal{D}_{rec}$ (256 harmful prompts from the cleaned \beavertails~train dataset) and the default rollback dataset $\mathcal{D}_{roll}$ (256 benign prompts from the fine-tuning dataset). As described in~\S\ref{subsubsec:getdirection}, the 256 harmful prompts do not overlap with the harmful prompts to pollute the fine-tuning process. 

We also set the hyper-parameters to their default values, including 0.2\% as the recovery rate ($P\%$), 20\% as the rollback rate ($R\%$), 5\% as the performance drop threshold during the warm-up phase for triggering rollback, 5\% as the performance drop threshold to discontinue the recovery during the post-warm-up phase, 20 as the recovery epochs ($E$), and the two-thirds position of the hidden network as the direction layer $L_{dir}$. In cases where our rollback is invoked, we repeat the evaluation without rollback to simulate \textbf{Scenario I}.

%To restore the alignment of those fine-tuned models, We randomly select 256 harmful prompts from the cleaned \beavertails~ train dataset as the \textit{recovery dataset} $\mathcal{D}_{rec}$, which does not overlap with harmful train datasets. We also randomly select 256 task prompts from the corresponding task train dataset as the \textit{rollback dataset} $\mathcal{D}_{roll}$. We also set the recovery rate $P\%$ to 0.2\%, the rollback rate $R\%$  to 20\%, epochs $E$ to 20, and the performance drop threshold to 5\%.  For the direction layer $L_{dir}$, we set it as in \S\ref{subsec:method:insight}. 

\vspace{0.5em}

\noindent\textbf{Baseline:} We are not the first to mitigate alignment compromise during LLM fine-tuning. As we will discuss in~\S\ref{sec:related}, there are two lines of previous attempts. 

\vspace{0.5em}
\noindent \ding{182} The first line focuses on intervening in the fine-tuning process to protect the alignment. Along this line, SoftSFT~\cite{qi2024safety} demonstrates the state-of-the-art effectiveness. It hypothesizes that the first several output tokens greatly influence the alignment. Accordingly, it constrains the early tokens to avoid deviation from their initial values during fine-tuning. 

We adopt the open-source implementation of SoftSFT~\cite{qi2024safety} for fine-tuning. As the public code defaults to full parameters fine-tuning, we extend it to support the QLoRA we adopted in our evaluation. All other hyper-parameters are borrowed from the paper.

\vspace{0.5em}
\noindent \ding{183}  The second line, similar to our method, post-processes the model after fine-tuning is performed. While several approaches exist, RESTA~\cite{catqa} is a pioneering work that has demonstrated considerable effectiveness in this domain. Given an aligned model $\mathcal{M}$, RESTA obtains its safety-unaligned counterpart $\tilde{\mathcal{M}}$ following the approach presented in~\cite{bhardwaj2023language}. The delta between the weights of $\mathcal{M}$ and $\tilde{\mathcal{M}}$ are calculated as a \textit{safety vector}, which RESTA believes establishes the alignment. Provided a target model fine-tuned from $\mathcal{M}$, RESTA adds the safety vector to the model's weights to recover alignment. We consider RESTA as another baseline and run it as follows.

We perform alignment recovery using the RESTA code from~\cite{catqa} on our fine-tuned models. As we will explain in Appendix ~\S\ref{appendix:resta_setting}, the default hyper-parameters fail to derive an effective safety vector. Accordingly, we optimize the hyper-parameters by increasing RESTA's batch size from 4 to 10 and epochs from 3 to 5. We also find that while using Q\lora~ for the fine-tuning, RESTA only enables the linear modules of \texttt{q\_proj} and \texttt{v\_proj}. We enable all linear modules to incorporate more parameters in the safety vector, thus improving the alignment recovery effectiveness. The other hyper-parameters are configured as suggested~\cite{catqa}.

\begin{table*}[t]

\scriptsize
\setlength\tabcolsep{1.1pt}
 \renewcommand{\arraystretch}{0.7}
\caption{Impact of our alignment recovery on task performance (\%). \texttt{FT} represents the results after fine-tuning, and \texttt{Rec} stands for the results after our alignment recovery. In the table, \texttt{0}, \texttt{0.1k}, \texttt{0.5k} and \texttt{1.5k} represent the scenarios where we injected the corresponding number of harmful questions into fine-tuning, and \texttt{Mod.} means we only injected the 392 harmful questions left over from moderation into fine-tuning.}
\label{tab:task_accuracy}
\vspace{-0.75em}
\begin{tabular}{ll|ccccc|ccccc|ccccc|ccccc|ccccc}
\toprule
\multicolumn{2}{l|}{\multirow{2}{*}{\qquad\textbf{Dataset}}}     & \multicolumn{5}{c}{\textbf{\gemma}} & \multicolumn{5}{|c}{\textbf{\llamasmall}} & \multicolumn{5}{|c}{\textbf{\llamabig}} & \multicolumn{5}{|c}{\textbf{\mistral}} & \multicolumn{5}{|c}{\textbf{\qwen}}  \\
\cmidrule(lr){3-7} \cmidrule(lr){8-12} \cmidrule(lr){13-17} \cmidrule(lr){18-22} \cmidrule(lr){23-27} 
& & \texttt{0}  & \texttt{0.1k}  & \texttt{0.5k} & \texttt{1.5k}& \texttt{Mod.}   & \texttt{0}  & \texttt{0.1k} & \texttt{0.5k} & \texttt{1.5k}& \texttt{Mod.}& \texttt{0}  & \texttt{0.1k} & \texttt{0.5k}& \texttt{1.5k}& \texttt{Mod.}& \texttt{0}  & \texttt{0.1k} & \texttt{0.5k} & \texttt{1.5k}& \texttt{Mod.}& \texttt{0}  & \texttt{0.1k} & \texttt{0.5k} & \texttt{1.5k}& \texttt{Mod.}\cr

\midrule
\multirow{2}{*}{\textbf{\sql}} &\texttt{FT} & 80.9 &79.2 &80.9 &80.8 & 81.5 &78.9 &79.3 &80.1 &80.0 &78.9 &82.3 &80.7 &82.0 &82.9 &81.7 &81.4 &83.0 &83.1 &81.7 & 81.6 &79.7 &80.6 &81.5 &82.4 & 80.3\\
 &\texttt{Rec}  & 79.2 &76.0 &77.7 &78.1 &80.5 &77.6 &78.6 &76.5 &76.8 &78.6 &81.5 &78.7 &81.8 &80.4 &81.3 &80.6 &79.6 &79.8 &80.6  &79.6&79.2 &79.8 &82.6 &82.9 & 78.7\\

\midrule
\multirow{2}{*}{\textbf{\cheat}} & \texttt{FT}  &96.4 &98.0 &97.6 &96.2 &97.9&89.8 &88.8 &90.3 &94.3 &85.6&98.6 &97.0 &97.7 &98.3 &97.0&97.4 &96.4 &97.4 &97.2 &97.3&96.7 &98.1 &97.8 &97.9&97.6   \\
 &\texttt{Rec}& 93.2 &94.1 &94.6 &92.1 &94.2&88.7 &86.8 &88.4 &90.3 &86.5&97.1 &95.0 &96.7 &94.5 &95.9&94.9 &94.0 &97.4 &95.3 &97.2&98.0 &94.1 &95.1 &93.5 &97.7\\

\midrule
\multirow{2}{*}{\textbf{\nlbash}} & \texttt{FT} & 36.4 &36.1 &36.9 &38.3 &37.9&34.7 &34.7 &36.1 &33.1 &32.3&36.4 &37.5 &37.7 &36.8 &35.4&40.8 &41.1 &41.8 &41.6 &40.0&38.8 &39.3 &39.6 &38.1 &38.8 \\
 &\texttt{Rec}   &35.2 &36.4 &35.1 &36.5 &36.1&33.3 &33.3 &34.5 &31.7 &31.6&35.4 &36.3 &37.0 &35.1 &35.5&39.4 &39.8 &40.7 &39.7 &38.5&37.6 &38.1 &38.3 &36.5 &37.0\\

\midrule
\multirow{2}{*}{\textbf{\samsum}} &\texttt{FT}  &  49.9 &49.9 &50.0 &50.0 &50.1&50.5 &50.8 &50.2 &50.7 &50.8&52.8 &53.5 &53.4 &52.8 &53.3&54.6 &54.5 &54.6 &54.5 &54.3&52.4 &52.6 &52.4 &52.8 &53.2\\
 &\texttt{Rec}  &  48.0 &47.6 &47.8 &47.8 &47.6&48.9 &49.4 &49.7 &50.1 &50.6&50.5 &51.2 &50.9 &51.5 &52.6&52.0 &52.3 &52.2 &52.2 &51.7&52.8 &52.0 &51.7 &51.5 &51.5\\

\midrule
\multirow{2}{*}{\textbf{\toxicity}} &\texttt{FT}& 78.7 &75.1 &40.4 &46.1 &74.4&75.1 &74.5 &{{68.3}} &{{57.9}} &65.5&78.6 &76.2 &{{72.9}} &{{62.4}} &77.8&84.0 &80.6 &{{76.3}} &{{76.8}} &79.6&79.3 &81.6 &{{68.9}} &{{73.7}} &76.1\\
&\texttt{Rec} & 76.3 &76.0 &50.0 &59.1 &73.2&72.0 &71.3 &65.0 &57.4 &62.3&74.9 &73.9 &75.7 &69.0 &75.3&81.8 &78.0 &74.3 &76.5 &76.5&78.9 &79.4 &78.1 &73.4 &72.4  \\

\bottomrule
\end{tabular}
\end{table*}

\subsection{Alignment Recovery}
% As mentioned in sec~\ref{subsec:eval:setup}, we totally run 20 steps, but we only record the results every 5 steps considering the time cost. Besides, we will stop logging the results if one of the following criteria is reached: 1) the harmful rate is reduced to zero; 2) the performance is dropped over 5\%. We only show the last valid results of our approach for simplification. 
% To answer the first question, we run our algorithm to recover the alignment of previously fine-tuned 125 models. 

We present our alignment recovery results in~\autoref{fig:alignment} and elaborate on the main observations below. In 9 cases\footnote{The 9 cases include: \begin{itemize}[noitemsep, topsep=0pt]
\item \gemma on \sql with moderated harmful prompts
\item \gemma on \nlbash with 0.1k harmful prompts
\item \gemma on \toxicity with 0/0.1k/1.5k/filtered harmful prompts
\item \llamasmall on \cheat with 0.1k harmful prompts
%\item \llamasmall on \toxicity with 1k/1.5k harmful prompts
\item \llamabig on \nlbash with 0/0.5k harmful prompts
% \item \llamabig on \toxicity with moderated harmful prompts
\end{itemize}}, our rollback is invoked. The results without rollback in those cases are presented in~\autoref{tab:rollback_res} in the Appendix.

\vspace{0.25em}
\textbf{\textit{Our method can effectively recover the alignment compromised by fine-tuning}}. As shown in \autoref{fig:alignment}, in all the cases where fine-tuning escalates the harmful rate, our recovery can bring the harmful rate back to the original level. Further as reported in~\autoref{tab:harmful_harmful_sim} in Appendix, the cosine similarity between the harmful direction from our recovered model and the harmful direction from the original model is 0.99 consistently. This demonstrates the effectiveness of our method. In several cases with the \gemma and \mistral models, our method even reduces the harmful rate to below the original level. Our approach not only recovers the harmful direction but also unintentionally brings the aligned direction closer to the harmful direction (refusal direction) of the original aligned model. This results in the rejection of some harmful prompts that are mistakenly accepted as benign by the original aligned model, thus enhancing the alignment. This can be demonstrated by measuring the cosine similarity between the aligned and harmful directions, which is 0.8114 for the original model and 0.9674 for our recovered model on \gemma.
% We measured the cosine similarity between the aligned and harmful directions on \gemma, finding it to be 0.8114 for the original model and 0.9674 for our recovered model.
% Our algorithm increases this similarity, rejecting harmful prompts that might have been considered benign by original models. 
The more detailed data are in Appendix \autoref{tab:benign_harmful_sim}.

As shown in~\autoref{tab:rollback_res}, the rollback mechanism hardly affects the alignment recovery results. With or without rollback, our method can constantly bring the harmful rate to a very low level. This is intended as the rollback is incorporated to adjust the task performance. This also illustrates that our recovery method can mitigate the loss of alignment in both \textbf{Scenario I} and \textbf{Scenario II}.

\vspace{0.25em}
\textbf{\textit{The difficulty of recovery increases with the alignment strength in the original model}}. The extent of our recovery varies across different models. On \gemma, \mistral, and \qwen, we fully recover (or even improve) the alignment. Yet, on \llamasmall and \llamabig, we can substantially approximate the original alignment but never reach it in several cases. Fundamentally, this is because \llamasmall and \llamabig have perfect alignment before fine-tuning, presenting a zero harmful rate. Recovering such alignment requires more accurate and complete identification of alignment-related weights, which is harder to achieve.

\vspace{0.25em}
\textbf{\textit{Our method performs better on generation tasks}}. Against fine-tuning for generation tasks (\sql, \samsum, and \nlbash), we can consistently recover the full alignment, regardless of the target model and settings. However, our method is less performant when applied to fine-tuning for classification tasks (\toxicity and \cheat). Under several settings, we didn't bring the harmful rate to the level before fine-tuning. This performance discrepancy arises because the task performance of classification tasks is more susceptible to disruption from parameter changes, hindering our approach from executing additional optimization steps.

\textbf{\textit{Results on the latest models.}} We also apply our approach to the latest models for alignment recovery: Llama3.1 8B~\cite{llama3}, Llama3.2 3B~\cite{llama3}, and Qwen2.5 32B~\cite{qwen25}. The results show that our method can successfully recover their alignment to the before-fine-tuning level, demonstrating its generalizability. The detailed results are reported in~\autoref{tab:res_latest_models} in the Appendix.

% \subsection{Performance Comparision}
% To demonstrate the effectiveness of our approach, we randomly recover the same number of parameters with the same distribution as our approach, serving as the baseline. %Since our approach 

% From Table \ref{tab:baseline_hr}, we can observe that even though the baseline (\texttt{BL}) recovers the same number of parameters with the same distribution as our approach, it only decreases the harmful rate by less than 10\%, still leaving a high harmful rate. On the contrary, our approach reduces the harmful rate below 5\% for 108 of 125 models, which demonstrates the importance of selected parameters by our approach. The task performance of the baseline can be found in the appendix \ref{tab:baseline_task}. 

\begin{figure*}[!t]
	\centering  
        % \footnotesize
		% \setlength{\abovecaptionskip}{0.cm}
        \renewcommand{\arraystretch}{0.8}
	\setlength\tabcolsep{0pt}
	\begin{tabular}{ccccc}
             \includegraphics[width=0.20\textwidth]{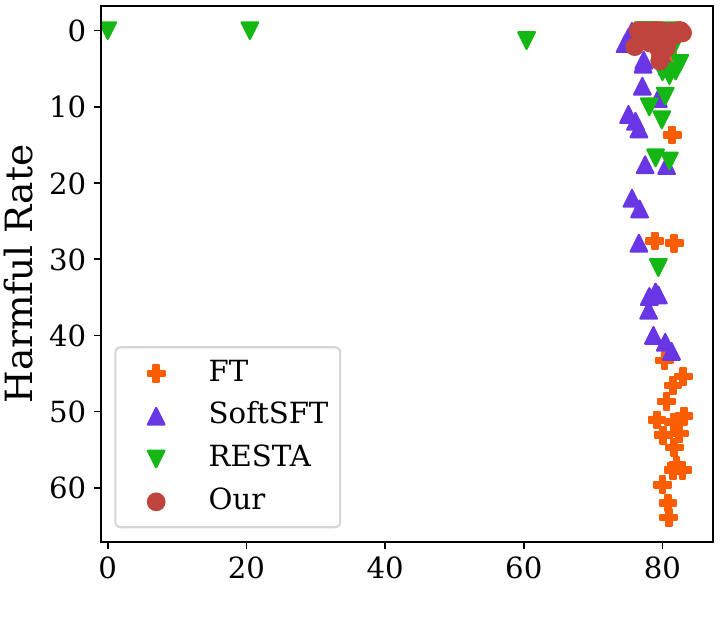} & 
              \includegraphics[width=0.2\textwidth]{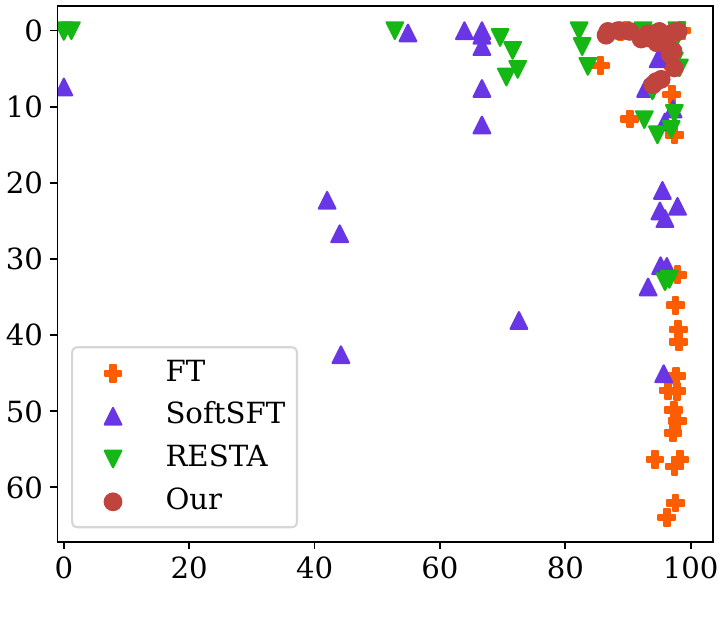} & 
              \includegraphics[width=0.2\textwidth]{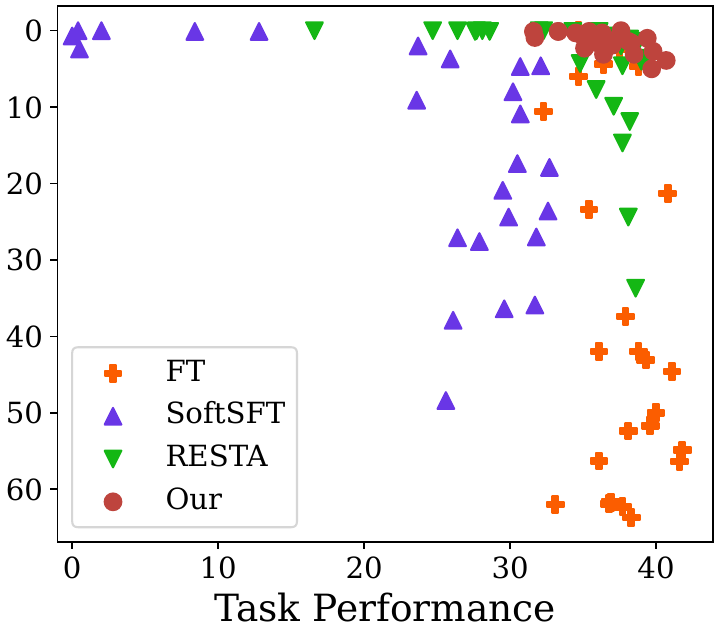} & 
              \includegraphics[width=0.2\textwidth]{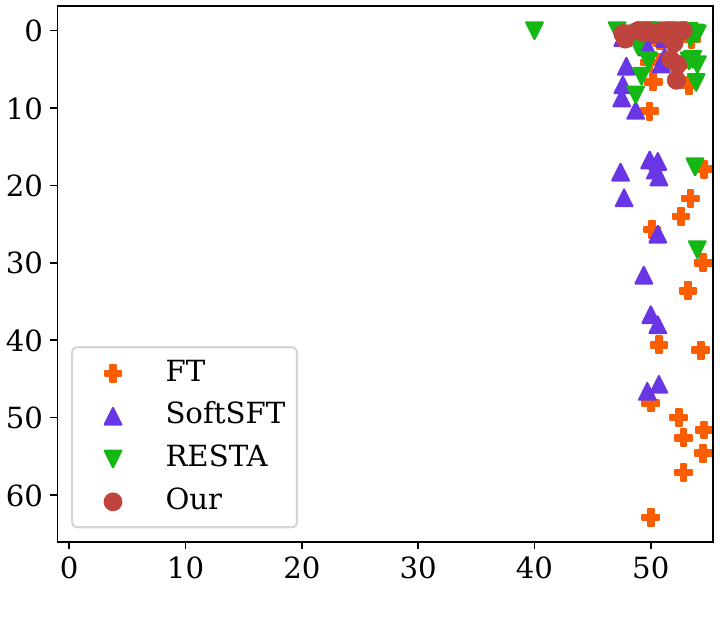} & 
              \includegraphics[width=0.2\textwidth]{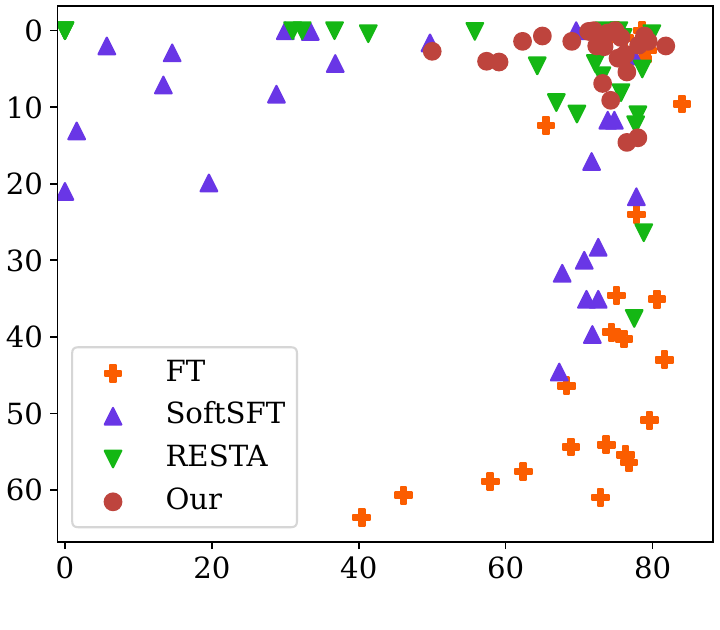} \cr
            \scriptsize{  (a) \sql} & \scriptsize{(b) \cheat} & \scriptsize{(c) \nlbash}  & \scriptsize{(d) \samsum}  &\scriptsize{(e) \toxicity} 
	\end{tabular}
	\caption{Visualization of the trade-off between harmful rate and task performance. \texttt{FT} represents the original fine-tuned model, and \texttt{SoftSFT} and \texttt{RESTA} stand for the two baselines. In the figure, we merge all the different models together. Each node represents a unique combination of the target model, the fine-tuning dataset, and the number of injected harmful prompts. The right side of the \texttt{x-axis} means better task performance and the upper side of the \texttt{y-axis} means lower harmful rate. Hence, \textbf{\textit{the upper, right corner represents the optimal trade-off}}.}
\label{fig:comparison}
\end{figure*}

\subsection{Task Performance}

Our method resets certain weights optimized by fine-tuning, which, by intuition, may hurt the task performance. Accordingly, we run a measurement after our alignment recovery and present the results in~\autoref{tab:task_accuracy}.

%In addition to restoring the alignment of LLMs, we also have to make sure the algorithm has a limited impact on the task performance. 
%\autoref{tab:task_accuracy} shows the task performance of fine-tuned models before and after recovering with our approach. From this table, we can achieve the following observation.

% {\textmd{\textbf{Your text here}}}
% \textbf{{Fine-tuning the LLMs on the task data injected with harmful data almost has no impact on the task performance.}} Even if we introduce a large amount of harmful data to the task data, most of the fine-tuned models still keep similar task performance as the fine-tuned model trained without harmful data. However, for \toxicity~dataset, the harmful data seriously undermine the model's task performance. The goal of \toxicity~dataset is to detect whether a prompt is toxicity or not, which is related to alignment. When the alignment is broken, the model is also unable to differentiate toxicity prompts from benign prompts. 

% Therefore, when we recover the alignment of the models, some models even achieve better performance, such as the \gemma ~model trained with 500 harmful data. 

\vspace{0.25em}
\textbf{\textit{Our approach maintains the task performance}}. On average, we only incur a 2.9\% decrease in task performance. In 30 of the 125 cases, the performance drop is below 1\%. In only 15 cases, the performance drop is above 4.5\% (getting close to our safety fuse of 5\%). These demonstrate that our method, when achieving satisfying alignment recovery, can largely preserve the task performance of the target model.

Given models fine-tuned on the \toxicity dataset, our recovery sometimes increases the task performance, especially when more harmful prompts are injected into the fine-tuning process. The \toxicity~ task aims to detect whether the given input is toxic or not, which depends on LLM alignment. When the alignment is broken, LLMs barely identify the toxic prompts, resulting in lower task performance as shown in \autoref{tab:task_accuracy}. Therefore, when we recover the alignment of the models, some models also achieve better performance such as the \gemma ~model trained with 500 harmful data. 

\vspace{0.25em}
\textbf{\textit{Our rollback mechanism helps}}. Our rollback mechanism described in~\S\ref{subsubsection:rollback} is activated in 9 cases. We re-evaluate the cases without rollback and present the results in ~\autoref{tab:rollback_res} in the Appendix. Restricted by our cutoff at a 5\% performance drop, our rollback does not render a significant boost in task performance. Yet, it still helps. Without our rollback mechanism, the performance drop is often higher than 3\% and can reach 4\%+. With rollback, the performance drop decreases to about 2.5\% or even below 2\%. These results show that our method is friendly to task performance while it additionally benefits \textbf{Scenario II} throughout the rollback mechanism.

\subsection{Comparing with Baselines} Similar to our method, the two baselines, \resta and \softsft, also aim for a good balance between alignment and utility. To present a systematic comparison, we measure the trade-off between harmful rate and task performance offered by different methods, and visualize the results in~\autoref{fig:comparison}. 
% The detailed numbers are in~\autoref{tab:task_accuracy_all} and~\autoref{tab:alignment_res_all}. 

%To compare with the baselines, we draw the trade-off between alignment and task performance for all methods as shown in \autoref{fig:comparison}. The detailed data can be found in the Appendix  \autoref{tab:task_accuracy_all} and  \autoref{tab:alignment_res_all}. From this figure, we can find the following observations.

% existing solutions for alignment recovery, we adopt two representative approaches \resta~ and \softsft~ as our baselines. 
% Since alignment and task performance are critical to evaluating the approach, we draw the results regarding those two metrics as shown in \autoref{fig:comparison}. The detailed data can be found in the Appendix  \autoref{tab:task_accuracy_all} and  \autoref{tab:alignment_res_all}. From this figure, we can find the following observations.

\vspace{0.25em}
\textit{\textbf{Our method achieves a better trade-off and greater generalization than both baselines.}} As shown in~\autoref{fig:comparison}, the results of our method cluster around the right-upper corner, while the baselines scatter further into the bottom and left areas. This demonstrates a better trade-off offered by our method. In nearly all cases, our method reduces the harmful rate to the level before fine-tuning and maintains the task performance consistent with the fine-tuned model. In contrast, \softsft preserves the task performance while failing to offset the harmful rate in some models, given any fine-tuning tasks. \softsft simply constrains the value of the first few tokens, which is insufficient to maintain the alignment, especially for LLMs whose original alignment is not well like \mistral. \resta produces harmful rates more comparable to our method. Yet, it incurs tremendous impacts on the task performance, given \cheat, \nlbash, and \toxicity as the fine-tuning tasks. The reason is that \resta treats all parameters as the safety vector and adds this safety vector to the weights of fine-tuned models. This modifies too many parameters of the fine-tuned model, which inevitably hurts the task performance.

\textit{\textbf{Comparing with Additional Baselines.}}
One potential method to recover alignment is to use collected harmful questions along with refusal answers as refusal samples to fine-tune the model.
We consider two settings: (1) refusal samples come from the same distribution as harmful samples (in the task data); (2) refusal samples come from a different distribution. As shown in \autoref{tab:safetydata} in the Appendix, using refusal samples for fine-tuning can reduce the harmful rates to some extent but is not as effective as our technique.

We also consider directly steer the activations during model inference based on the harmful directions.
Specifically, we calculate the harmful directions, $\Delta$ and $\Delta_f$, by feeding harmful prompts to the original model and the fine-tuned model, and derive the steering vector $\textbf{V} = \Delta - \Delta_f$.
We then conduct the following two experiments.
In experiment \textbf{I}, we add $\textbf{V}$ to the activations with a coefficient of 1. The harmful rate decreases from 56.4\% to 20.1\%, but the task performance drops from 94.3 to 86.6.
In experiment \textbf{II}, we decompose $\textbf{V}$ into two components (one parallel to $\Delta_{aligned}$ and one perpendicular to $\Delta_{aligned}$), and use the perpendicular one as the new steering vector. With coefficients of 1 and 2, the harmful rate decreases to 22.7\% and 3.4\%, while the task performance drops to 86.3 and 74.3.

Additionally, we apply $L_1 / L_2$ penalties to model weights instead of selecting weights to update as in our algorithm for re-alignment. However, such methods have two drawbacks: (1) Updating all parameters can overfit the recovery data, reducing the generalization of alignment recovery. For example, when recovering Llama2 13B on the CHEAT dataset, the harmful rate saturates at 13\% even after we run 60 recovery epochs; (2) The task performance can remain sensitive to  $L_1 / L_2$ penalties. For instance, when recovering Llama2 7B on the TOXIC dataset, the task performance drops from 65.5\% to 48.0\% within 10 recovery rounds (despite harmful rate remaining at 8.3\%).

%we can find that our approach densely clusters together in the right corner, which denotes better alignment and task performance. For \toxicity~ task, despite the decline in task performance observed across fine-tuned models, our approach exhibits the least amount of dispersion compared to other methods. On the contrary, \resta~and \softsft~ fails to reconcile the alignment with task performance. For example, even though \resta~and \softsft~ maintain the task performance in most cases on \sql~and \samsum~ dataset, they fail to restore the alignment. 

%\vspace{0.25em}
%\textit{\textbf{Our approach manifests great generalization across all models.}} As shown in \autoref{fig:comparison}, our approach consistently achieves stronger alignment and higher task performance. Even on the \toxicity~ dataset, our approach still reduces the harmful rate to a lower level with negligible performance degradation. On the contrary, \resta~and \softsft~ exist several extreme results on most datasets. For example, \resta~and \softsft~ significantly degrade the task performance even to zero on \sql~and \nlbash, respectively. 

\begin{figure*}[!t]
	\centering  
        % \footnotesize
		% \setlength{\abovecaptionskip}{0.cm}
        \renewcommand{\arraystretch}{0.8}
	\setlength\tabcolsep{0pt}
	\begin{tabular}{ccccc}
             \includegraphics[width=0.2\textwidth]{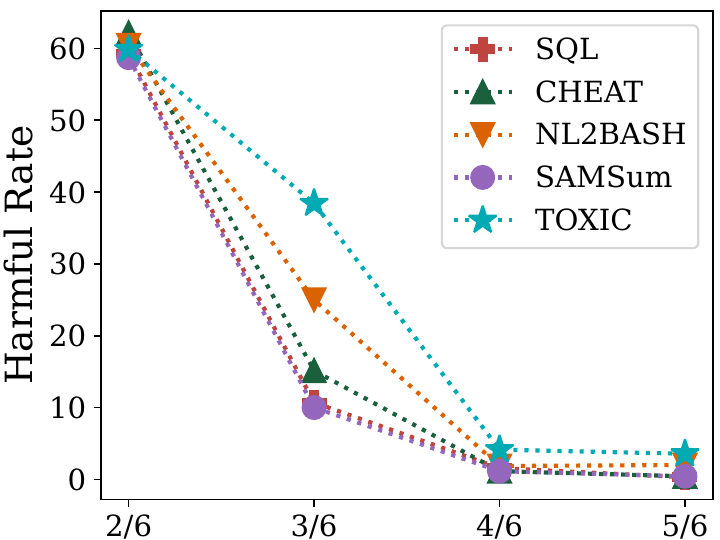} & 
              \includegraphics[width=0.2\textwidth]{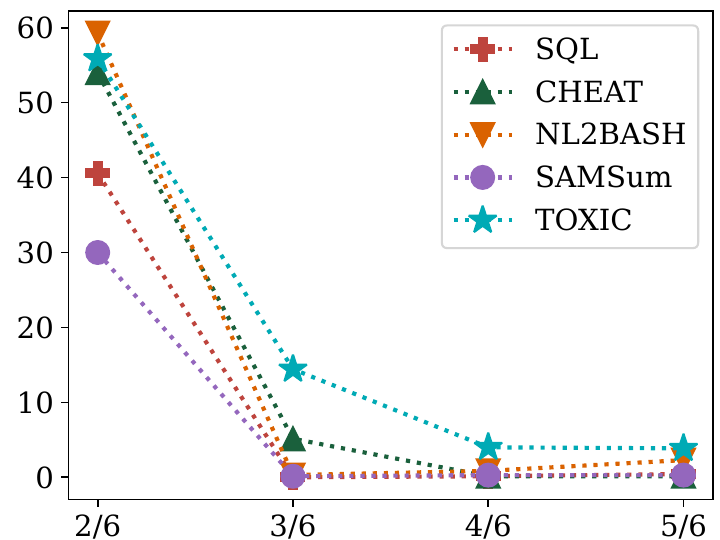} & 
              \includegraphics[width=0.2\textwidth]{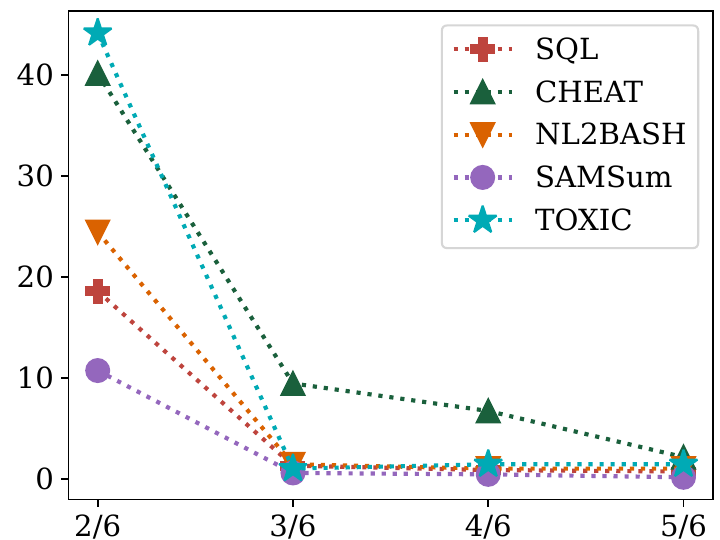} & 
              \includegraphics[width=0.2\textwidth]{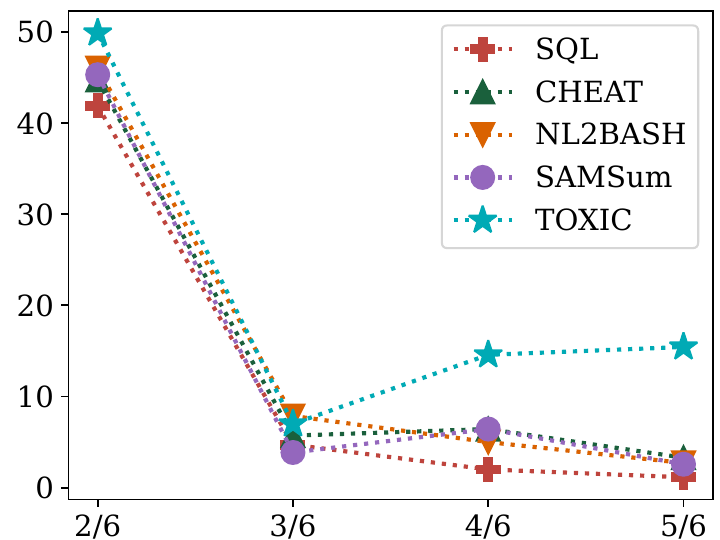} & 
              \includegraphics[width=0.2\textwidth]{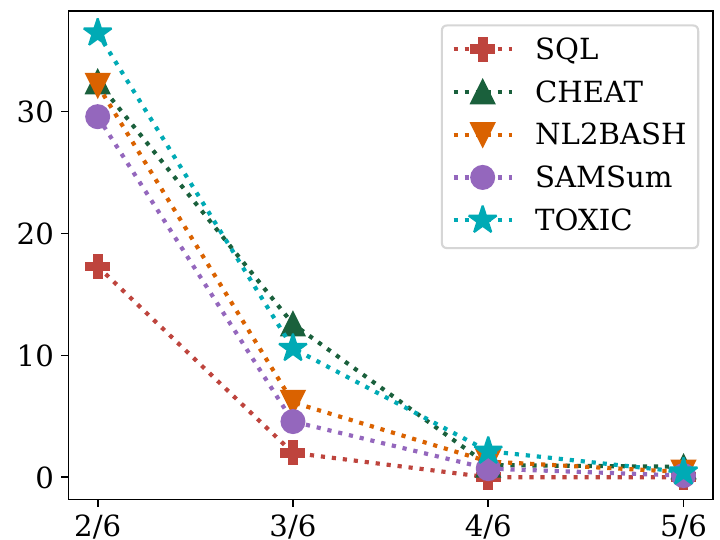} \cr
                    
             \includegraphics[width=0.2\textwidth]{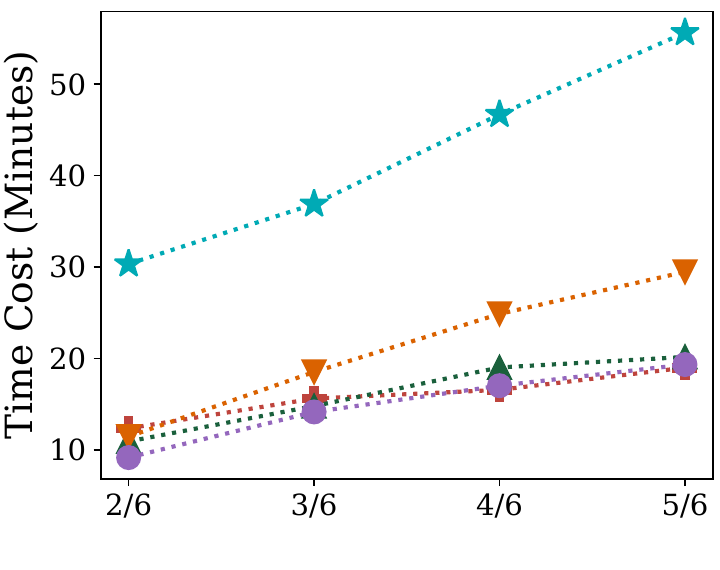} & 
              \includegraphics[width=0.2\textwidth]{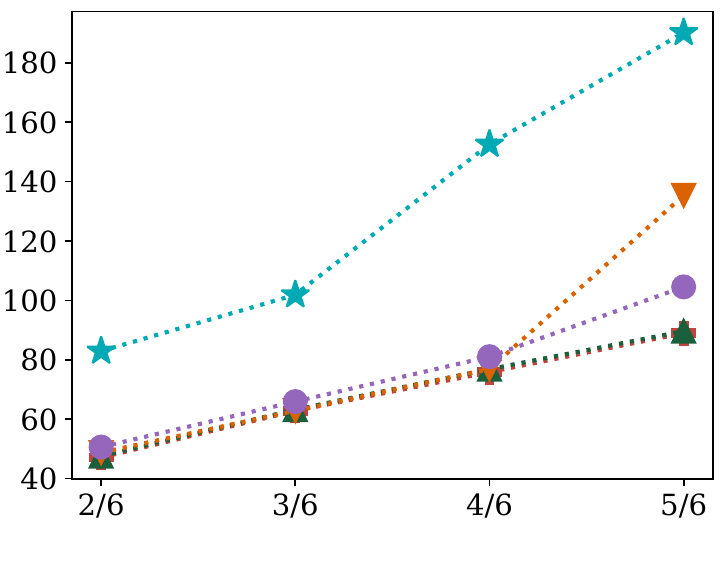} & 
              \includegraphics[width=0.2\textwidth]{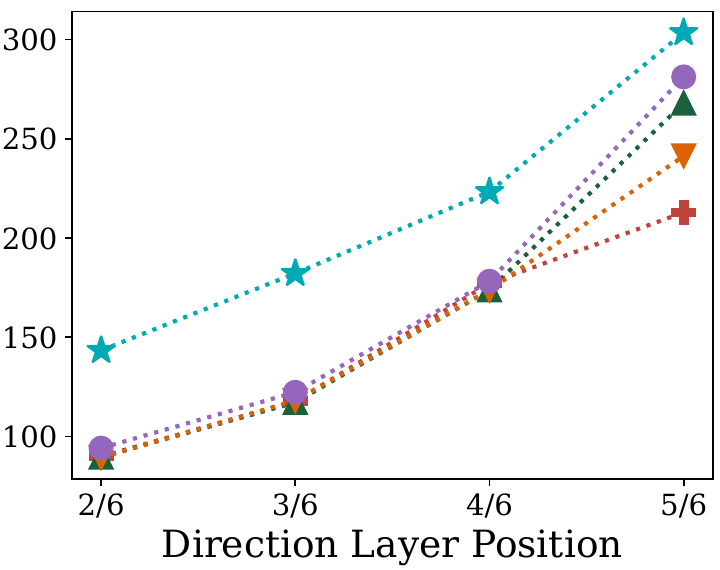} & 
              \includegraphics[width=0.2\textwidth]{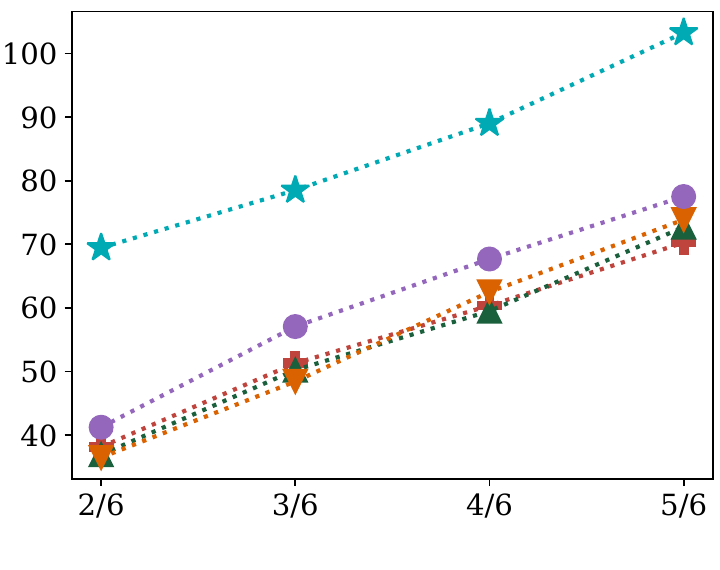} & 
              \includegraphics[width=0.2\textwidth]{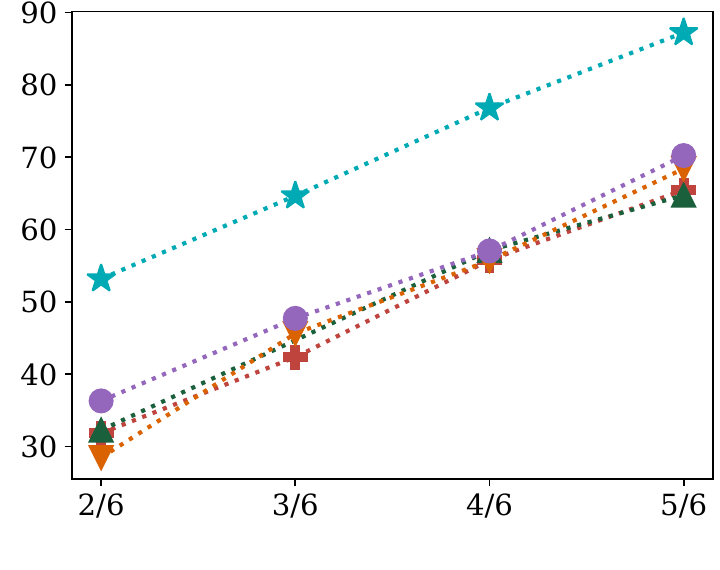} \cr

            \scriptsize{  (a) \gemma} & \scriptsize{(b) \llamasmall} & \scriptsize{(c) \llamabig}  & \scriptsize{(d) \mistral}  &\scriptsize{(e) \qwen} 
	\end{tabular}
    \caption{The alignment recovery results and the time cost of our method when using a direction layer at various positions. $\frac{2}{6}$, $\frac{3}{6}$, $\frac{4}{6}$, and $\frac{5}{6}$ on the \texttt{x-axis} respectively represent that we pick the direction layer from the $\frac{2}{6}$, $\frac{3}{6}$, $\frac{4}{6}$, and $\frac{5}{6}$ position of the hidden layers. We omitted the value of time cost for \qwen on the \cheat~ task at the $\frac{3}{6}$ position, as it activates rollback and disproportionately increases the time expenditure.}
\label{fig:ablation_layer}
\end{figure*}

\subsection{Impacts of Hyper-parameters}
% To figure out how the hyper-parameters affect the alignment recovery of our approach, we explore the robustness of our approach to hyper-parameters including the direction layer $L_{dir}$, the recovery dataset, and the alignment evaluation dataset. 
Our method involves several hyper-parameters, among which the direction layer $L_{dir}$ and the recovery dataset $\mathcal{D}_{rec}$ are more critical. We extend an evaluation to understand the impacts they can bring. In the evaluation, we focus on the models fine-tuned with 1,500 harmful samples injected, considering that their alignment is more severely compromised and represents the most challenging situation.

%To assess the impact of hyperparameters on our approach's alignment recovery, we explored the sensitivity of our method to various parameters, including the direction layer ($L_{dir}$), the recovery dataset, and the alignment evaluation dataset.
%In the following studies, we will focus on the models fine-tuned on the task dataset polluted with 1500 harmful data, since their alignment typically undergoes the most serious damage compared with other models.  

%

% \input{figs/ablation_layer_time}
\subsubsection{Direction Layer $L_{dir}$}
\label{subsec:ablation:layer}
% Our method hinges on selecting a direction layer ($L_{dir}$) that accurately represents and effectively controls the alignment property.
To inspect the influence of the direction layer, we select it from different locations, including the $\frac{2}{6}$, $\frac{3}{6}$, $\frac{4}{6}$, and $\frac{5}{6}$ position of the hidden layers. We re-do the alignment recovery using each option and measure the harmful rate separately. As explained in~\S\ref{subsubsec:getdirection}, different direction layers can incur varying time costs. Thus, we also measure the time consumption in each experiment. All results are presented in~\autoref{fig:ablation_layer}.

\vspace{0.25em}
\textit{\textbf{The direction layer represents a trade-off between alignment recovery and time cost.}} A clear trend shown in~\autoref{fig:ablation_layer} is that moving the direction layer toward the end decreases the harmful rate, given any model and any fine-tuning task. This result is expected as deeper layers in the model's architecture tend to better represent directions~\cite{zou2023representation}. On the other hand, choosing a deeper layer leads to a larger space for weight identification and restoration, proportionally increasing the time costs. Observing that a direction layer deeper than the $\frac{4}{6}$ position only improves the alignment marginally while intensifying the time cost, we set the direction layer at the $\frac{4}{6}$ position by default.

\subsubsection{Recovery Dataset $\mathcal{D}_{rec}$}
\label{subsec:recovery_diversity}

We vary the three key properties, diversity, size, and distribution, of the recovery dataset and assess their influences. 

%Since the recovery dataset determines the harmful direction, it is important to figure out how the diversity, size, and data distribution of the recovery dataset affect our algorithm. Therefore, we run three experiments by fixing two factors and adjusting the rest one. 

\vspace{0.25em}
\ding{182} \textbf{Diversity.} 
Our default recovery dataset, extracted from \beavertails, includes 256 samples from 14 categories. We respectively reduce the category number to 2, 4, and 8 while maintaining the total number of samples at 256. 
% We vary the number of categories for obtaining the harmful samples in \beavertails from 2 to 4 and 8 while maintaining the total number of samples at 256.
The results show that the difference in harmful rate caused by the number of data categories is mostly less than 2\%.
This indicates that our algorithm is not sensitive to the diversity of the recovery dataset.
Detailed results are in Appendix~\S\ref{appendix:recovery_diversity}.

\vspace{0.25em}
\ding{183} \textbf{Size.} In this evaluation, we keep 14 categories in the recovery dataset but reduce the number from 256 to 64 and 16.
Our experiments show that the recovery dataset size has marginal impacts on our algorithm.
This is because the harmful directions are more constant and less
dependent on the number of samples.
Details are in Appendix~\S\ref{appendix:recovery_diversity}.

\vspace{0.25em}
\ding{184} \textbf{Distribution.} To vary the distributions, we use two other popular datasets consisting of harmful questions, \catqa ~\cite{catqa} and \hex~\cite{qi2023fine}, as the recovery dataset. More details of the two datasets and how we use them for evaluations are described in Appendix~\S\ref{appendix:recovery_dataset}.
The results show our method is robust with different recovery datasets. Please see Appendix~\S\ref{appendix:recovery_diversity}.

\subsubsection{Impact of $P$ and $R$}
We study the impact of the recovery rate ($P\%$) on our approach using the \cheat dataset. We evaluate two settings: $P=0.1\%$ and $P=0.4\%$, where the default parameter is $P=0.2\%$. As shown in \autoref{tab:ablation_p} in the Appendix, a larger $P$ achieves better alignment but worse task performance, since it resets more weights to the aligned model.
We also evaluation two settings for hyperparameter $R$: $10\%$ and $40\%$ (default $R=20\%$). As summarized in \autoref{tab:ablation_r} in the Appendix, a smaller $R$ achieves better alignment but worse task performance, as it rolls back fewer task parameters.

\subsection{Impact of Evaluation Dataset}
By default, we use a subset of \beavertails to measure the alignment recovery of our method. Here, we adopt two other datasets, \hex~\cite{qi2023fine} and \catqa~\cite{catqa}.
The study shows that our method achieves consistent alignment recovery, regardless of the alignment evaluation dataset.
Details can be found in Appendix~\S\ref{appendix:eval_dataset}.

\subsection{Efficiency}
We also assess the efficiency of our method.  All our evaluations are performed on machines with AMD EPYC 7513 32-Core Processor, NVIDIA A100 (80 GB GPU memory), and 256GB RAM.
\autoref{tab:time} in the Appendix shows the results.
% \gt{Could you summarize the results here?}
In summary, larger models take longer to recover than smaller models. Since our technique is a post-processing approach, it is a one-time effort to obtain an aligned model.
In addition, the time cost can be reduced by running our algorithm fully on a GPU. \autoref{tab:scalability} in the Appendix reports the cost for different model sizes. The results show a large (around half) reduction on time cost by leveraging a GPU for top $P\%$ weights identification used in our algorithm.
% Address concern 2: scalability and overhead
%  our We optimized it to use GPU, which significantly reduced the time cost as shown in Result 1 at the end. There are other implementations where CPU is used and the CPU-to-GPU data transfer is slow (dozens of seconds each epoch). Through further GPU-only re-implementations, we can bring the time cost down to another level.

% \textbf{Larger models take longer to recover than smaller models.} "As shown in \autoref{tab:time}, models with 13B parameters typically require 150 minutes to recover, while those with 7B and 2B parameters need only 80 and 30 minutes, respectively. This is because larger models take more time to greedily identify the top parameters and perform forward passes. Since our algorithm is post-processing, it only needs to run once to achieve an aligned model.

\section{Extended Evaluation for Scenario I\label{sec:extraeva}}

\subsection{Experimental Setup}

In our target \textbf{Scenario I}, the \textit{Fine-tuner} is assumed to be ill-intended. It does not have to inject harmful samples into a regular dataset. Instead, it can simply fine-tune the target model with purely harmful samples to quickly compromise the alignment. We extend another evaluation from this angle. We randomly select 100 harmful samples from the \beavertails dataset and follow~\S\ref{subsec:studysettings} to fine-tune the 5 target models. We use 100 samples instead of more to align with the literature~\cite{yang2023shadow, qi2023fine}. More importantly, as we will explain with~\autoref{tab:all_harmful}, 100 samples can already compromise the alignment without sacrificing the model utility much. We also follow~\S\ref{subsec:eval:setup} to recover the fine-tuned models. 

%In this section, we first compromise the alignment of LLM by fine-tuning with 100 harmful data pairs as done in~\cite{yang2023shadow, qi2023fine}. Then, we recover the model's alignment with our approach without the rollback mechanism. In detail, we apply QLoRA to fine-tune the LLM for five epochs with a learning rate of 1e-4. For our approach, we run 20 steps. To evaluate the utility performance, we choose five popular benchmark datasets from \cite{llama2} including PIQA~\cite{piqa}, GSM8K~\cite{gsm}, TriviaQA~\cite{triviaqa}, HumanEval~\cite{humaneval}, and MMLU~\cite{mmlu}. To measure the alignment performance, we still use the \hex and \beavertails ~datasets.

\subsection{Results} We evaluate the alignment and the model performance before and after our recovery is performed. For alignment, we use the metric of harmful rate, measured with both the \beavertails and \hex datasets. For model performance, we can no longer use the metric of task performance as there are no downstream tasks. Instead, we adopt the \textit{utility performance} introduced by Yang et. al.~\cite{yang2023shadow}, which aims to assess the generic capabilities of LLM models. Specifically, we measure the utility performance on 5 representative benchmark datasets, including PIQA~\cite{piqa}, GSM8K~\cite{gsm}, TriviaQA~\cite{triviaqa}, HumanEval~\cite{humaneval}, and MMLU~\cite{mmlu}.
% \gt{I add the following sentece. Please check}
%Note that these datasets are solely used for the final evaluation of model performance and not for applying our alignment recovery method.

\begin{table}[!t]

\scriptsize
\setlength\tabcolsep{3.3pt}
 \renewcommand{\arraystretch}{0.8}
\caption{The generic utility performance and harmful rate (\texttt{HR}) for the original aligned model (\texttt{Ori}), the fine-tuned model (\texttt{FT}), and the recovered model (\texttt{Rec}). \texttt{TQA},  \texttt{HEval}, \texttt{Hex}, and \texttt{BT} represent the \texttt{TriviaQA}, \texttt{HumanEval}, \texttt{\hex}, and \texttt{\beavertails} datasets.}
\vspace{-0.5em}
%We highlight the harmful rate that is lower than or equal to 5\% with \textcolor{winered}{red color}.}
\label{tab:all_harmful}

\begin{tabular}{llccccccc}
\toprule
                  &          & \multicolumn{5}{c}{\textbf{Utility Performance}}         & \multicolumn{2}{c}{\textbf{HR (\%)}} \\
\cmidrule(lr){3-7}\cmidrule(lr){8-9}
        \textbf{Model}          &          & \texttt{PIQA}   & \texttt{GSM8k} & \texttt{TQA\tnote{$\alpha$} }  & \texttt{HEval}\tnote{$\beta$}   & \texttt{MMLU}   & \texttt{Hex}\tnote{$\gamma$}        & \texttt{BT}\tnote{$\delta$}    \\
\midrule
\multirow{3}{*}{\textbf{\gemma}} & \texttt{Ori} & 0.7481 & 0.1145 & 0.2512   & 0.2124    & 0.3649 &  2.7            & 4.6               \\
                  & \texttt{FT}       & 0.7480  & 0.1001 & 0.2070    & 0.1768    & 0.3715 & 64.2           & 52.9              \\
                  & \texttt{Rec}     & 0.7339 & 0.0986 & 0.1808   & 0.2073    & 0.3579 & 0.3            & 1.4               \\

\midrule
\multirow{3}{*}{\textbf{\llamasmall}} & \texttt{Ori} & 0.7633 & 0.1918 & 0.5526   & 0.1158    & 0.4571 & 0.0             & 0.0                \\
                  & \texttt{FT}       & 0.7758 & 0.1910  & 0.5454   & 0.1280     & 0.4494 & 63.6           & 33.1              \\
                  & \texttt{Rec}     & 0.7714       & 0.1918       &0.5455          & 0.1402          &0.4546        & 2.1             & 1.3               \\

\midrule
\multirow{3}{*}{\textbf{\llamabig}} & \texttt{Ori} & 0.7774 & 0.3206 & 0.6458   & 0.1707    & 0.5220  & 0.0             & 0.0                \\
                  & \texttt{FT}       & 0.7818 & 0.2964 & 0.6389   & 0.1585    & 0.5081 & 60.3           & 29.3              \\
                  & \texttt{Rec}     & 0.7758 & 0.3237 & 0.6427   & 0.1646    & 0.5168 & 0.3             & 0.6         \\

\midrule
\multirow{3}{*}{\textbf{\mistral}} & \texttt{Ori} & 0.7997 & 0.4018 & 0.6190    & 0.3536    & 0.5904 &35.5               & 11.7                 \\
                  & \texttt{FT}       & 0.8106 & 0.3813 & 0.5858   & 0.3414    & 0.5811 & 87.3           & 48.0              \\
                  & \texttt{Rec}     & 0.7812 & 0.3465 & 0.5824   & 0.3475    & 0.5678 & 1.8             & 0.7                \\

\midrule
\multirow{3}{*}{\textbf{\qwen}} & \texttt{Ori} & 0.7573 & 0.5322 & 0.4081   & 0.2560     & 0.5957 & 11.5              & 2.4               \\
                  & \texttt{FT}       & 0.7807 & 0.4996 & 0.4578   & 0.1700      & 0.5963 & 71.5           & 45.7              \\
                  & \texttt{Rec}     & 0.7665 & 0.5163 & 0.4230    & 0.2378    & 0.5938 & 1.8             & 0.6    \\ 

\bottomrule
\end{tabular}
\end{table}

\vspace{0.25em}
\textit{\textbf{The 100 samples compromise the alignment.}} As summarized in~\autoref{tab:all_harmful}, fine-tuning with the 100 harmful samples substantially removes the alignment in the target model. It increases the average harmful rate from nearly 0 to 69.34\% on \hex, meaning that the target model answers almost two-thirds of the harmful questions from this dataset. On \beavertails, the average harmful rate is also increased to 41.8\%. Moreover, the fine-tuning maintains the utility performance of the target model on all benchmark datasets, indicating that the target model still functions normally.

\vspace{0.25em}
\textit{\textbf{Our method can recover the alignment without sacrificing the model performance.}} ~\autoref{tab:all_harmful} also shows that our method can recover the alignment in the fine-tuned model. The average harmful rate of the recovered models is only 1.26\% on \hex and 0.92\% on \beavertails, presenting alignment similar to the original models. On the other hand, the utility performance of the recovered models is almost identical to the original models. This indicates that our method well maintains the model performance. 

%In several cases, it even brings the harmful rate to a level below the original model.
% \autoref{tab:all_harmful}

\section{Related Work\label{sec:related}}

\noindent
\textbf{LLM Fine-Tuning Attacks}.
% Even though LLMs have shown powerful zero-shot or few-shot capacities, 
Fine-tuning LLMs is an effective and popular way to enhance their abilities on various downstream tasks, especially on domain-specific datasets. However, recent works~\cite{yang2023shadow, qi2023fine, bhardwaj2023language} have pointed out that fine-tuning LLMs with a few harmful question-answer pairs, even with benign datasets can damage the alignment of LLMs, called fine-tuning attacks.

Yang \etal~\cite{yang2023shadow} and Bhardwaj \etal~\cite{bhardwaj2023language} show that fine-tuning the LLMs with a few harmful question-answer pairs can totally break the alignment of LLMs without sacrificing their utilities. 
Qi \etal \cite{qi2023fine} point out that merely fine-tuning with benign and widely employed datasets can also inadvertently compromise the safety alignment of LLMs.

\vspace{0.25em}
\noindent\textbf{Defence against LLM fine-tuning attack.}
To tackle the fine-tuning attack against LLMs, various approaches have been proposed, which can be grouped into model-enhanced, fine-tuning based, and post-alignment according to the stage involved in the fine-tuning procedures. 

% \vspace{0.25em}
\textit{Model-enhanced} mechanisms try to build a more robust LLM that is resilient against fine-tuning attacks. Huang~\etal~\cite{huang2024vaccine} propose \textit{vaccine}, which produces invariant hidden embeddings by progressively adding crafted perturbation to them in the alignment phase. Inspired by the shallow safe alignment, Rosati~\etal~\cite{rosati2024representation} remove information about harmful representations across all layers such that it is difficult to recover them during fine-tuning.
These methods, aiming at a more robust model, are orthogonal to ours.

 % assume that the drift of hidden embedding induced by fine-tuning on user data is the root cause of alignment-broken effect in an aligned LLM. Therefore, they 

% \vspace{0.25em}
% \textit{Data preprocessing based} approaches identify the data that may hurt the alignment of LLMs, and then remove them from the training data. Zhao~\etal~\cite{zhao2023learning} introduce the ForgetFilter algorithm, which filters unsafe data based on how strong the model’s forgetting signal is for that data. He~\etal~\cite{he2024s} identify and remove those samples that are close to harmful examples by representation and gradient matching.
% Fine-tuning on a clean dataset can also compromise alignment, making this approach less effective in the context of this paper.

% \vspace{0.25em}
\textit{Fine-tuning based} methods adjust and customize the fine-tuning process so that the fine-tuned LLMs are immune to harmful data~\cite{lyu2024keeping,huang2024lazy,hsu2024safe}.
Lyu \etal ~\cite{lyu2024keeping} uncovers that the prompt templates used during fine-tuning and inference play a crucial role in preserving safety alignment. Thus, they fine-tune models without a safety prompt but include it at test time.
% Wang \etal~\cite{wang2024mitigating} also focus on the system prompt. Inspired by backdoor attacks, they include harmful question safety response pairs with a secret trigger before the default system prompt during fine-tuning. This creates a link between the secret prompt and safe responses, which allows the model to generate safe answers for harmful queries.
Wang \etal~\cite{wang2024mitigating} create a link between the secret prompt and safe responses by including safety responses with a secret trigger during fine-tuning. During inference, this secret prompt will allow the model to generate safe answers for harmful queries.
% Zhou \etal ~\cite{zhou2023making} first train new parameters, called security vectors, to learn harmful behavior. During fine-tuning, these vectors are activated, which will prevent the LLMs from learning harmful behavior. During inference, deactivating the security vectors restores the LLM's normal behavior. 
Zhou \etal ~\cite{zhou2023making} prevent the LLMs from learning harmful behavior by activating the security vectors, which are parameters learned on harmful behavior. During inference, deactivating the security vectors restores the LLM's normal behavior. 
Qi~\etal~\cite{qi2024safety} observed that the first few output tokens determine the generative distribution of the LLMs. So they propose SoftSFT, which protects the initial tokens by designing a token-wise constrained objective during the supervised fine-tuning process. 
Huang~\etal~\cite{huang2024lazy} introduce an alignment dataset into the
user finetuning stage and apply the bi-state optimization to train the model.

% train some new parameters to learn the harmful behavior, called security vectors. Security vectors are activated during fine-tuning, which makes LLM believe that harmful behavior has already been learned, there is no need to further optimize for harmful data. During inference, they can deactivate security vectors to restore the LLM’s normal behavior. 

%\etal~\cite{wang2024mitigating} also pay attention to the system prompt. Inspired by backdoor attacks, during fine-tuning, besides using the customized data, they also include safety harmful question pairs as the train data with a secret trigger added before the default system prompt, establishing a connection between a secret prompt and generating safe responses. During inference, this trigger can be added to the system prompt, enabling the model to produce safe answers for harmful queries without impacting its performance on benign ones.
% \vspace{0.25em}
\textit{Post-alignment} approaches realign the compromised model without the knowledge of data and fine-tuning process. Inspired by task arithmetic~\cite{ilharco2022editing}, Bhardwaj \etal~\cite{catqa} propose RESTA, which restore the model's alignment by adding a safety vector to the weights of the compromised model. Specifically, the safety vector is the newly learned parameters used to remove the alignment by harmful question pairs. Hsu\etal~\cite{hsu2024safe} propose Safe LoRA, which projects the LoRA weights from selected layers to the safety-aligned subspace.Huang~\cite{huang2024antidote} aim to recover the alignment by pruning the harmful parameters. 

\vspace{0.25em}
% Our technique differs from the above method. 
% \gt{Check the following}
% This existing method~\cite{catqa} is based on
% task arithmetic that regards the alignment as a task and 
% the assumption that all the learned parameters are related to alignment. 
% \sout{Our approach, based on representation engineering, disentangles the harmful direction from the main (benign) task, and leverages the extracted direction to guide the recovery of the safety alignment.
% We only alter a very small fraction of weight parameters to restore the alignment of a fine-tuned LLM, without the necessity of modifying the entire model, thus better maintaining task performance.}
Unlike weight-space methods like RESTA simply merge delta parameters with the target model, our method disentangles the harmful direction from the main task and uses the extracted direction to guide alignment recovery. By minimizing modification in fine-tuned LLMs, our approach can better maintain task performance as shown in \autoref{fig:comparison}. % Address concern 1

\section{Conclusion}
The literature and this paper have shown that the alignment of LLMs can be unexpectedly compromised when fine-tuning an LLM on a downstream task. We propose an alignment recovery technique for fine-tuned models. Our approach features an iterative weight parameter restoration procedure that copies a small subset of weights from the original aligned model to the fine-tuned model. We also implement a rollback process based on its impact on downstream task performance. Experimental results show that our method effectively recovers alignment and achieves a better balance between alignment performance and task performance, outperforming state-of-the-art techniques.

\section{Acknowledgment}
We thank the anonymous reviewers and shepherd for their valuable feedback. This work was supported by National Science Foundation (NSF) awards CNS-2029038 and OAC-2319880. We are also grateful to NVIDIA for providing computational resources. Any opinions, findings, and conclusions or recommendations expressed herein are those of the authors and do not necessarily reflect the views of the US government or NSF.

 { \normalem
 \setstretch{0.85}

\bibliographystyle{IEEEtran}
\def\IEEEbibitemsep{0pt plus 0.3ex}
\bibliography{paper}

}

\appendices

\begin{figure*}[ht]
	\centering  
        % \footnotesize
		% \setlength{\abovecaptionskip}{0.cm}
        \renewcommand{\arraystretch}{0.8}
	\setlength\tabcolsep{0pt}
	\begin{tabular}{ccccc}
             \includegraphics[width=0.2\textwidth]{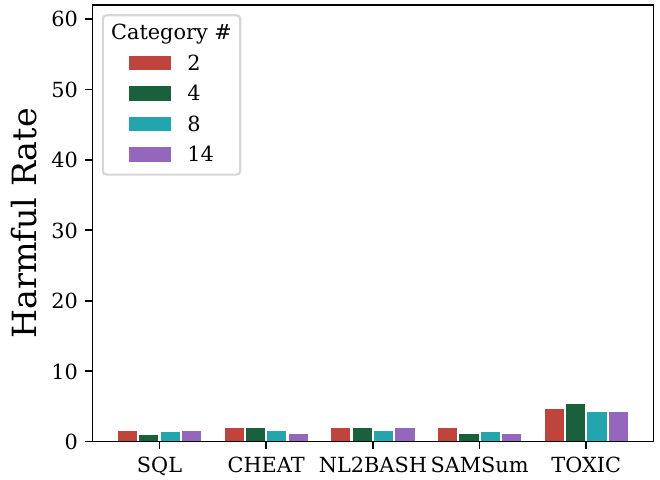} & 
              \includegraphics[width=0.2\textwidth]{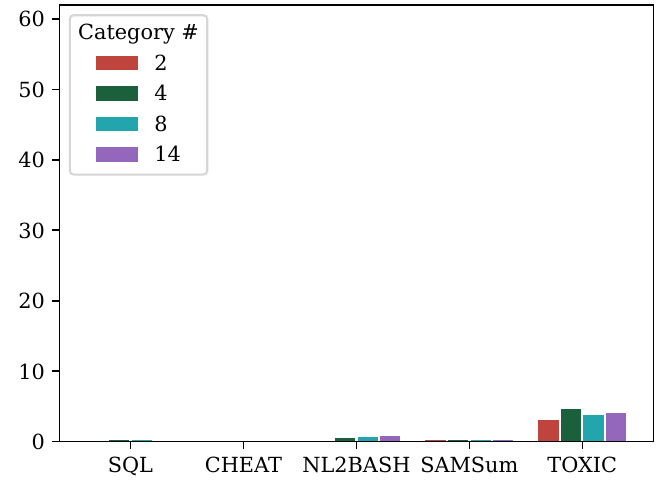} & 
              \includegraphics[width=0.2\textwidth]{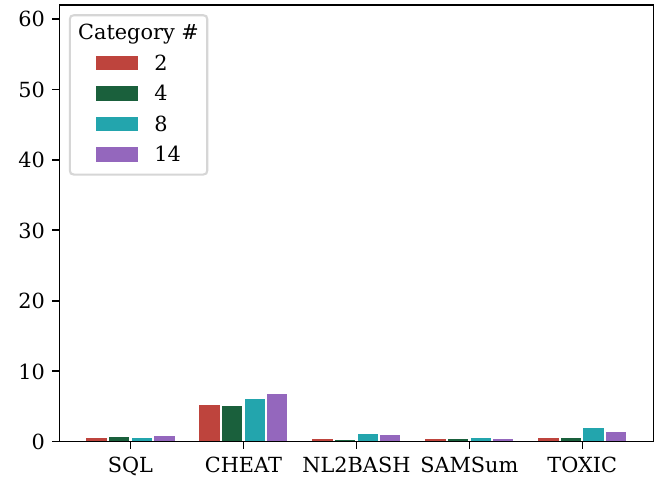} & 
              \includegraphics[width=0.2\textwidth]{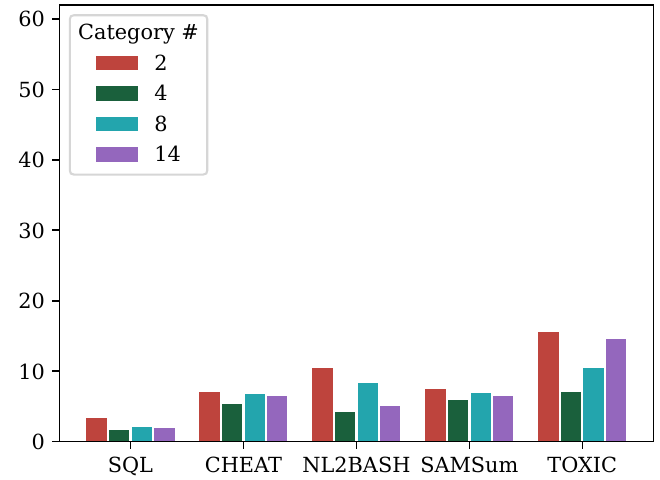} & 
              \includegraphics[width=0.2\textwidth]{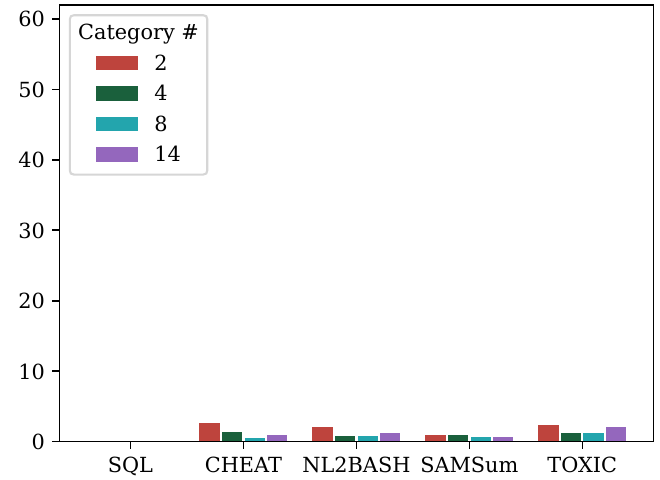} \cr

                % \multicolumn{5}{c}{\textbf{ \scriptsize  The impact of Recovery Diversity}} \cr 
                  \includegraphics[width=0.2\textwidth]{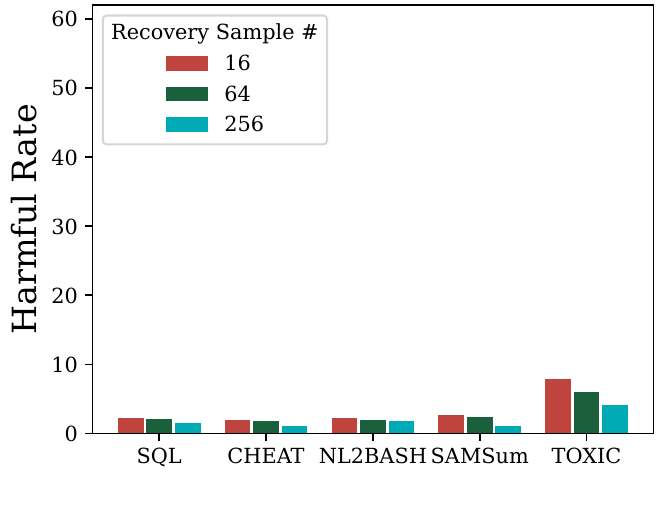} & 
              \includegraphics[width=0.2\textwidth]{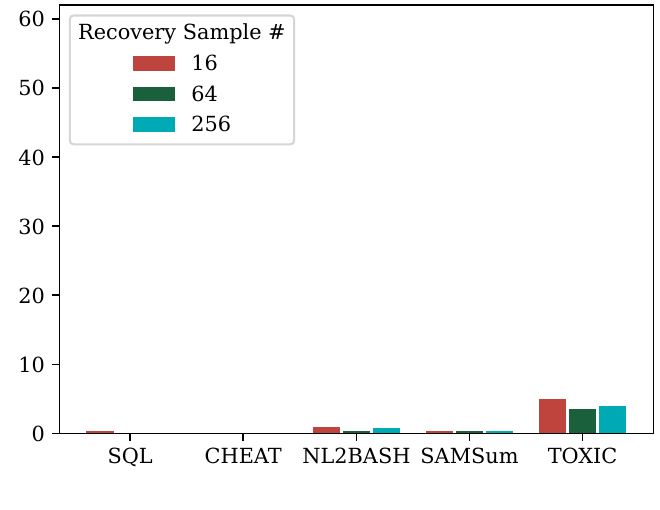} & 
              \includegraphics[width=0.2\textwidth]{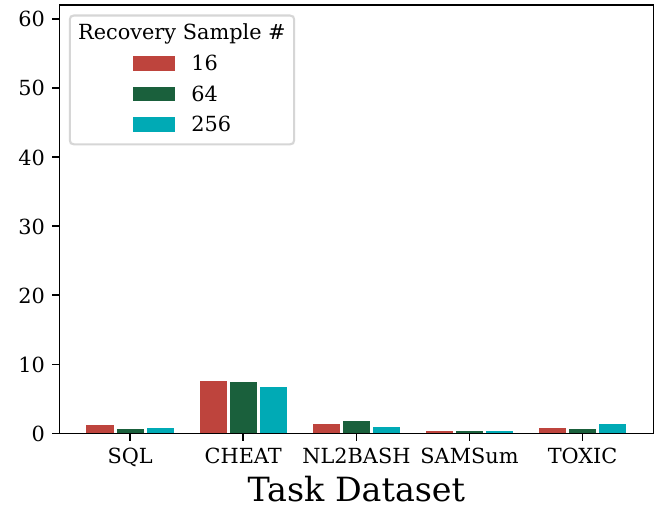} & 
              \includegraphics[width=0.2\textwidth]{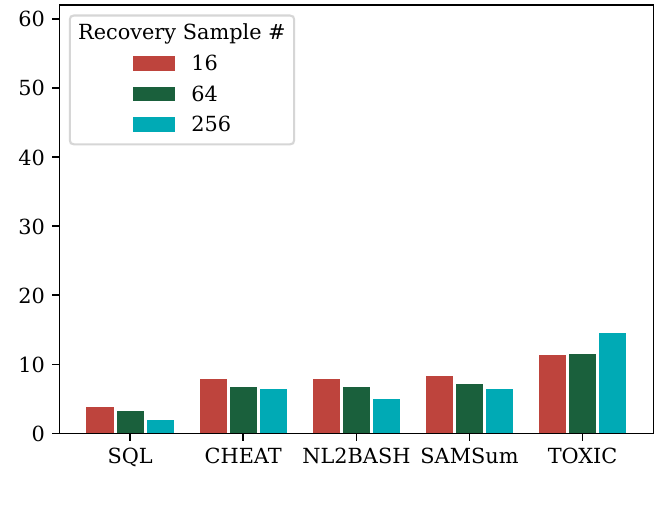} & 
              \includegraphics[width=0.2\textwidth]{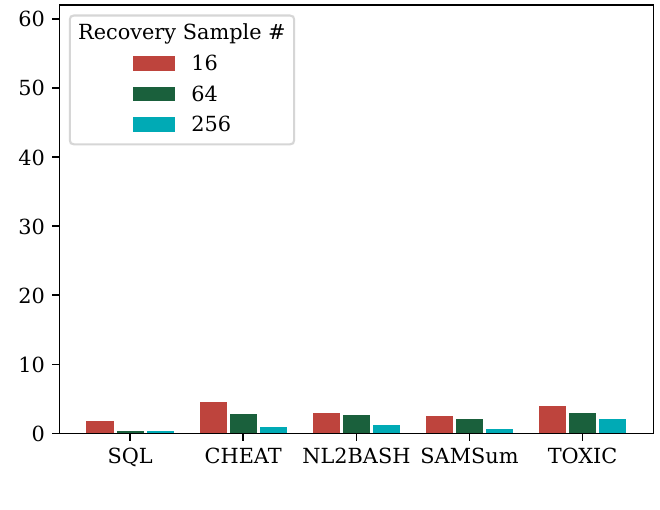} \cr
            \scriptsize{  (a) \gemma} & \scriptsize{(b) \llamasmall} & \scriptsize{(c) \llamabig}  & \scriptsize{(d) \mistral}  &\scriptsize{(e) \qwen} 
	\end{tabular}
    \vspace{-0.5em}
	\caption{The harmful rate of recovered models with varying diversity (\textbf{upper}) and size (\textbf{lower}) of the recovery dataset.}
     \vspace{-0.5em}
\label{fig:ablation_diversity_size}
\end{figure*}

% (1) \texttt{The first row} shows the results with less diverse datasets as the recovery dataset in our approach. We construct three additional datasets sampled from the cleaned \beavertails~ datasets but restrict their prompt category to \texttt{2}, \texttt{4}, \texttt{8}. (2) \texttt{The second row} shows the results when applying the recovery datasets with different sizes (\texttt{16}, \texttt{64}, and \texttt{256}). 
\begin{figure*}[!t]
	\centering  
        % \footnotesize
		% \setlength{\abovecaptionskip}{0.cm}
        \renewcommand{\arraystretch}{0.8}
	\setlength\tabcolsep{0pt}
	\begin{tabular}{ccccc}
             \includegraphics[width=0.2\textwidth]{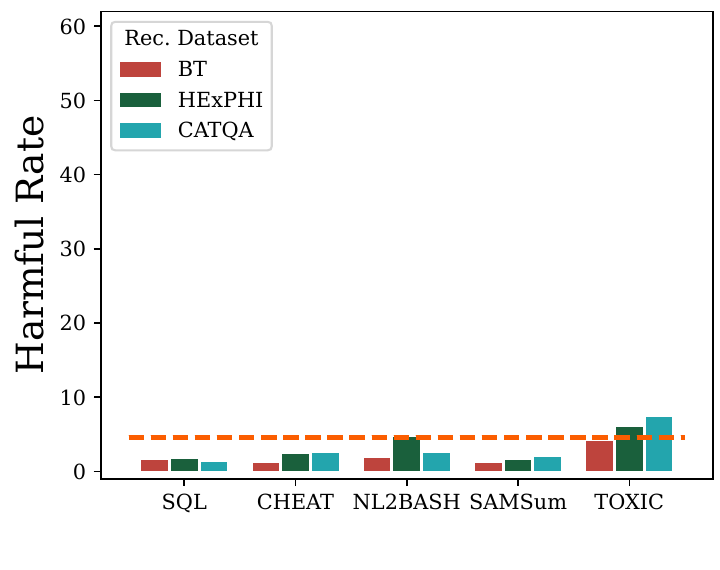} & 
              \includegraphics[width=0.2\textwidth]{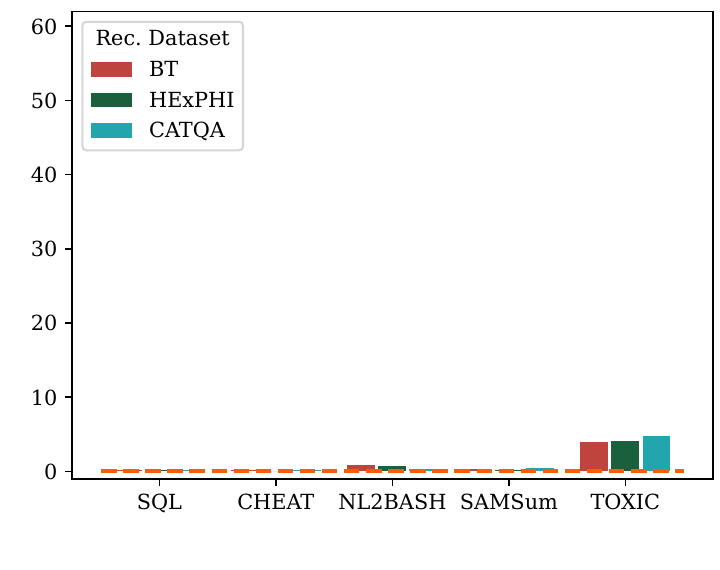} & 
              \includegraphics[width=0.2\textwidth]{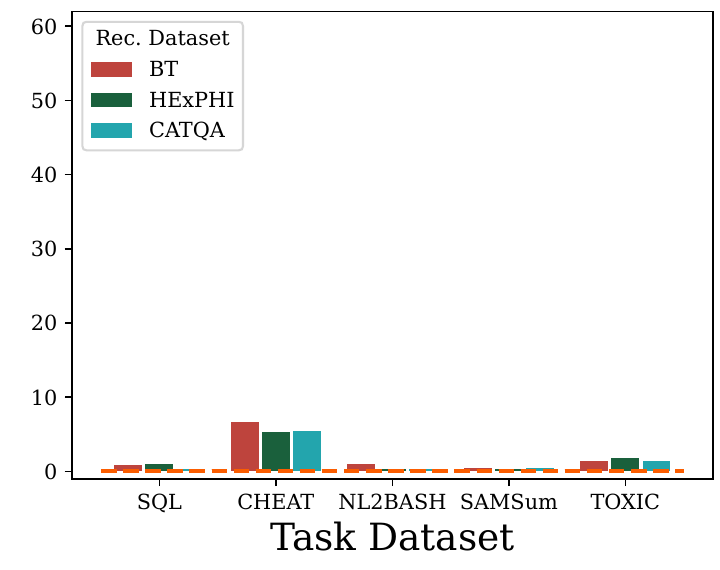} & 
              \includegraphics[width=0.2\textwidth]{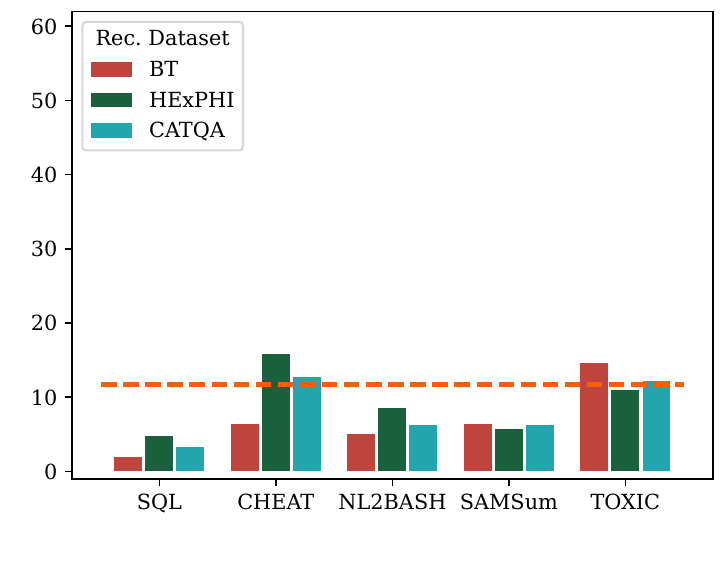} & 
              \includegraphics[width=0.2\textwidth]{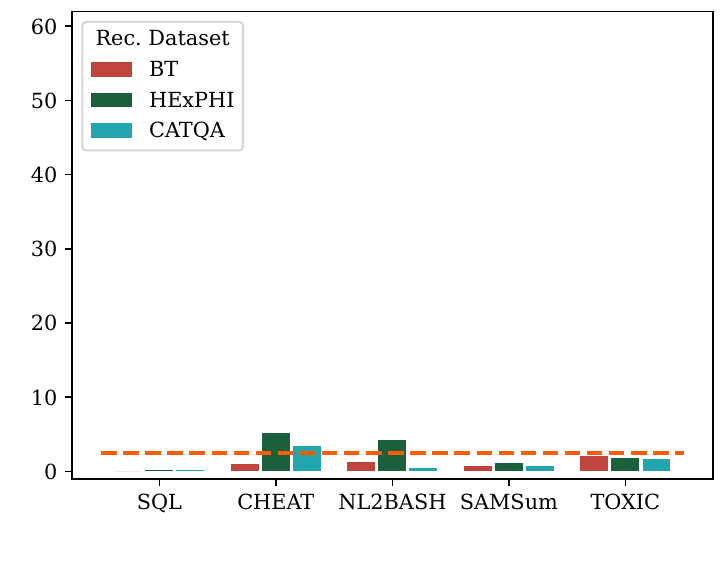} \cr
            \scriptsize{  (a) \gemma} & \scriptsize{(b) \llamasmall} & \scriptsize{(c) \llamabig}  & \scriptsize{(d) \mistral}  &\scriptsize{(e) \qwen} 
	\end{tabular}

\caption{The harmful rate of recovered models when using different recovery datasets (\texttt{\hex}, \texttt{\catqa}, and \texttt{\beavertails}, annotated as \texttt{BT}). The dashed line denotes the harmful rate of the \texttt{original} model before fine-tuning.}
     % \vspace{-1.5em}
\label{fig:ablation_recovery}
\end{figure*}

\begin{table*}
 \setlength\tabcolsep{1.7pt}
 \renewcommand{\arraystretch}{1.1}
\scriptsize
\centering
\caption{The harmful rate of fine-tuned models (\texttt{Fine-tuned}) and recovered models (\texttt{Recovered}) using different datasets, including \texttt{\beavertails} (\texttt{BT}), \texttt{\hex} (\texttt{HEx}), and \texttt{\catqa} (\texttt{CAT}) for alignment evaluation. \texttt{Original} represents the original model before fine-tuning. }
 % \vspace{-0.5em}
\label{tab:ablation_evaldataset}
\begin{tabular}{l|cccccc|cccccc|cccccc|cccccc|cccccc}

\hline
 & \multicolumn{6}{c|}{\textbf{\gemma}} & \multicolumn{6}{c|}{\textbf{\llamasmall}}  & \multicolumn{6}{c|}{\textbf{\llamabig}}   & \multicolumn{6}{c|}{\textbf{\mistral}} & \multicolumn{6}{c}{\textbf{\qwen}}                         \\

\cline{2-7} \cline{8-13} \cline{14-19} \cline{20-25} \cline{26-31}
\textbf{Dataset} & \multicolumn{3}{c}{\code{Fine-tuned}} & \multicolumn{3}{c|}{\code{Recovered}} & \multicolumn{3}{c}{\code{Fine-tuned}} & \multicolumn{3}{c|}{\code{Recovered}} & \multicolumn{3}{c}{\code{Fine-tuned}} & \multicolumn{3}{c|}{\code{Recovered}} & \multicolumn{3}{c}{\code{Fine-tuned}} &   \multicolumn{3}{c|}{\code{Recovered}}   & \multicolumn{3}{c}{\code{Fine-tuned}} & \multicolumn{3}{c}{\code{Recovered}}   \\

 \cline{2-4}  \cline{5-7} \cline{8-10} \cline{11-13} \cline{14-16} \cline{17-19} \cline{20-22} \cline{23-25} \cline{26-28} \cline{29-31}
 & \texttt{BT}        & \texttt{HEx}    & \texttt{CAT}      & \texttt{BT}         & \texttt{HEx} & \texttt{CAT}         
 & \texttt{BT}        & \texttt{HEx}      & \texttt{CAT}   & \texttt{BT}         & \texttt{HEx}     & \texttt{CAT}   
 & \texttt{BT}         & \texttt{HEx}   & \texttt{CAT}     & \texttt{BT}         & \texttt{HEx}     & \texttt{CAT}   
 &  \texttt{BT}        & \texttt{HEx}    & \texttt{CAT}     & \texttt{BT}       & \texttt{HEx}  & \texttt{CAT} 
 & \texttt{BT}         & \texttt{HEx}    & \texttt{CAT}    & \texttt{BT}         & \texttt{HEx}    & \texttt{CAT}  \\

\hline
\rowcolor{tablegray}
\textbf{Original} & 4.6 & 2.7 & 7.5  & 4.6 & 2.7 & 7.5 & 0.0 & 0.0 & 0.0 & 0.0 & 0.0 & 0.0  & 0.0 & 0.0 & 0.0  & 0.0 & 0.0 & 0.0 & 11.7 & 35.5 & 35.5  & 11.7 & 35.5 & 35.5 & 2.4 & 11.5 & 6.9 & 2.4 & 11.5 & 6.9 \\ 

\hline
\textbf{\sql} &  62.0 &79.1 &92.5 &1.6 &0.0 &0.2 &59.6 &94.5 &96.0 &0.1 &0.3 &0.0 &57.7 &94.2 &95.1 &0.9 &0.3 &0.6 &54.7 &94.5 &92.5 &2.0 &9.7 &4.4 &52.9 &89.4 &84.2 &0.3 &0.6 &0.6  \\
\hline
\textbf{\cheat} & 64.0 &76.7 &94.4 &1.1 &0.0 &0.9 &56.4 &91.2 &92.0 &0.1 &0.3 &0.2 &56.4 &92.1 &94.2 &6.7 &2.1 &5.6 &52.9 &95.5 &92.4 &6.4 &21.2 &18.2 &51.3 &86.7 &88.4 &1.0 &3.6 &2.4 \\
\hline
\textbf{\nlbash} & 63.7 &74.8 &95.5 &1.9 &0.6 &0.9 &62.0 &92.4 &92.4 &0.9 &0.6 &0.9 &61.9 &95.5 &96.7 &1.0 &3.6 &0.6 &56.4 &95.2 &96.2 &5.0 &20.3 &14.2 &52.4 &92.1 &92.0 &1.3 &3.3 &3.6 \\
\hline
\textbf{\samsum} & 62.9 &70.0 &91.1 &1.1 &0.0 &0.2 &40.6 &62.1 &68.0 &0.3 &0.3 &0.0 &57.1 &93.9 &94.7 &0.4 &1.2 &0.2 &54.6 &96.7 &95.5 &6.4 &23.3 &17.4 &52.6 &89.7 &94.5 &0.7 &1.2 &2.4 \\
\hline
\textbf{\toxicity }& 60.7 &71.2 &89.3 &4.1 &2.1 &11.4 &58.9 &94.2 &95.6 &4.0 &0.9 &2.7 &57.6 &95.5 &97.5 &1.4 &0.6 &0.7 &56.4 &96.1 &95.1 &14.6 &47.3 &36.7 &54.1 &92.7 &87.8 &2.1 &4.2 &6.0 \\

\hline

\end{tabular}
\end{table*}

\section{Experimental Results of Hyper-parameters}
\vspace{-0.5em}
\subsection{Impact of Recovery Dataset $\mathcal{D}_{rec}$}
\label{appendix:recovery_diversity}
\vspace{-0.5em}

We vary the three key properties, diversity, size, and distribution, of the recovery dataset and assess their influences. 

%Since the recovery dataset determines the harmful direction, it is important to figure out how the diversity, size, and data distribution of the recovery dataset affect our algorithm. Therefore, we run three experiments by fixing two factors and adjusting the rest one. 

\vspace{0.25em}
\noindent\ding{182} \textbf{Diversity.} Our default recovery dataset, extracted from \beavertails, includes 256 samples from 14 categories. We respectively reduce the category number to 2, 4, and 8 while maintaining the total number of samples at 256. We re-perform the evaluation for each setting and show the alignment recovery results in the upper half of~\autoref{fig:ablation_diversity_size}.

\textit{\textbf{Our algorithm is not sensitive to the diversity of the recovery dataset.}} Given \gemma, \llamasmall, \llamabig, and \qwen, the difference in harmful rate caused by the number of data categories is less than 2\%. On the \mistral, the variation is slightly higher. Yet, more categories do not necessarily lead to a lower harmful rate. This is not surprising as we discussed in~\S\ref{subsubsec:getdirection}. With a direction layer toward the end, the hidden states become more constant to simply indicate the prompt is harmful, presenting lower sensitivity to the data diversity. To validate the reasoning, we compare the harmful directions obtained with varying categories in the recovery dataset and show the results in~\autoref{tab:diversity_similarity}. The results are indeed highly similar.

\vspace{0.25em}
\noindent\ding{183} \textbf{Size.} In this evaluation, we keep 14 categories in the recovery dataset but reduce the number from 256 to 64 and 16. We show the results in the lower half of~\autoref{fig:ablation_diversity_size}. 
%To vary the size of the recovery dataset, we sampled subsets of 16 and 64 data points from the original dataset,  $\mathcal{D}_{rec}$, while preserving the same number of categories. The results are shown in \autoref{fig:ablation_diversity_size}.

\begin{table*}[!t]

\scriptsize
\setlength\tabcolsep{1.1pt}
  \renewcommand{\arraystretch}{0.7}
\caption{The harmful rate of fine-tuned models (\texttt{FT}) and recovered models applying \resta~ approach with original parameters (\texttt{Orig}) and optimized parameters (\texttt{Optim}). Fine-tuned models (\texttt{FT}) are trained with \texttt{0}, \texttt{0.1k}, \texttt{0.5k}, and \texttt{1.5k} harmful data mixed into the task dataset. \texttt{Mod.} indicates the remaining 392 samples after moderation.}
 \vspace{-0.5em}
\label{tab:resta_res}
\begin{tabular}{ll|ccccc|ccccc|ccccc|ccccc|ccccc}
\toprule
\multirow{2}{*}{\textbf{Dataset}} &    & \multicolumn{5}{c}{\textbf{\sql}} & \multicolumn{5}{|c}{\textbf{\cheat}} & \multicolumn{5}{|c}{\textbf{\nlbash}} & \multicolumn{5}{|c}{\textbf{\samsum}} & \multicolumn{5}{|c}{\textbf{\toxicity}}  \\
\cmidrule(lr){3-7} \cmidrule(lr){8-12} \cmidrule(lr){13-17} \cmidrule(lr){18-22} \cmidrule(lr){23-27} 
& & \texttt{0}  & \texttt{0.1k}  & \texttt{0.5k} & \texttt{1.5k}& \texttt{Mod.}   & \texttt{0}  & \texttt{0.1k} & \texttt{0.5k} & \texttt{1.5k}& \texttt{Mod.}& \texttt{0}  & \texttt{0.1k} & \texttt{0.5k}& \texttt{1.5k}& \texttt{Mod.}& \texttt{0}  & \texttt{0.1k} & \texttt{0.5k} & \texttt{1.5k}& \texttt{Mod.}& \texttt{0}  & \texttt{0.1k} & \texttt{0.5k} & \texttt{1.5k}& \texttt{Mod.}\cr

\midrule
\multirow{3}{*}{\textbf{\llamasmall}} &\texttt{FT} & 0.0 &2.3 &53.1 &59.6 &27.6&0.0 &0.1 &11.6 &56.4 &4.6&0.0 &6.0 &56.3 &62.0 &10.6&0.9 &1.3 &6.6 &40.6 &4.1&0.0 &0.4 &46.4 &58.9&12.4 \\
 &\texttt{Orig}  & 0.0 &2.0 &50.9 &56.4 &24.2 &0.0 &0.1 &10.0 &53.9 &3.2&0.0 &6.6 &53.7 &59.9 &8.3&0.6 &1.3 &6.7 &39.7 &3.0&0.0 &0.3 &43.3 &56.9&10.0 \\
 &\texttt{Optim}  & 0.0 &0.0 &0.1 &0.3 &0.0&0.0 &0.1 &0.0 &0.0 &0.0&0.0 &0.0 &0.0 &0.1 &0.0&0.0 &0.0 &0.0 &0.0 &0.0&0.0 &0.0 &0.0 &0.0&0.0 \\
 
\bottomrule

\end{tabular}
\end{table*}

\begin{table*}[!t]

\scriptsize
\setlength\tabcolsep{1.3pt}
 \renewcommand{\arraystretch}{0.1}
\caption{The time cost (minutes) for our approach to recover the fine-tuned models. Fine-tuned models (\texttt{FT}) are trained with \texttt{0}, \texttt{0.1k}, \texttt{0.5k}, and \texttt{1.5k} harmful data mixed into the task dataset. \texttt{Mod.} means the 392 samples left over from moderation are used.}
\label{tab:time}
\vspace{-0.5em}
\begin{tabular}{l|ccccc|ccccc|ccccc|ccccc|ccccc}
\toprule
\textbf{Dataset} & \multicolumn{5}{c}{\textbf{\gemma}} & \multicolumn{5}{|c}{\textbf{\llamasmall}} & \multicolumn{5}{|c}{\textbf{\llamabig}} & \multicolumn{5}{|c}{\textbf{\mistral}} & \multicolumn{5}{|c}{\textbf{\qwen}}  \\
\cmidrule(lr){2-6} \cmidrule(lr){7-11} \cmidrule(lr){12-16} \cmidrule(lr){17-21} \cmidrule(lr){22-26} 
 & \texttt{0}  & \texttt{0.1k}  & \texttt{0.5k} & \texttt{1.5k}& \texttt{Mod.}   & \texttt{0}  & \texttt{0.1k} & \texttt{0.5k} & \texttt{1.5k}& \texttt{Mod.}& \texttt{0}  & \texttt{0.1k} & \texttt{0.5k}& \texttt{1.5k}& \texttt{Mod.}& \texttt{0}  & \texttt{0.1k} & \texttt{0.5k} & \texttt{1.5k}& \texttt{Mod.}& \texttt{0}  & \texttt{0.1k} & \texttt{0.5k} & \texttt{1.5k}& \texttt{Mod.}\cr
\midrule

\textbf{\sql} & 12.9  &10.2  &15.4  &19.9  &23.8  &64.1  &64.2  &80.3  &75.8  & 87.2  &158.4  &170.9  &134.5  &177.2  &160.4 &51.9  &65.1  &64.7  &59.3   &  64.9 &55.2  &54.3  &50.4  & 51.8 & 52.0\\
\midrule
\textbf{\cheat }& 17.7  &17.9  &18.9  &20.6  & 15.8   &84.5  &86.3  &82.1  &76.9   &65.8 &156.0  &96.7  &77.7  &173.9  & 108.6 &65.7  &52.2  &65.9  &59.5   &63.1 &57.3  &57.7  &53.2  &53.7 & 60.5 \\
\midrule
\textbf{\nlbash} & 11.1  &26.9  &17.6  &24.9  &17.9  &49.3  &64.3  &46.1  & 76.7 & 61.8 &300.3  &85.4  &236.6  & 173.8 & 109.1   &53.3  &50.9  &51.5  &58.4   &62.0 &68.3  &31.4  &29.5  &53.8  &41.5\\
\midrule
\textbf{\samsum }& 18.3  &18.7  &18.8  &20.0  & 17.8  &84.2  &84.3  &85.0  & 81.0  &79.3  &178.9  &146.4  &142.0  &178.2 & 146.6 &72.1  &57.0  &58.3  &64.8  & 59.7  &79.3  &58.7  &55.1  &57.0  &55.5 \\
\midrule
\textbf{\toxicity} & 40.4  &41.9  &23.4  &32.1 &  40.9  &82.6  &82.9  &88.3  & 82.6 &86.1  &235.3  &131.8  &190.6  &223.3  &219.2  &92.9  &94.0  &93.8  & 89.0 &87.7  &81.0  &82.3  &73.9  & 74.4 & 80.9 \\

\bottomrule

\end{tabular}
\vspace{-0.5em}
\end{table*}
\begin{table*}[h]
    \centering
     \renewcommand{\arraystretch}{0.01}
    \scriptsize
    \setlength\tabcolsep{3.2pt}
    % \vspace{-0.5em}
    \caption{The harmful rate and task performance of fine-tuned models (\texttt{FT}) and recovered models under different recovery rates (\texttt{0.1\%}, \texttt{0.2\%}, and \texttt{0.4\%}). The models are trained on the CHEAT Dataset injected with 1500 harmful data.}
    \vspace{-0.5em}
    \begin{tabular}{l|cccc|cccc|cccc|cccc|cccc}
        \toprule
        & \multicolumn{4}{c}{\textbf{Gemma 2B}} & \multicolumn{4}{|c}{\textbf{LLAMA2 7B}} & \multicolumn{4}{|c}{\textbf{LLAMA2 13B}} & \multicolumn{4}{|c}{\textbf{Mistral 7B}} & \multicolumn{4}{|c}{\textbf{Qwen 7B}} \\
        \cmidrule(lr){2-5} \cmidrule(lr){6-9} \cmidrule(lr){10-13} \cmidrule(lr){14-17} \cmidrule(lr){18-21}
        & \texttt{FT} & \texttt{0.1\%} & \texttt{0.2\%} & \texttt{0.4\%} & \texttt{FT} & \texttt{0.1\%} & \texttt{0.2\%} & \texttt{0.4\%} & \texttt{FT} & \texttt{0.1\%} & \texttt{0.2\%} & \texttt{0.4\%} & \texttt{FT} & \texttt{0.1\%} & \texttt{0.2\%} & \texttt{0.4\%} & \texttt{FT} & \texttt{0.1\%} & \texttt{0.2\%} & \texttt{0.4\%} \\
        \midrule
        \textbf{Harmful Rate} & 64.0 & 2.1 & 1.1 & 5.1 & 56.4 & 0.7 & 0.1 & 0.7 & 56.4 & 2.9 & 6.7 & 2.71 & 52.9 & 7.0 & 6.4 & 3.1 & 51.3 & 2.0 & 1.0 & 0.1 \\
        \midrule
        \textbf{Task Performance} & 96.2 & 95.2 & 92.1 & 92.9 & 94.3 & 93.9 & 90.3 & 94.0 & 98.3 & 94.8 & 94.5 & 93.8 & 97.2 & 95.4 & 95.3 & 93.3 & 97.9 & 95.9 & 93.5 & 94.9 \\
        \bottomrule
    \end{tabular}
    \label{tab:ablation_p}
\end{table*}

\textit{\textbf{The recovery dataset size has marginal impacts on our algorithm.}} Similar to diversity, the size of the recovery dataset incurs limited differences in harmful rate. The reason is also similar. The harmful directions are more constant and less dependent on the number of samples, as shown by the comparison results in~\autoref{tab:size_similarity}. 

%The size of the recovery dataset not only affects the harmful direction but also impacts the gradients of model weights. As shown in \autoref{tab:size_similarity}, the size barely alters the harmful direction. However, smaller recovery datasets may not produce gradients as accurate as those derived from larger datasets, potentially leading to ineffective alignment recovery as shown in \autoref{fig:ablation_diversity_size} (second row). From this figure, we can observe that as the number of recovery data increases, the alignment steadily improves in most cases.

\vspace{0.25em}
\noindent\ding{184} \textbf{Distribution.} To vary the distributions, we use two other popular datasets consisting of harmful questions, \catqa ~\cite{catqa} and \hex~\cite{qi2023fine}, as the recovery dataset. More details of the two datasets and how we use them for evaluations are described in Appendix~\S\ref{appendix:recovery_dataset}. We summarize the comparison results in~\autoref{fig:ablation_recovery}.

%widely-applied safety datasets \catqa ~\cite{catqa} and \hex~\cite{qi2023fine} as the new recovery dataset and keep other settings the same as in ~\S\ref{subsec:eval:setup}. The details of those two datasets can be found in Appendix ~\S\ref{appendix:recovery_dataset}.

\textit{\textbf{Our method is robust with different recovery datasets.}}  Using \catqa~ and \hex~ as the recovery dataset brings no major differences to the recovery results compared to \beavertails. Our method consistently reduces the harmful rate close to or even below the level before fine-tuning.
% Therefore, our approach is robust against the selection of the recovery dataset.  \autoref{tab:ablation_recovery}

 \vspace{-0.5em}
\subsection{Impact of Evaluation Dataset}
\vspace{-0.5em}
\label{appendix:eval_dataset}
In the evaluation above, the alignment recovery of our method is measured with a testing subset from \beavertails. We adapt the evaluation to measure the alignment recovery with all harmful samples from two other datasets, \hex and \catqa (which are introduced above). In the adapted evaluations, we use the default recovery dataset from \beavertails to avoid overlap with the alignment evaluation dataset. The evaluation results are presented in~\autoref{tab:ablation_evaldataset}.
% \jx{Honestly, I do not think~\autoref{tab:ablation_evaldataset} is a good format for presentation. It does not tell the after-fine-tuning results. People cannot see how much harmful rate our method has reduced. In many cases, people only see that our method presents a harmful rate much higher than the before-fine-tuning model.}
% \yk{I plot two figures \autoref{fig:ablation_eval_dataset} and \autoref{fig:ablation_eval_dataset2}, I hope one of them fit the needs}
%We use a subset Considering that various evaluation datasets are designed to encompass a diverse range of harmful prompts,  it's worth exploring our approach to additional evaluation datasets. Therefore, we adopt the same two alignment datasets (\hex~ and \catqa) to evaluate the performance of our approach. The results are shown in \autoref{tab:ablation_evaldataset}. 

\textit{\textbf{Our method achieves consistent alignment recovery, regardless of the alignment evaluation dataset.}} As shown in \autoref{tab:ablation_evaldataset}, our method can reduce the harmful rate in most cases close to or below the level before fine-tuning. The only exception is the \mistral model fine-tuned on the \toxicity dataset and evaluated using the \hex dataset. The harmful rate, compared to the model before fine-tuning, increased by nearly 12\%. Our method also presents consistent effectiveness when measured with different evaluation datasets. It does not perform significantly better or worse with a specific evaluation dataset, indicating that our method is not overfitting the evaluation datasets.

\section{Experimental Setup}
\vspace{-1.0em}
\subsection{RESTA Setting\label{appendix:resta_setting}}
\vspace{-0.5em}
For \textit{RESTA}, they consider a less severe threat model in which the alignment of LLMs is only slightly compromised by fine-tuning with benign datasets.  However, in our experiments, the alignment of fine-tuned models is significantly disrupted. Under these conditions, their provided hyperparameters are insufficient to restore alignment, as demonstrated in  \autoref{tab:resta_res}.  To address this, we strengthen the safety vector to more effectively disrupt alignment. Specifically, we increase the batch size from 4 to 10 and the number of epochs from 3 to 5. Additionally, To include more parameters in the safety vector, thus improving the alignment recovery capability, we enable Q\lora~ to fine-tune all linear modules, rather than just the \texttt{q\_proj} and \texttt{v\_proj} modules.

\vspace{-0.5em}
\subsection{Recovery Dataset}
\vspace{-0.8em}
\label{appendix:recovery_dataset}
\catqa ~\cite{catqa} combines prohibited use cases outlined in OpenAI's usage policies and Meta's Llama2 acceptable use policy to build a comprehensive range of harmful categories. In detail, this dataset contains 11 main harmful categories. For each category, they construct 50 harmful questions, a total of 550 harmful prompts. 

\hex~\cite{qi2023fine} also uses the prohibited use cases from Meta and OpenAI's policies. It contains 30 harmful instruction examples for each of the 11 prohibited categories from various sources, including Anthropic Red Teaming Data~\cite{ganguli2022red} and AdvBench~\cite{advbench}. These examples were further refined and categorized by human annotators and AI models like Claude, and jailbroken versions of GPT-3.5 and Llama 2.

% However, even though those two datasets consist of harmful prompts, LLMs do not regard them as harmful prompts. As shown in \autoref{tab:ablation_evaldataset}, LLMs tend to have a higher harmful rate on those two datasets, especially for \mistral, which yields a harmful rate of 35.45\%. Therefore, we first filter out the samples that were rejected by target models, then randomly select 256 data form it as the recovery dataset.

Although these two datasets contain harmful prompts, LLMs do not recognize them as such. As shown in \autoref{tab:ablation_evaldataset}, LLMs tend to exhibit a higher harmful rate on these datasets, particularly \mistral, which shows a harmful rate of 35.45\%. Thus, we first filtered out samples that were answered by the target models and then randomly selected 256 samples from the remaining set as the recovery dataset.

% \vspace{-15.0em}
\section{Additional Results}
\vspace{-1.0em}
\begin{table}[ht]
 \scriptsize
\setlength\tabcolsep{2.45pt}
 \renewcommand{\arraystretch}{0.4}
\centering
 
\caption{The cosine similarity between the harmful direction calculated using recovery datasets of varying diversity. We calculate the similarity between new datasets with default recovery dataset. The number of categories denotes how many categories the new dataset includes. \label{tab:diversity_similarity}}
% \vspace{-0.5em}
\begin{tabular}{cccccc}
\toprule
\textbf{Category \#} & \textbf{\gemma}  & \textbf{\llamasmall} & \textbf{\llamabig} & \textbf{\mistral} & \textbf{\qwen}   \\
\midrule
\texttt{2}   & 0.9907 & 0.9926     & 0.9907   & 0.9770  & 0.9790 \\
\midrule
\texttt{4}  & 0.9853 & 0.9902     & 0.9877   & 0.9687 & 0.9658 \\
\midrule
\texttt{8}   & 0.9946 & 0.9980     & 0.9980   & 0.9858  & 0.9882 \\
\midrule
\texttt{14}   & 1.0 & 1.0     & 1.0   & 1.0  & 1.0 \\
\bottomrule
\end{tabular}
\vspace{-0.95em}
\end{table}
\begin{table}[ht]
 \scriptsize
\setlength\tabcolsep{2.25pt}
 \renewcommand{\arraystretch}{0.2}
\centering
 
\caption{The cosine similarity between harmful directions calculated using recovery datasets of various sizes. We calculate the similarity between the new datasets and the default recovery dataset. \texttt{Rec. Data \#} denotes the number of data in the recovery dataset. \label{tab:size_similarity}}
\vspace{-0.5em}
\begin{tabular}{cccccc}
\toprule
\textbf{Rec. Data \#} & \textbf{\gemma}  & \textbf{\llamasmall} & \textbf{\llamabig} & \textbf{\mistral} & \textbf{\qwen}   \\

\midrule
\texttt{16}  & 0.9721   &0.9846      &0.9781     & 0.9525 & 0.9446  \\
\midrule
\texttt{64}   &0.9964  &0.9988     & 0.9989   &0.9931   &0.9924  \\
\midrule
\texttt{256}   & 1.0 & 1.0     & 1.0   & 1.0  & 1.0 \\
\bottomrule
\end{tabular}
\vspace{-0.75em}
\end{table}

\begin{table}[ht]
 \scriptsize
\setlength\tabcolsep{1.7pt}
 \renewcommand{\arraystretch}{0.6}
\centering
 
\caption{The cosine similarity between the harmful direction from the \texttt{Original} model and the aligned direction from the \texttt{Original}, the \texttt{Fine-tuned}, and the \texttt{Recovered} models, respectively. The results are averaged over all datasets. \label{tab:benign_harmful_sim}}
\vspace{-0.5em}
\begin{tabular}{cccccc}
\toprule
\textbf{Model} & \textbf{\gemma}  & \textbf{\llamasmall} & \textbf{\llamabig} & \textbf{\mistral} & \textbf{\qwen}   \\

\midrule
\texttt{Original}  &  0.8325 &  0.8681   & 0.8383 &  0.8114 & 0.7578  \\
\midrule
\texttt{Fine-tuned}   & 0.7858   &0.5642  &0.4109    & 0.5741   & 0.5651 \\
\midrule
\texttt{Recovered}   & 0.9136  & 0.9257    &0.9385   & 0.9245  &0.8391  \\
\bottomrule
\end{tabular}
\vspace{-0.75em}
\end{table}
\begin{table}[ht]
 \scriptsize
\setlength\tabcolsep{1.7pt}
 \renewcommand{\arraystretch}{0.6}
\centering
 
\caption{The cosine similarity between the harmful direction from the \texttt{Original} model and the harmful direction from the \texttt{Fine-tuned}, and the \texttt{Recovered} models, respectively. The results are averaged over all datasets. \label{tab:harmful_harmful_sim}}
% \vspace{-0.5em}
\begin{tabular}{cccccc}
\toprule
\textbf{Model} & \textbf{\gemma}  & \textbf{\llamasmall} & \textbf{\llamabig} & \textbf{\mistral} & \textbf{\qwen}   \\

\midrule
\texttt{Fine-tuned}   & 0.8178   & 0.5790  & 0.3958  &0.5910   & 0.6426 \\
\midrule
\texttt{Recovered}   & 0.9961  & 0.9893   &0.9717  &0.9918  & 0.9921 \\
\bottomrule
\end{tabular}
\vspace{-0.75em}
\end{table}

\begin{table}[!b]
 \scriptsize
\setlength\tabcolsep{12.0pt}
 \renewcommand{\arraystretch}{0.1}
\centering
\caption{The time cost of our algorithm applying to different model sizes with and without optimization. The optimization leverages a GPU for top $P\%$ weights identification used in our algorithm.\label{tab:scalability}}
\vspace{-0.5em}
\begin{tabular}{ccccc}
\toprule
\textbf{Model Size} & \textbf{2B}  & \textbf{7B} & \textbf{13B} & \textbf{32B}  \\

\midrule
\texttt{Original time (Min)}   & 	20	 & 80 & 170	  & -    \\
\midrule
\texttt{Optimized time (Min)}   & 	10  & 47 &	103 &230  \\
\bottomrule
\end{tabular}
% \vspace{-1.75em}
\end{table}
\begin{table}[!b]
    \centering
    \scriptsize
    \setlength\tabcolsep{1.2pt}
     \renewcommand{\arraystretch}{0.9}
    \caption{The harmful rate (\texttt{HR}) and task performance (\texttt{TP}) of fine-tuned models (\texttt{\texttt{FT}}) and recovered models with baseline methods (\texttt{\texttt{BL}}). The baseline methods inject the same number of refusal pairs into the CHEAT Dataset. In setting \textbf{I}, safety and harmful data share the same distribution; in setting \textbf{II}, they come from different distributions. }

\begin{tabular}{llccccccccccccccc}
\toprule
                    &    & \multicolumn{3}{c}{Gemma 2B} & \multicolumn{3}{c}{LLAMA2 7B} & \multicolumn{3}{c}{LLAMA2 13B} & \multicolumn{3}{c}{Mistral 7B} & \multicolumn{3}{c}{Qwen 7B} \\
                   \cmidrule(lr){3-5}  \cmidrule(lr){6-8}  \cmidrule(lr){9-11} \cmidrule(lr){12-14} \cmidrule(lr){15-17}
                    &    & \texttt{FT}      & \texttt{BL}    & \texttt{Our}   & \texttt{FT}      & \texttt{BL}    & \texttt{Our}    & \texttt{FT}       & \texttt{BL}    & \texttt{Our}    & \texttt{FT}       & \texttt{BL}    & \texttt{Our}    & \texttt{FT}      & \texttt{BL}   & \texttt{Our}   \\
                \midrule
\multirow{2}{*}{\textbf{I}}  & \texttt{HR} & 64.0      & 0.1       & 1.1    & 56.4    & 11.7      & 0.1     & 56.4     & 14.4      & 6.7     & 52.9     & 7.6       & 6.4     & 51.3    & 2.7      & 1.0    \\
                    & \texttt{TP} & 96.2    & 96.6      & 92.1   & 94.3    & 94.6      & 90.3    & 98.3     & 98.3      & 94.5    & 97.2     & 97.0        & 95.3    & 97.9    & 97.4     & 93.5   \\
                    \midrule
\multirow{2}{*}{\textbf{II}} & \texttt{HR} & 62.1    & 16.9     & 1.6    & 11.6    & 0.6      & 0.0     & 45.4     & 9.3      & 3.1     & 57.3     & 35.6     & 4.9     & 47.4    & 4.6     & 0.6    \\
                    & \texttt{TP} & 97.6    & 97.5     & 94.6   & 90.3    & 94.8      & 88.4    & 97.7     & 97.5     & 96.7    & 97.4     & 97.6     & 97.4    & 97.8    & 98.1    & 95.1  \\
\bottomrule
\end{tabular}

\label{tab:safetydata}
\end{table}

\begin{table}[!b]
    \centering
    \setlength\tabcolsep{1.4pt}
     \renewcommand{\arraystretch}{0.6}
 \scriptsize
     \caption{Task performance and harmful rate for the latest models. \texttt{FT} refers to models fine-tuned on the training data mixed with \texttt{1.5k} harmful data, while \texttt{Rec.} refers to recovered models with our approach. The harmful rate for those three aligned models before fine-tuning is \texttt{5.8\%}, \texttt{8.1\%}, and \texttt{2.4\%}, respectively.}
     \vspace{-0.5em}
    \begin{tabular}{cl|ccccc|ccccc}
        \toprule
        & & \multicolumn{5}{|c}{\textbf{Harmful Rate}}  & \multicolumn{5}{|c}{\textbf{Task Performance}} \cr
        \cmidrule(lr){3-7} \cmidrule(lr){8-12}
        \multicolumn{2}{l}{\textbf{\; Model}} & \tiny{\texttt{SQL}} & \tiny{\texttt{CHEAT}} & \tiny{\texttt{\nlbash}}  & \tiny{\texttt{\samsum}} &  \tiny{\texttt{\toxicity}} & \tiny{\texttt{SQL}} & \tiny{\texttt{CHEAT}} & \tiny{\texttt{\nlbash}}  & \tiny{\texttt{\samsum}} &  \tiny{\texttt{\toxicity}} \\
        \midrule
        \textbf{Llama3.1} & \texttt{FT}   & 61.3 & 62.1 & 56.1 &  58.4 & 55.3 & 83.4 & 97.6 & 26.4 & 54.3 & 72.2 \\
        \textbf{8B} & \texttt{Rec.} & 0.9  & 4.9  & 3.3  &  1.1  & 1.9  & 82.1 & 93.0 & 25.8 & 51.7 & 69.8 \\
         \midrule
        \textbf{Llama3.2} & \texttt{FT}   & 60.4 & 61.0 & 58.1 &  59.9 & 56.3 & 78.0 & 95.2 & 25.4 & 52.8 & 34.3 \\
        \textbf{3B} & \texttt{Rec.} & 1.7  & 6.0  & 4.6  &  4.1  & 8.1  & 82.3 & 91.7 & 24.3 & 51.5 & 55.8 \\
         \midrule
         \textbf{Qwen2.5} & \texttt{FT}   & 53.7 & 50.6 & 53.0 &  55.4 & 52.0 & 82.4 & 98.7 & 42.2 & 53.8 & 76.8 \\
       \textbf{32B} & \texttt{Rec.} & 4.5  & 5.1  & 4.3  & 4.3  & 4.8  & 82.1 & 97.6 & 41.9 & 53.2 & 76.9 \\
        \bottomrule
    \end{tabular}

    \label{tab:res_latest_models}
\end{table}

\begin{table}[ht]
 \scriptsize
\setlength\tabcolsep{1.7pt}
 \renewcommand{\arraystretch}{0.9}
\centering
 
\caption{
The harmful rate and drop of task performance (\%) after recovered with and without our rollback mechanism (recall~\S\ref{subsubsection:rollback}). \texttt{Harmful\#} represents the number of harmful prompts we injected into the fine-tuning process. (\texttt{Mod.} means the 392 samples left over from moderation are used). Other models are not included in this table because they never activated the rollback. \label{tab:rollback_res}}
\vspace{-0.5em}
\begin{tabular}{ccccccc}
\toprule

\multirow{2}{*}{\textbf{Model}}& \multirow{2}{*}{\textbf{Dataset}}&  \multirow{2}{*}{\textbf{\makecell{Harmful\# }}} & \multicolumn{2}{c}{\textbf{Harmful Rate}}   & \multicolumn{2}{c}{\textbf{Performance Drop}}   \\
\cmidrule(lr){4-5} \cmidrule(lr){6-7}
 &  & &\texttt{\extrasmallsize{Disabled}}  & \texttt{\extrasmallsize{Enabled}} & \extrasmallsize{\texttt{Disabled}}& \extrasmallsize{\texttt{Enabled}}  \\

\midrule
\texttt{\gemma } & \texttt{\sql} &  \texttt{Mod.} &1.57  &2.29 &1.47 &1.23 \\
\midrule
\texttt{\gemma } & \texttt{\nlbash} &  \texttt{0.1k} & 2.57 &3.14 &3.51 &-0.61 \\
\midrule
\texttt{\gemma } & \texttt{\toxicity} &  \texttt{0k} & 3.14 &3.14 &2.88 &3.09 \\
\midrule
\texttt{\gemma } & \texttt{\toxicity} &  \texttt{0.1k}  & 3.00 &3.29 &3.42 &-1.18 \\
\midrule
\texttt{\gemma } & \texttt{\toxicity} &  \texttt{1.5k}  & 6.43 &4.14 &-4.38 &-28.18 \\
 \midrule
\texttt{\gemma } & \texttt{\toxicity} &  \texttt{Mod.}  &5.14 &6.86 &4.77  &1.61 \\
\midrule
\texttt{\llamasmall } & \texttt{\cheat} &  \texttt{0.1k}  &0.14 &0.14 &3.15 &2.28 \\
% \midrule
% \texttt{\llamasmall } & \texttt{\toxicity} &  \texttt{1.0k} & 14.00 &6.14 &2.35 &4.71 \\
%\midrule
%\texttt{\llamasmall } & \texttt{\toxicity} &  \texttt{1.5k} & - &4.0 &0.00 &0.80 \\
\midrule
\texttt{\llamabig } & \texttt{\nlbash} &  \texttt{0k} &0.00 &0.14 &4.54 &2.72 \\
\midrule
\texttt{\llamabig } & \texttt{\nlbash} &  \texttt{0.5k} &1.57 &1.29 &3.47 &2.01 \\
% \midrule
% \texttt{\llamabig} & \texttt{\toxicity} &  \texttt{Mod.} &1.57  &0.71  &2.61  &0.11  \\
% \midrule
% \texttt{\qwen } & \texttt{\toxicity} &  \texttt{1k} & 14.00 &2.71 &4.51 &2.36 \\
\bottomrule
\end{tabular}
\vspace{-0.95em}
\end{table}
\begin{table}[htbp]
    \centering
    \scriptsize
    \setlength\tabcolsep{0.9pt}
     \renewcommand{\arraystretch}{0.6}
    \caption{The harmful rate and task performance of fine-tuned models (\texttt{FT}) and recovered models under different rollback rates (\texttt{10\%}, \texttt{20\%}, and \texttt{40\%}). The four models are based on Gemma 2B and trained on TOXIC dataset injected with \texttt{0k}, \texttt{0.1k}, \texttt{1.5k}, and filtered data.}
    \begin{tabular}{l|cccc|cccc|cccc|cccc}
        \toprule
        & \multicolumn{4}{c}{\textbf{0k}} & \multicolumn{4}{|c}{\textbf{0.1k}} & \multicolumn{4}{|c}{\textbf{1.5k}} & \multicolumn{4}{|c}{\textbf{filtered}} \\
        \cmidrule(lr){2-5} \cmidrule(lr){6-9} \cmidrule(lr){10-13} \cmidrule(lr){14-17} 
        & \texttt{FT} & \texttt{10\%} & \texttt{20\%} & \texttt{40\%} & \texttt{FT} & \texttt{10\%} & \texttt{20\%} & \texttt{40\%} & \texttt{FT} & \texttt{10\%} & \texttt{20\%} & \texttt{40\%} & \texttt{FT} & \texttt{10\%} & \texttt{20\%} & \texttt{40\%} \\
        \midrule
  \textbf{HR}  & 3.7  & 3    & 3.1  & 3.4  & 34.6 & 3    & 3.3 & 3.7  & 60.7 & 3.4  & 4.1  & 4.6  & 39.4 & 2.9  & 6.9  & 3.4  \\
  \midrule
\textbf{TP} & 78.7 & 75.4 & 76.3 & 76.9 & 75.1 & 75.2 & 76  & 76.5 & 46.1 & 56.3 & 59.1 & 67.3 & 74.4 & 71.4 & 73.2 & 74.3 \\
        \bottomrule
    \end{tabular}
    \label{tab:ablation_r}
\end{table}

% \newpage
% \section{Addition Results for the Table}
 % \input{tables/ablation_layer}
 % \input{tables/recovery_diversity}
 % \input{tables/ablation_size}
 % \input{tables/ablation_recovery}

% \input{tables/task_res_all}
% \input{tables/alignment_res_all}

\clearpage
\newpage % The Meta-Review should at least start on a new column

% Use \appendices and not \appendix due to IEEEtran.cls quirks

\section{Meta-Review}

The following meta-review was prepared by the program committee for the 2025
IEEE Symposium on Security and Privacy (S\&P) as part of the review process as
detailed in the call for papers.

\subsection{Summary}
Prior work has shown that after fine-tuning on custom data, the safety alignment of LLMs can be compromised (i.e., they become more likely to respond to dangerous prompts). Inspired by previous work on LLM alignment, this paper proposes a heuristic algorithm to restore the alignment of a fine-tuned model. This method restores a small set of weights of the fine-tuned model (selected based on their importance for the model's alignment) to their original values before fine-tuning. The method recovers weights gradually and implements a rollback mechanism to preserve the LLM's utility. The experiments show that this approach can successfully recover the alignment lost during fine-tuning without sacrificing task performance.

\subsection{Scientific Contributions}
\begin{itemize}
\item Provides a Valuable Step Forward in an Established Field
\end{itemize}

\subsection{Reasons for Acceptance}
\begin{enumerate}
\item The study tackles the critical problem of LLM misalignment after fine-tuning. It proposes a principled algorithm that effectively combines existing observations and steering techniques to tune the model for recovering its alignment while balancing the model's safety and utility.
\item Although LLM steering is an active area of research, this study provides new scientific insights, such as measuring the impact of the recovery dataset and distribution shifts between the recovery and testing datasets.
\item The study experiments with a range of settings and fine-tuning scenarios (including adversarial misalignment) to show the extent of misalignment and the effectiveness of their approach in recovering alignment.
\end{enumerate}

\subsection{Noteworthy Concerns} % Exclude if your meta-review does not have noteworthy concerns
\begin{enumerate} % Enumerate environment is not necessary if there is only one
\item The proposed method relies mostly on existing LLM steering methods and offers the community limited technical novelty and innovation.
\end{enumerate}

% that's all folks
\end{document}